\documentclass[a4paper,11pt]{article}
\usepackage{jheppub}
\usepackage{amsmath}
\usepackage[numbers,sort&compress]{natbib}
\usepackage{soul}
\usepackage{xcolor}
\usepackage{float}
\usepackage{dsfont}
\usepackage{tikz}
\usetikzlibrary{arrows.meta, positioning}
\usepackage{amssymb}
\usepackage{physics}
\usepackage{hyperref}
\usepackage{bm,graphicx,graphics,epsfig}
\usepackage{dcolumn}
\usepackage{lineno}
\usepackage{subfigure} 
\usetikzlibrary{calc}
\usetikzlibrary{shapes}
\usetikzlibrary{calc,decorations.markings,decorations.pathmorphing}
\tikzset{cross/.style={cross out, draw=black, minimum size=2*(#1-\pgflinewidth), inner sep=0pt, outer sep=0pt}, cross/.default={1pt}}
\usepackage{lmodern} 
\usetikzlibrary{arrows.meta, positioning, decorations.pathreplacing, calligraphy}

\tikzset{
    time_axis/.style={-Latex, thick},
    time_tick/.style={draw, fill=black, circle, inner sep=1.5pt},
    boundary_node/.style={draw=red!80!black, fill=red!60, circle, inner sep=2.5pt, thick},
    internal_node/.style={draw=blue!80!black, fill=blue!60, circle, inner sep=2pt, thick},
    path_segment/.style={blue!80!black, thick},
    integration_axis/.style={gray, dashed},
    annotation_text/.style={font=\small},
    info_box/.style={
        draw=gray, 
        rounded corners, 
        inner sep=8pt, 
        text width=11cm, 
        align=center
    }
}

\makeatletter
\gdef\@fpheader{} 
\makeatother
\newcommand{\be}{\begin{equation}}
\newcommand{\ee}{\end{equation}}
\newcommand{\benn}{\begin{equation*}}
\newcommand{\eenn}{\end{equation*}}
\newcommand{\bea}{\begin{eqnarray}}
\newcommand{\eea}{\end{eqnarray}}

\newcommand{\mA}{\mathcal{A}}

\newcommand{\mD}{\mathcal{D}}

\newcommand{\mZ}{\mathcal{Z}}

\DeclareMathOperator{\im}{Im}

\DeclareMathOperator{\sgn}{sgn}

\title{Constrained Symplectic Quantization I:\\ the Quantum Harmonic Oscillator}

\author[a,b]{Martina Giachello}
\author[c,d]{Francesco Scardino}
\author[e]{Giacomo Gradenigo}
\affiliation[a]{Gran Sasso Science Institute, Viale F. Crispi 7, 67100 L'Aquila, Italy}
\affiliation[b]{INFN-Laboratori Nazionali del Gran Sasso, Via G. Acitelli 22, 67100 Assergi (AQ), Italy}
\affiliation[c]{Physics Department, INFN Roma1, Piazzale A. Moro 2, Roma, I-00185, Italy}
\affiliation[d]{Physics Department, Sapienza University, Piazzale A. Moro 2, Roma, I-00185, Italy}
\affiliation[e]{Physics and Astronomy Department "Galileo Galilei", Universit\`a di Padova, Via Marzolo 8, 35131 Padova, Italy}

\emailAdd{martina.giachello@gssi.it}
\emailAdd{francesco.scardino@roma1.infn.it}
\emailAdd{giacomo.gradenigo@gssi.it}

\abstract{
Symplectic quantization is a functional approach to quantum field theory that allows sampling of quantum fluctuations directly in Minkowski space-time by means of a generalized microcanonical ensemble similar to the one of the standard microcanonical approach to lattice field theory. In a previous paper~\cite{Giachello:2024wqt} we showed that, for an interacting scalar field theory in 1+1-dimensions, this formalism allows to capture numerically some crucial real-time features inaccessible to any Euclidean approach to lattice field theory. Yet, the new approach was plagued by two main limitations: an ill-defined non-interacting limit and the absence of a direct formal correspondence between its correlation functions and those generated by the Feynman path integral approach.\\
In this paper, we introduce the new \emph{"constrained symplectic quantization"} approach, for which the perfect equivalence with the Feynman path integral is proved and which is perfectly well defined for the free theory. This new approach is characterized by the analytic continuation of all fields and of the action from $\mathbb{R}$ to $\mathbb{C}$ and the presence of some constraints which guarantee the stability of the generalized Hamiltonian dynamics and the convergence of the corresponding generalized microcanonical partition function, hence the name of the theory. \\
We show the application of this formalism to the quantum harmonic oscillator on a Minkowskian-time lattice, finding perfect agreement between one- and two-point numerical correlators and the exact quantum-mechanical results.  We observe genuine real-time features such as the oscillatory propagator and the discrete excited-state energy spectrum. Our results provide strong numerical evidence that constrained symplectic quantization can sample real-time quantum-mechanical observables, offering a concrete route to overcome the limitations of Euclidean-time importance sampling.
}

\begin{document}

\maketitle
\flushbottom

\section{Introduction}
\label{sec:intro}

Symplectic quantization is the proposal of a new functional approach
to quantum field theory put forward in a recent series of conceptual
papers~\cite{Gradenigo:2021sck,Gradenigo:2021orx,Gradenigo:2024pwy}
and backed by some encouraging numerical evidence, presented
in~\cite{Giachello:2024wqt,Giachello:2024twu,Giachello:2024mtl},
showing that symplectic quantization allows for a numerical
sampling of the real-time dynamics of a field theory closely related to
ordinary scalar field theory on a lattice. The present paper aims to fully clarify from the theoretical viewpoint the relation between
symplectic quantization and the Feynman path integral approach to
quantum mechanics, showing how, at least in the case of the harmonic
oscillator, it allows one to reproduce features of the quantum dynamics
that are inaccessible to importance sampling algorithms based on imaginary
time quantum mechanics \cite{Creutz:1980gp,Parisi:1980ys,Damgaard:1987rr}.
Symplectic quantization, which is closely
inspired by the microcanonical approach to quantum field theory~\cite{deAlfaro:1983zz,Strominger:1982xu,Iwazaki:1984kx}, is a
functional deterministic approach built on an extended variable space,
where the quantum fluctuations of fields are parametrized by an
additional time parameter $\tau$, referred to as the {\it intrinsic
  time}
in~\cite{Gradenigo:2021sck,Gradenigo:2021orx,Gradenigo:2024pwy}, and
governed by an intrinsic Hamiltonian dynamics. For instance, in the
symplectic quantization approach, the position quantum operator
$\hat{q}(t)$ is replaced by the fluctuating field $q(t,\tau)$ \cite{Parisi:1980ys,Parisi:1983mg}, such
that the quantum fluctuations at a given Minkowskian time $t$ are those
generated by the Hamiltonian equations
\begin{equation}\label{eq:hamiltonbasic}
\frac{\partial q(t, \tau)}{\partial \tau} = \frac{\delta \mathbb{H}[q, \pi]}{\delta \pi(t, \tau)}, \quad \frac{\partial \pi(t, \tau)}{\partial \tau} = -\frac{\delta \mathbb{H}[q, \pi]}{\delta q(t, \tau)},
\end{equation}
where $\mathbb{H}[q, \pi]$ is a generalized Hamiltonian of the system of the form 
\begin{equation}\label{eq:oldansaltz}
\mathbb{H}[\pi, q] = \mathbb{K}[\pi] - S[q],
\end{equation}
with $S[q]$ the standard action of the problem considered, playing the
role of a potential, and $\mathbb{K}[\pi]$ is a generalized kinetic
energy, quadratic in the additional momenta $\pi(t,\tau) \propto
\partial q(t,\tau)/\partial\tau$. It is immediate to see that if one
replaces quantum mechanics with quantum field theory, $q(t,\tau)
\rightarrow \phi(x,\tau)$, where $x$ is a coordinate in Minkowski
space-time, the above construction is completely analogous to the
microcanonical approach to quantum fields used in numerical
lattice field theory~\cite{Callaway:1983, Iwazaki:1984kx, Duane:1987de}, with the only difference that the potential
term is not the Euclidean action $S_E[\phi]$ but minus the
Minkowskian action $-S[\phi]$. The striking result
of~\cite{Giachello:2024wqt,Giachello:2024twu,Giachello:2024mtl} is
that the above procedure, when realized as a numerical protocol to
sample the quantum fluctuations of a $\lambda \phi^4$ theory on a
$1+1$ lattice allows for a qualitative reproduction of the shape of the
free Feynman propagator when a small non-linear coefficient
$\lambda$ is employed.\par 

Despite this numerical success,
in~\cite{Giachello:2024wqt} we
have also explained the theoretical limitations of this approach.
Assuming a bona fide ergodic hypothesis for the trajectories
generated by Eq.~\eqref{eq:hamiltonbasic}, ubiquitous in the
microcanonical approach to quantum field theory, one claims
that the quantum fluctuations sampled are those corresponding
to the microcanonical measure:
\begin{align}
\varrho_{\text{micro}}[q,\pi] = \frac{1}{\Omega(\mA)}~\delta(\mathbb{H}[q, \pi] - \mA),
\end{align}
where the normalization is
\begin{align}
\Omega[\mA] = \int \mathcal{D}q\mathcal{D}\pi~\delta(\mathbb{H}[q, \pi] - \mA).
\end{align}
In~\cite{Giachello:2024wqt,Giachello:2024twu},
as a corollary of the numerical results, it is proved that, for a
system with $M$ degrees of freedom, in the large-$M$ limit the
microcanonical measure corresponds to a canonical one of the kind
\begin{align}
  M\rightarrow\infty ~~\Longrightarrow~~\varrho_{\text{micro}}[q,\pi] \longrightarrow \exp{S[q]/\hbar},
\label{eq:micro-equivalence}
\end{align}
where $\hbar = A/M$ is the unit of action. The equivalence of
Eq.~\eqref{eq:micro-equivalence} contains two important pieces of
information: first, it is perfectly consistent with the numerical
results contained in the same paper, where we found a Hamiltonian
dynamics which is stable only for a lower-bounded non-linear potential
in the action, while it is unstable for a free action: it must be in fact recalled that a free relativistic action $S[\phi]$ is not positive definite. Second, the
procedure does not directly sample the Feynman path integral
measure, which is proportional to $\exp(iS[\phi]/\hbar)$, but a different object, $\exp(S[\phi]/\hbar)$.\par 

The purpose of the present discussion is to demonstrate how
the symplectic quantization approach can be further extended with
respect to the form in which we have proposed it
in~\cite{Giachello:2024wqt,Giachello:2024twu,Giachello:2024mtl}. This
extension furnishes a stable dynamics capable of numerically sampling
quantum fluctuations on a lattice, even in the case of a free action
and, most importantly, provides expectation values which are precisely
those obtained from the Feynman path integral approach \cite{Feynman1948}. The logic
followed by the extension of the symplectic quantization formalism
proposed here is simple and in agreement with recent developments to circumvent the sign problem in numerical field theory \cite{Alexandru:2020wrj}: to sample the correct quantum
fluctuations we consider an analytic prolongation to the complex plane of all fields. For instance, for the position field in quantum mechanics, we consider $q(t,\tau)\in \mathbb{C}$ rather than $q(t,\tau)\in
\mathbb{R}$, so that we also have $S[q] \in \mathbb{C}$.  Then, we
consider a {\it real} generalized Hamiltonian and a {\it constrained}
deterministic dynamics such that stability is guaranteed, as well as a
direct correspondence with the path integral measure. It is worth
noticing that the logic of the constrained symplectic quantization approach
has profound analogies with the evaluation of path integrals with the
Lefschetz-Thimbles strategy~\cite{Cristoforetti:2012su, Alexandru:2020wrj, Scorzato:2015qts, Behtash:2015loa}, which considers the analytic
continuation of all fields/variables in the complex plane \cite{Witten:2010cx}, and then
performs the integration along constrained paths to guarantee the convergence of functional integrals. The main difference between Symplectic Quantization and the 
Lefschetz-thimble approach is that while in the latter one puts constraints on the path integral
evaluation, where the path integral plays the role of a partition function,
in the symplectic quantization approach, one puts constraints directly on the microscopic dynamical processes which generate the path integral ``measure''. As such, and by virtue of its striking ability to reproduce quantum mechanical
expectation values, we believe that the present approach could be a
new interesting proposal to handle the sign problem in many instances in numerical field
theoretical and many-body quantum dynamics~\cite{Gattringer:2016kco, Troyer:2004ge, Loh:1990zz}.\\

This paper is therefore dedicated to illustrating all the analytical
subtleties of our new {\it constrained symplectic quantization}
approach and to numerically validate it in a prototypical reference
case: the quantum harmonic oscillator.\\

The theoretical characterization of the constrained symplectic quantization approach proceeds in two main steps. First, we define the analytic continuation of fields and action to the complex plane and introduce a generalized Hamiltonian, proving that the associated microcanonical measure generates connected field correlators that are equivalent to those obtained from the Feynman path integral. These points are discussed in Sec.~\ref{sec:formulation} and demonstrated in Sec.~\ref{sec:equivalence}. Second, we identify the class of microscopic constraints that bound the Hamiltonian dynamics, which define the corresponding integration contours allowing the sampling of the microcanonical partition function and by extension the Feynman path-integral. This is discussed in detail in Sec.~\ref{sec:Harm-Osc-dynamics}, with a focus--both for clarity and for practical purposes--on the system studied in this work: the quantum harmonic oscillator. Sec.~\ref{sec:numerical-algorithm} then explains how the constrained dynamics introduced in Sec.~\ref{sec:Harm-Osc-dynamics} is implemented within the numerical symplectic algorithm used to integrate the intrinsic-time evolution. Finally, Sec.~\ref{sec:numerical-results} presents numerical results showing that the symplectic quantization approach yields an algorithm capable of capturing the real-time dynamics of the system and reproducing its excited-state structure. In particular, we compute the two-point correlation function $\langle \hat{q}(t)\hat{q}(t')\rangle$, we numerically verify the canonical commutation relation $\langle [\hat{q}(t),\hat{p}(t)]\rangle = i\hbar$, and we study $\langle \hat{q}^{2n}(t)\rangle$. The Fourier transform of the latter with respect to Minkowskian time exhibits a peak structure that matches precisely the excited states of the harmonic oscillator.

\section{Fields and Action: analytic continuation from $\mathbb{R}$ to \texorpdfstring{$\mathbb{C}$}.}
\label{sec:formulation}

We begin by motivating the choice of the new generalized Hamiltonian
$\mathbb{H}_{\text{SQ}}$, which is required once all dynamical variables
are analytically continued to the complex plane. Three main reasons
motivate, at the same time, both the need for a new generalized Hamiltonian
$\mathbb{H}_{\text{SQ}}$---different from the one considered in our previous
work~\cite{Giachello:2024wqt}---and the necessity to analytically continue
all variables to $\mathbb{C}$.

In~\cite{Giachello:2024wqt} we showed that, by fixing the generalized action
$\mA$ to the real value $\mA=M\hbar$, for an interacting field on a $1+1$
lattice with $M$ degrees of freedom it is possible to reproduce
numerically (at least qualitatively) the shape of the free Feynman
propagator by averaging the two-point correlation function along the
Hamiltonian dynamics. Nevertheless, an \emph{exact} correspondence between
quantum field theory in the path-integral formulation and the microcanonical
approach considered in~\cite{Giachello:2024wqt} cannot be established as long
as one employs a real generalized action, $\mathcal{A}\in\mathbb{R}$,
together with real fields, $\phi\in\mathbb{R}$. Indeed, the calculation
presented in~\cite{Giachello:2024wqt} shows that, in the continuum limit
$M\to\infty$, the microcanonical partition function $\Omega[\hbar M]$
corresponds to a Minkowskian statistical mechanics, rather than to quantum
field theory.

From the expression of the microcanonical partition function computed
in~\cite{Giachello:2024wqt} in the large-$M$ limit,
\begin{align}
\Omega[\hbar/z,J] = \int \mathcal{D}\phi~\exp\left( \frac{z}{\hbar} S[\phi] +\frac{z}{\hbar} J\phi \right),
\end{align}
one sees that an exact equivalence with quantum field theory would be recovered
only by fixing the generalized action of the microcanonical approach to the
purely imaginary value $\mA=-i\hbar M$, so that
\begin{align}
M\rightarrow\infty ~~~\Longrightarrow~~~\log\Omega[\mA =- i\hbar M] \sim \log\mathcal{Z}[\hbar].
\end{align}
However, a generalized action of the form $\mathbb{H}[\pi,q]=\mathbb{K}[\pi]-S[q]$
cannot be fixed by construction to purely imaginary values, since it is a real
functional of real fields. Therefore, the microcanonical Minkowskian approach of~\cite{Giachello:2024wqt}, despite allowing for remarkable advances with respect
to the standard canonical Euclidean approach, must be extended in order to define a
statistical ensemble in direct correspondence with quantum field theory---and, in the
non-relativistic limit, with quantum mechanics.
To pursue this goal we show here, focusing on the paradigmatic case of the quantum
harmonic oscillator, that an analytic continuation from $\mathbb{R}$ to $\mathbb{C}$
is needed both for the action $S$ and for all fields entering the theory. Despite
the analytic continuation of fields and of the action $S$ to the complex domain, the
theory must still fix the generalized Hamiltonian to a \emph{real} value
$\mathbb{H}_{\text{SQ}}=\mA$, since only in this case it is possible to impose a
physical ``quantization'' constraint of the form $\mA=\hbar M$. For this reason, a
new form of the generalized action/Hamiltonian $\mathbb{H}_{\text{SQ}}$ must be
chosen with respect to the one introduced in~\cite{Giachello:2024wqt}.

Only one previous work discussed an extension of the symplectic quantization approach
in the context of quantum mechanics~\cite{Gradenigo:2024pwy}. There it was argued that,
in order to reproduce quantum fluctuations with a deterministic dynamics, the quantum
observable $\hat q(t)$ should be promoted to a \emph{position} field $q(t,\tau)$,
accompanied by its \emph{conjugate momentum} field $\pi(t,\tau)$. However, as shown
numerically and analytically in~\cite{Giachello:2024wqt}, this procedure does not, by
construction, yield a direct correspondence between quantum-mechanical fluctuations
and the fluctuations sampled by the symplectic quantization approach. Here we propose
that such a correspondence can be achieved only by also considering the analytic
continuation from $\mathbb{R}$ to $\mathbb{C}$ of the fields $q(t,\tau)$ and
$\pi(t,\tau)$:
\begin{align}
  q(t,\tau) \in \mathbb{R} ~~~&\longrightarrow~~~~q_R(t,\tau) + i\,q_I(t,\tau) ~\in~\mathbb{C}, \nonumber \\
  \pi(t,\tau) \in \mathbb{R} ~~~&\longrightarrow~~~~\pi_R(t,\tau) + i\,\pi_I(t,\tau) ~\in~\mathbb{C}.
\end{align}
In addition, we introduce a new generalized action for the symplectic quantization
approach, which from here on we denote by $\mathbb{H}_{\text{SQ}}$,
\begin{align}
  \mathbb{H}_{\text{SQ}}[\pi, \bar{\pi}, q, \bar{q}] = \int dt\, \bar{\pi}(t,\tau) \pi(t,\tau) + 2\,\Im S[q,\bar{q}].
  \label{eq:csq-hamiltonian}
\end{align}
We initially present the generalized Hamiltonian in Eq.~\eqref{eq:csq-hamiltonian}
as an ansatz; its correctness will be proven in Sec.~\ref{sec:equivalence}, where we
show that it guarantees a direct correspondence between the microcanonical partition
function of symplectic quantization and the Feynman path integral. In
Eq.~\eqref{eq:csq-hamiltonian}, $\Im S[q,\bar{q}]$ is shorthand for
\begin{equation}
  \Im S[q,\bar{q}] = \frac{S[q] - \bar{S}[\bar{q}]}{2i}.
  \label{eq:Sqq-1}
\end{equation}
In Sec.~\ref{sec:Harm-Osc-dynamics} we discuss the equations of motion generated by
the Hamiltonian in Eq.~\eqref{eq:csq-hamiltonian}. By imposing suitable symmetries
between the real and imaginary parts of the fields, these equations allow one to
sample precisely the quantum fluctuations encoded in the Feynman path integral.

For generic systems, the generalized action/Hamiltonian of symplectic quantization---as
in Eq.~\eqref{eq:csq-hamiltonian}---will be built out of two key ingredients: a
generalized \emph{real} kinetic energy term $\mathbb{K}[\pi]$, which is a quadratic
form in the momenta $\pi$, and the imaginary part of the analytically continued action
$S[q,\bar{q}]$. We will often refer to the latter as the \emph{holomorphic} action,
since its analytic continuation from $\mathbb{R}$ to $\mathbb{C}$ implicitly assumes
that it defines a holomorphic functional of the corresponding fields.

\section{Microcanonical action and the Feynman path integral}
\label{sec:equivalence}

In this section, we prove the equivalence in the continuum limit
between the correlation functions generated by the microcanonical partition function built on the
conservation of the generalized action $\mathbb{H}_{\text{SQ}}$
introduced in Sec.~\ref{sec:formulation} and the Feynman path
integral. Differently from our previous work~\cite{Giachello:2024wqt},
we specialize here from the beginning of our discussion to quantum
mechanics, which in the present formalism is treated as
$0+1$-dimensional quantum field theory. Our purpose is to prove that,
by fixing the value of the generalized action to $\mA = \hbar M$ in
the microcanonical partition function $\Omega(\mA)$, with $M$ the
number of points in the discretization of the time axis $t$, in the
continuum limit $M\rightarrow\infty$ the correlation functions
generated by $\Omega[\mA,J]$ are identical to the correlation
functions generated by $\mathcal{Z}[\hbar,J]$, where
$\mathcal{Z}[\hbar,J]$ is the Feynman path integral of the
corresponding theory. Namely, what we are going to prove is that 
\begin{align}
  \lim_{M\rightarrow\infty}\frac{\delta^n\Omega[\hbar M,J]}{\delta q(t_{1})\ldots\delta q(t_{n})}\Bigg|_{J=0} =
  \left\langle  q(t_1) \ldots q(t_n) \right\rangle = 
  \frac{\delta^n\mathcal{Z}[\hbar,J]}{\delta q(t_{1})\ldots\delta q(t_{n})}\Bigg|_{J=0}.
  \label{eq:equivalence-sq-fey}
\end{align}
For this purpose, let us introduce the generalized microcanonical
generating functional of correlators of symplectic quantization as
\begin{align}
\Omega[\mA,J] = \frac{1}{\Omega[\mA,0]} \int_{\boldsymbol{\Gamma}(t)} \mathcal{D}\bar{\pi}\,\mathcal{D}\pi\,\mathcal{D}\bar{q}\,\mathcal{D}q\;
\delta\!\Big(\mA - \mathbb{H}[q,\bar{q},\pi,\bar{\pi}] + i\, J\cdot q\Big)\,,
\label{eq:micro-part-func}
\end{align}
where $\boldsymbol{\Gamma}(t)$ denotes the set of one-dimensional integration contours in the complex plane along which the functional integral is convergent,
parametrized by the Minkowskian time $t$, a contour for each of the fields to be integrated,
$q(t)$, $\bar{q}(t)$, $\pi(t)$ and $\bar{\pi}(t)$:
\begin{align}
\boldsymbol{\Gamma}(t) = \Gamma_q(t) \cup \Gamma_{\bar{q}(t)} \cup \Gamma_\pi(t) \cup \Gamma_{\bar{\pi}}(t), 
\label{eq:contours}
\end{align}
where, for instance, we have
\begin{align}
\Gamma_q(t) = \left\lbrace \gamma_q(t) \in \mathbb{C}: t \in [t_{i},t_f] \right\rbrace,
\end{align}
and where $\gamma_q(t)$ is the integration contour for the position
field $q(t)$ at time $t$. Let us stress an
important point: the analytic continuation of the fields from $\mathbb{R}$ to $\mathbb{C}$ does not imply any doubling of degrees of freedom: the
integration domain for any given field at any given point of spacetime remains a one-dimensional path, with the only difference that the path is immersed in $\mathbb{C}$, rather than being simply the real line. The only additional degrees of freedom are represented by the
conjugate momenta $\pi(t)$, but these will be integrated out at a
certain point along the calculation.\\

In the following discussion on quantum mechanics we start by
discretizing the coordinate time $t$ in $M = (t_f-t_i)/a$ points $t_i$, with $i=1,\ldots,M$, so that also the
functional integral is discretized in $M$ ordinary integrals. The $M$ degrees of freedom to be integrated can be considered either as the value taken by the position field at each coordinate time, $q(t_i)$, or as the component of any orthonormal basis which can be used to
represent the position field. In analogy with the treatment discussed
in~\cite{Strominger:1982xu}, let us bring forward the argument by
considering a generic orthonormal basis, which for simplicity can be
thought as the Fourier components of discretized position field:
\begin{equation}
q(t_i) = \sum_{n=1}^M q_n(t_i)\, c_n, 
\qquad
\bar{q}(t_i) = \sum_{n=1}^M \bar{q}_n(t_i)\, \bar{c}_n,
\end{equation}
with 
\begin{equation}
\sum_{i=1}^M q_n(t_i) q_m(t_i) = \delta_{nm}, 
\qquad
\sum_{i=1}^M \bar{q}_n(t_i) \bar{q}_m(t_i) = \delta_{nm}.
\end{equation}
Having discretized the time axis $t$, the functional integration of
Eq.~\eqref{eq:micro-part-func} can be interpreted in terms of a finite
measure of the form
\begin{equation}
\int \mathcal{D} q\,\mathcal{D} \bar{q} \approx \int \mathcal{D}_M q\,\mathcal{D}_M \bar{q} \equiv~\prod_{n=1}^M \left(\int_{\Gamma_n} dc_n d\bar{c}_n\right) \equiv \prod_{i=1}^M \left( \int_{\Gamma_i} dq(t^{\text{i}}) d\bar{q}(t^{\text{i}})\right).
\end{equation}

Unlike the standard path-integral convention, this measure does not include an explicit factor of $\hbar$. In a time interval of length $T=(t_f-t_i)$ with lattice spacing $a$, the number of basis functions is~\cite{Strominger:1982xu}:

\begin{equation}\label{eq:Mlambda-complex}
M = \frac{T}{a} = T~\frac{\Lambda}{\pi},
\end{equation}
where $\Lambda = \pi/a$ acts as a momentum cutoff. The limit $M \to
\infty$ can be approached either via the continuum limit
$\Lambda\to\infty$ or the thermodynamic limit $T\to\infty$. From here on, the large-$M$ limit is understood in this sense.
Finally, adopting the notational shorthand
\begin{align}
\pi\cdot\bar{\pi} &\equiv \int dt\, \pi(t)\bar{\pi}(t)\qquad J\cdot q \equiv \int dt\, J(t)q(t)
\end{align}
we can express the microcanonical generating functional on the lattice
in a form suitable for the subsequent analysis in the large-$M$
limit.
As a first step, one can perform the integration over the momenta
$\pi$ and $\bar{\pi}$. This can be carried out either via a spherical
integration or, more conveniently, by expressing the delta function as
a Fourier integral:
\begin{align}
\Omega[\mA,J] = \frac{1}{\Omega[\mA,0]} \int 
\mathcal{D}_M\bar{\pi}\,\mathcal{D}_M\pi\,\mathcal{D}_M\bar{q}\,\mathcal{D}_M q\,d\lambda \;
e^{-i\lambda\, \bar{\pi}\cdot \pi + i\lambda(\mA - 2\,\im\, S[q,\bar{q}] + i J\cdot q )}\,.
\end{align}
The $\pi$'s complex Gaussian integral yields
\begin{align}
	\Omega[\mA,J] = \frac{1}{\Omega[\mA,0]}\int \mathcal{D}_M\bar{q}\mathcal{D}_Mq\,\mathcal{D}\lambda\, \lambda^{-M}\, e^{i\lambda(\mA-2\im S[q,\bar{q}]+i J\cdot q)}\,,
\end{align} 
where the $\lambda$ integral can be performed up to irrelevant constants that cancel by the denominator
\begin{align}
	\Omega[\mA,J] =  \frac{1}{\Omega[\mA,0]}\int \mathcal{D}_M\bar{q}\mathcal{D}_Mq\, \left(\mA-2\im S[q,\bar{q}]+i J\cdot q\right)^{M-1}\,.
\end{align} 

At this point we fix the generalized Hamiltonian to
$\mA=\hbar M$, which corresponds to assigning a value $\hbar$ to each
degree of freedom. By then plugging this value of the action into the expression of $\Omega[\mA,J]$ one gets 
\begin{align}
	\Omega[\mA = \hbar M,J]
	&=  \frac{1}{\Omega[\hbar M,0]}\int \mathcal{D}\bar{q}_M\mathcal{D}q_M\, \left(1-\frac{2}{\hbar M}\im S[q,\bar{q}]+\frac{i}{\hbar M}J\cdot q\right)^{M-1}\nonumber\\
	&= \frac{1}{\Omega[\hbar M,0]}\int \mathcal{D}\bar{q}_M\mathcal{D}q_M\,e^{(M-1)\log\left(1-\frac{2}{\hbar M}\im S[q,\bar{q}]+\frac{i}{\hbar M}J\cdot q\right)} 
\label{eq:Omega-integr-pi}
\end{align} 
where again all the multiplicative constants simplify with the
denominator.
Let us now expand Eq.~\eqref{eq:Omega-integr-pi} in powers of the source $J$, in order to recover the standard source expansion of a generating functional. Writing
\begin{equation}
\left(1-\frac{2}{\hbar M}\Im S[q,\bar q]+\frac{i}{\hbar M}J\!\cdot\! q\right)^{M-1}
=
\sum_{n=0}^{\infty}\binom{M-1}{n}\left(\frac{i}{\hbar M}J\!\cdot\! q\right)^n
\left(1-\frac{2}{\hbar M}\Im S[q,\bar q]\right)^{M-1-n},
\end{equation}
and using $(J\!\cdot\! q)^n=\int dt_1\cdots dt_n\,J(t_1)\cdots J(t_n)\,q(t_1)\cdots q(t_n)$, we obtain
\begin{align}
\Omega[\hbar M,J]
=
\frac{1}{\Omega[\hbar M,0]}
\sum_{n=0}^{\infty}\frac{1}{n!}\left(\frac{i}{\hbar M}\right)^n
\frac{\Gamma(M)}{\Gamma(M-n)}
\int dt_1\cdots dt_n\,J(t_1)\cdots J(t_n)\,
\Big\langle q(t_1)\cdots q(t_n)\Big\rangle_{M-n},
\label{eq:Omega-series}
\end{align}
where we used $\binom{M-1}{n}=\Gamma(M)/\big[n!\,\Gamma(M-n)\big]$ and where we have introduced the ``dressed'' correlator $\langle q(t_1)\ldots q(t_n) \rangle_{M-n}$:
\begin{align}
  \langle q(t_1)\ldots q(t_n) \rangle_{M-n} =
  \int \mathcal{D}_M\bar{q}\mathcal{D}_Mq~q(t_1)\ldots q(t_n)
  \left(1-\frac{2}{\hbar M}\im S[q,\bar{q}]\right)^{M-n}
\end{align}

and where we have neglected the $-1$ on the exponent since we are interested in the large-$M$ limit.
The calculation of $\langle q(t_1)\ldots q(t_n) \rangle_{M-n}$ will be
our next goal, with the purpose of substituting the result back into Eq.~\eqref{eq:Omega-series}. We proceed to expand $\left(1-\frac{2}{\hbar M}\im S[q,\bar{q}]\right)^{M-n}$
around $\frac{1}{M}$. 
Let $x=\frac{2}{\hbar}\im S[q,\bar{q}]$, then by using the notable limit of $\lim_{M\to \infty}(1-\frac{x}{M})^{M-n} = e^{-x}$ we separate the argument of the path integral as
\begin{equation}
	\left(1-\frac{x}{M}\right)^{M-n} = e^{-x}\times\left(1-\frac{x}{M}\right)^{M-n} e^x\,.
\end{equation}
Therefore, we now expand
\begin{equation}
	\left(1-\frac{x}{M}\right)^{M-n} e^x
\end{equation}
by convoluting the two series expansions
\begin{align}
  &\left(1-\frac{x}{M}\right)^{M-n}
   = \sum_{j=0}^{\infty}\binom{M-n}{j}\left(-\frac{x}{M}\right)^j \nonumber\\
  &e^x =\sum_{l=0}^\infty \frac{1}{l!} x^l
\end{align}
and by using the standard Cauchy formula, we get
\begin{equation}
	(1-\frac{x}{M})^{M-n} e^x = \sum_{j=0}^{\infty}x^j \sum_{k=0}^j \binom{M -n}{k}\frac{1}{(j-k)!}\frac{(-1)^k}{M^k}\,.
\end{equation}
Since $x= \frac{2}{\hbar} \Im S[q,\bar{q}]$, we define the coefficients $c_j(M,n)$ as
\begin{equation}\label{eq:coeff}
	c_j(M,n) = \left(\frac{2}{\hbar}\right)^j\sum_{k=0}^j \binom{M-n}{k}\frac{1}{(j-k)!}\frac{(-1)^k}{M^k}\,,
\end{equation}
and so
\begin{equation}
 \left(1-\frac{2}{\hbar M}\im S[q,\bar{q}]\right)^{M-n} =e^{-\frac{2}{\hbar}\im S[q,\bar{q}]} \sum_{j=0}^{\infty}c_j(M,n)\,\left(\im S[q,\bar{q}]\right)^j\,.
\end{equation}
We rewrite the expression for the {\it dressed} correlator as 
\begin{align}
	\langle q(t_1)\ldots q(t_n) \rangle_{M-n} &= \sum_{j=0}^\infty c_j(M,n)	\int \mathcal{D}_M\bar{q}\mathcal{D}_Mq~q(t_1)\ldots q(t_n)\,	e^{-\frac{2}{\hbar}\im S[q,\bar{q}]}  \left(\im S[q,\bar{q}]\right)^j\,\nonumber \\
	&= \sum_{j=0}^\infty	c_j(M,n)	\int \mathcal{D}_M\bar{q}\mathcal{D}_Mq~q(t_1)\ldots q(t_n)\,	e^{\frac{i}{\hbar} S[q]}e^{-\frac{i}{\hbar} \bar{S}[\bar{q}]}    \left(\frac{S[q]-\bar{S}[\bar{q}]}{2i}\right)^j\,.
	\label{eq:corr-expanded}
\end{align}
where again all the multiplicative constants simplify with the
denominator. As proven in App.~\ref{app:coeff}, in the large-$M$ limit decrease as
\begin{align}
c_j(M,n) \sim \frac{1}{M}\qquad \forall j\ge1\,.
\end{align}
In analogy with the Lefschetz-Thimbles approach discussed
in~\cite{Blau:2016vfc,Witten:2010zr,Witten:2010cx}, we define two independent integration contours $\Gamma_q$ and
$\Gamma'_{\bar{q}}$, respectively for $q$ and $\bar{q}$, which we therefore treat as separate variables (one can always define paths which guarantee convergence of functional integration, for
instance by choosing for $\Gamma_{q}$ the $i\epsilon$ rotated path
into the complex plane that recovers the Feynman prescription and for
$\Gamma'_{\bar{q}}$ the anti-holomorphic version for the $i\epsilon$
rotated path). It is then convenient to exploit the binomial theorem
\begin{align}
\left(\frac{S[q]-\bar{S}[\bar{q}]}{2i}\right)^j = \frac{1}{(2i)^j}\sum_{k=0}^{j}\binom{j}{k} S[q]^{j-k}S[\bar{q}]^k,
\end{align}
which allows to see that the integration with respect to $q$ and $\bar{q}$
inside the expression of Eq.~\eqref{eq:corr-expanded} can be neatly factorized, yielding 
\begin{align}
  & \langle q(t_1)\ldots q(t_n) \rangle_{M-n} = \nonumber \\
  & \sum_{j=0}^\infty c_j(M,n) \frac{1}{(2i)^j}\sum_{k=0}^{j}\binom{j}{k}~\int_{\Gamma_{q}} \mathcal{D}_Mq~q(t_1)\ldots q(t_n)~S[q]^{j-k}~e^{\frac{i}{\hbar}S[q]}~\times~\int_{\Gamma'_{\bar{q}}}\mathcal{D}_M\bar{q}~\bar{S}[\bar{q}]^{k}\,e^{-\frac{i}{\hbar} \bar{S}[\bar{q}]}.\nonumber \\
  \label{eq:dcorr-summation}
\end{align}
Let us then claim the finiteness of the holomorphic action $S[q]$ in the
continuum limit, which is a quite reasonable assumption for quantum mechanics. As a consequence, all terms of the sum in Eq.~\eqref{eq:dcorr-summation}
vanish in the continuum limit:
\begin{align}
M\rightarrow\infty~~~\Longrightarrow~~~c_j(M,n) S[q]^{j} \sim \frac{1}{M}~~~\forall~j\geq 1.
\label{eq:single-term-asympt}
\end{align}
The asymptotic behaviour of Eq.~\eqref{eq:single-term-asympt} then allows us to drastically simplify the behavior of the dressed correlator in the
large-$M$ limit: 
\begin{align}
\langle q(t_1)\ldots q(t_n) \rangle_{M-n} \cong \int_{\Gamma_{q}} \mathcal{D}_Mq~q(t_1)\ldots q(t_n)\,e^{\frac{i}{\hbar}S[q]}\times\int_{\Gamma'_{\bar{q}}}\mathcal{D}_M\bar{q}\, \,e^{-\frac{i}{\hbar} \bar{S}[\bar{q}]}   + O\left(\frac{1}{M}\right).
\end{align}
By then recalling the expression of the symplectic quantization action written as a function
of connected correlators in Eq.~\eqref{eq:Omega-series} and noticing that in the continuum
limit we also have 
\begin{equation}
	\lim_{M\to\infty}\left(\frac{1}{M}\right)^n\frac{\Gamma(M)}{\Gamma(M-n)}\to 1,
	\label{eq:coeffto1}
\end{equation}
we can then simplify as follows the expression of the symplectic quantization microcanonical partition function
\begin{align}
  &\Omega[\hbar M,J]= \nonumber \\
  & = \frac{1}{\Omega[\hbar M,0]}  \sum_{n=0}^{\infty}\frac{1}{n!}\left(\frac{i}{2}\right)^n\left(\frac{2}{\hbar M}\right)^n\frac{\Gamma(M)}{\Gamma(M-n)}\nonumber \times\int dt_1\ldots dt_n~J(t_1)\ldots J(t_n) \left\langle q(t_1)\ldots q(t_n)\right\rangle_{M-n} \nonumber \\
  & \cong \frac{1}{\Omega_\infty[\hbar,0]} \sum_{n=0}^{\infty}\frac{1}{n!}\left(\frac{i}{\hbar}\right)^n
  \int dt_1\ldots dt_n~J(t_1)\ldots J(t_n)\times\int_{\Gamma_{q}} \mathcal{D}q~q(t_1)\ldots q(t_n)\,e^{\frac{i}{\hbar}S[q]}\,\int_{\Gamma'_{\bar{q}}}\mathcal{D}\bar{q}\, \,e^{-\frac{i}{\hbar} \bar{S}[\bar{q}]}\,, \nonumber \\
&= \,\frac{1}{\Omega_{\infty}[\hbar,0]}  \int_{\Gamma_{q}} \mathcal{D}q~\,e^{\frac{i}{\hbar}S[q]}\,\sum_{n=0}^{\infty}\frac{1}{n!}\left(\frac{i}{\hbar}\right)^n\int dt_1\ldots dt_n~J(t_1)\ldots J(t_n)\,q(t_1)\ldots q(t_n)\int_{\Gamma'_{\bar{q}}}\mathcal{D}\bar{q}\, \,e^{-\frac{i}{\hbar} \bar{S}[\bar{q}]} \nonumber\\
&= \,\frac{1}{\Omega_{\infty}[\hbar,0]}  \int_{\Gamma_{q}} \mathcal{D}q~\,e^{\frac{i}{\hbar}S[q]+\frac{i}{\hbar}\int dt\, J(t)q(t)}\,\int_{\Gamma'_{\bar{q}}}\mathcal{D}\bar{q}\, \,e^{-\frac{i}{\hbar} \bar{S}[\bar{q}]} \nonumber\\
  &= \,\frac{ \int_{\Gamma_{q}} \mathcal{D}q~\,e^{\frac{i}{\hbar}S[q]+\frac{i}{\hbar}\int dt\, J(t)q(t)}\,\int_{\Gamma'_{\bar{q}}}\mathcal{D}\bar{q}\, \,e^{-\frac{i}{\hbar} \bar{S}[\bar{q}]}}{\int_{\Gamma_{q}} \mathcal{D}q~\,e^{\frac{i}{\hbar}S[q]}\,\int_{\Gamma'_{\bar{q}}}\mathcal{D}\bar{q}\, \,e^{-\frac{i}{\hbar}\bar{S}[\bar{q}]}}.
\end{align}
It must be then noticed that the term containing the anti-holomorphic part of the action is factorized both at numerator and denominator, so that it cancels out, yielding the final result
\begin{align}
  \lim_{M\rightarrow\infty} \Omega[\hbar M,J] = \Omega_\infty[\hbar,J] = \frac{ \int_{\Gamma_{q}}\mathcal{D}q~\,e^{\frac{i}{\hbar}S[q]+\frac{i}{\hbar}\int dt\,J(t)q(t)}}{\int_{\Gamma_{q}} \mathcal{D}q~\,e^{\frac{i}{\hbar}S[q]}}.
\end{align} 
Thus, by choosing $\Gamma_{q}$ as the $i\epsilon$ "Feynman" path, we have proved that 
\begin{equation}  \lim_{M\rightarrow\infty}\frac{\delta^n\Omega[\hbar M,J]}{\delta q(t_{1})\ldots\delta q(t_{n})}\Bigg|_{J=0}
  = \frac{\delta^n\Omega_\infty[\hbar,J]}{\delta q(t_{1})\ldots\delta q(t_{n})}\Bigg|_{J=0} =
  \frac{\delta^n\mathcal{Z}[\hbar,J]}{\delta q(t_{1})\ldots\delta q(t_{n})}\Bigg|_{J=0}
\end{equation}
namely that in the continuum limit $M\rightarrow\infty$ the
microcanonical generating functional is equivalent to the Feynman path
integral. The proof just presented is crucial, as it places on firm grounds the correspondence between the microcanonical partition function of symplectic quantization and the Feynman path integral. At the same time, the argument is somewhat formal, since it does not indicate how to choose the integration contours in the original microcanonical expression in order to evaluate it directly, without reverting to the Feynman path integral.
In the path-integral framework, an appropriate deformation of integration contours in the complex plane underlies the modern Lefschetz-thimble strategy; an extensive account can be found, for instance, in~\cite{Alexandru:2018gsd}. In our approach, a direct and constructive way to compute (or, more precisely, to sample) the microcanonical partition function is provided instead by the underlying microscopic Hamiltonian dynamics. This dynamics offers a natural mechanism to select the relevant integration cycles in the complexified field space.
We discuss this point in detail in the next section, Sec.~\ref{sec:Harm-Osc-dynamics}, focusing on the quantum harmonic oscillator. As we shall show, the key requirement is to \emph{constrain} the Hamiltonian flow to the appropriate stable manifold. This is precisely the reason for the name “constrained symplectic quantization”.

\section{Harmonic Oscillator: constrained equations of motion}
\label{sec:Harm-Osc-dynamics}

In the previous section we proved that the symplectic-quantization microcanonical functional and the Feynman path integral are equivalent as generating functionals for quantum correlators. In this section we introduce a \emph{constrained} Hamiltonian dynamics—generated by the generalized Hamiltonian in Eq.~\eqref{eq:csq-hamiltonian}—as a natural way to sample the measure associated with the corresponding microcanonical partition function.
To this purpose let us now specify the system: a one-dimensional
non-relativistic particle of mass $m$ moving in a fixed potential
$V(q)$. As for $V(q)$ we start our investigation from the harmonic oscillator
potential
\begin{align}
V(q) = \frac{1}{2} m \Omega^2 q^2.
\end{align}
The harmonic oscillator is chosen as a benchmark due to its exact
solvability, well-known spectral properties, and central role in both
quantum mechanics and quantum field theory. It provides a controlled
setting to test our formalism and to illustrate how real-time
evolution can be simulated deterministically without encountering the
sign problem that typically plagues path integrals with a Minkowskian
action. In order to avoid any ambiguity when speaking about {\it
  ``time evolution''} and {\it ``dynamics''}, let us stress that in
the following such concepts will be always be referred to the
intrinsic time $\tau$ flow, whereas the Minkowskian, or coordinate,
time $t$, in order to avoid confusion will be always denoted from
hereafter as $x_0$ and treated as a spatial coordinate, an approach
which also emphasizes the role of quantum mechanics as a
$0+1$-dimensional field theory. We therefore write the action of the
one-dimensional harmonic oscillator as
\begin{align}
  S[q] &= \int_{t_i}^{t_f} dt \left[ \frac{m}{2} \left( \frac{dq(t)}{dt} \right)^2 -  \frac{1}{2} m \Omega^2 q^2(t) \right] \nonumber \\
  &=\int_{x_0^{i}}^{x_0^{\text{f}}} dx_0 \left[ \frac{m}{2} \left( \frac{dq}{dx_0} \right)^2 -  \frac{1}{2} m \Omega^2 q^2(x_0) \right].
\label{eq:action}
\end{align}
The novelty of the symplectic quantization approach presented in this paper is not only to replace the quantum operator $\hat{q}(x_0)$ with a field $q(x_0,\tau)$—so that quantum fluctuations at fixed coordinate time $x_0$ are parametrized by the intrinsic time $\tau$—but also to analytically continue the field itself to the complex plane:
\begin{align}
q(x_0,\tau) \in \mathbb{R}~~~\longrightarrow~~~ q(x_0,\tau) \in \mathbb{C}
\end{align}
so that,
\begin{equation}
    q(x_0,\tau) = q_R(x_0,\tau) + i q_I(x_0,\tau),
\end{equation}
where $q_R(x_0,\tau)$ and $q_I(x_0,\tau)$ denote respectively the real and imaginary parts of the field. 
Let us stress that, on the contrary, the expression of the action in
Eq.~\eqref{eq:action} must not be changed despite the analytic prolongation
of the field $q(x_0,\tau)$ from real to imaginary values, so that for the action itself we must consider the analytic continuation to $\mathbb{C}$:
\begin{align}
S[q] \in \mathbb{R}~~\longrightarrow~~S[q] \in \mathbb{C}.
\end{align}
We then assume for the generalized Hamiltonian which generates the dynamics
in $\tau$ the following form
\begin{equation}\label{eq:hamiltonian}
    \mathbb{H}[\pi, \bar{\pi}, q, \bar{q}] = \bar{\pi} \cdot \pi + 2\,\im{S}[q,\bar{q}],
\end{equation}
where the $\im{S}[q,\bar{q}]$ is a formal notation for  
\begin{equation}
  \im{S}[q,\bar{q}] = \frac{S[q] - \bar{S}[\bar{q}]}{2i},
  \label{eq:Sqq-2}
\end{equation}
where $S[q]$ is the holomorphic complex action and $\bar{S}[\bar{q}]$
its anti-holomorphic counterpart. Let us stress that
$\im{S}[q,\bar{q}]$ is solely defined from Eq.~\eqref{eq:Sqq-2} since in
the theory there is no action $S[q,\bar{q}]$ which depends on both $q$
and $\bar{q}$. Things are different for the generalized
Hamiltonian $\mathbb{H}[\pi, \bar{\pi}, q, \bar{q}]$, which is the
only functional of the theory depending on both olomorphic fields
$(q,\pi)$ and their anti-olomorphic counterparts
$(\bar{q},\bar{\pi})$. The Hamiltonian equations of motion along the
intrinsic time $\tau$ read then
\begin{align}\label{eq:hamilton-eqs}
    \dot{q} &= \frac{\partial \mathbb{H}}{\partial \pi} = \bar{\pi}, & 
    \dot{\bar{q}} &= \frac{\partial \mathbb{H}}{\partial \bar{\pi}} = \pi, \nonumber\\
    \dot{\pi} &= -\frac{\partial \mathbb{H}}{\partial q} = i \frac{\partial S[q]}{\partial q}, & 
    \dot{\bar{\pi}} &= -\frac{\partial \mathbb{H}}{\partial \bar{q}} = -i \frac{\partial \bar{S}[\bar{q}]}{\partial \bar{q}}.
\end{align}
The corresponding equations of motion mix the two fields $q$ and
$\bar{q}$:
\begin{equation}\label{eq:complex-eom}
    -i \frac{d^2}{d\tau^2}\bar{q}(x_0,\tau) = - m \frac{\partial^2}{\partial x_0^2} q(x_0,\tau) - m \Omega^2 q(x_0,\tau).
\end{equation}
If we then assume that the coordinate time $x_0$ belongs to a finite
interval, $x_0\in[x_0^{\text{i}},x_0^{\text{f}}]$, it is possible to diagonalize the
equations of motion by decomposing the field $q(x_0,\tau)$ into
Fourier modes
\begin{equation}\label{eq:fourier}
    q(x_0,\tau) = \sum_{\lbrace k_0(\ell):~\ell=1,\ldots,M\rbrace} \, e^{i k_0(\ell) x_0} \, q(k_0(\ell),\tau),
\end{equation}
where $k_0(\ell) = \ell 2\pi/(x_0^{\text{f}}-x_0^{\text{i}})$ with $\ell\in\mathbb{Z}$, so that 
\begin{align}\label{eq:harmonic-fourier}
    \frac{d^2}{d\tau^2}\bar{q}(k_0,\tau) + i \omega^2(k_0) q(k_0,\tau) = 0, \qquad 
    \omega^2(k_0) = m (\Omega^2 - k_0^2).
\end{align}
The further step is to write Eq.~\eqref{eq:harmonic-fourier} as a set
of two coupled real equations for the real and imaginary part of $q(k_0,\tau)$:
\begin{align}
    \ddot{q}_R(k_0,\tau) - \omega^2(k_0) q_I(k_0,\tau) &= 0, \nonumber\\
    \ddot{q}_I(k_0,\tau) - \omega^2(k_0) q_R(k_0,\tau) &= 0.
    \label{eq:coupled-real-imag}
\end{align}
At this point, when approaching the study of the Hamiltonian
dynamics generated by $\mathbb{H}[\pi, \bar{\pi}, q, \bar{q}]$ as a mean to sample the partition function $\Omega[\hbar M,J]$ in
Eq.~\eqref{eq:micro-part-func}, we must solve the problem of unbounded solutions of Eq.~\eqref{eq:harmonic-fourier}, which parallels the problem of choosing the appropriate integration contours in order to guarantee convergence in the functional integral of $\Omega[\hbar M,J]$. A quite
straightforward inspection of Eq.~\eqref{eq:coupled-real-imag} reveals
that in order to keep solutions on a bounded manifold it is sufficient
to impose the following constraints:
\begin{align}
  \text{if}~\omega^2(k_0)>0~~~&\Longrightarrow~~~q_R(k_0,\tau) = - q_I(k_0,\tau)~~~\&~~~\pi_R(k_0,\tau) = - \pi_I(k_0,\tau) \nonumber \\
  \text{if}~\omega^2(k_0)<0~~~&\Longrightarrow~~~q_R(k_0,\tau) =  q_I(k_0,\tau) ~~~~~\&~~~\pi_R(k_0,\tau) =  \pi_I(k_0,\tau)
  \label{eq:stable-surface0}
\end{align}
which can be also written in the more compact form
\begin{align}\label{eq:stable-surface}
q_I(k_0,\tau) &= -\sgn[\omega^2(k_0)]~q_R(k_0,\tau) \nonumber \\
    \pi_I(k_0,\tau) &= -\sgn[\omega^2(k_0)]~\pi_R(k_0,\tau),
\end{align}
leading to the following equations of motion
\begin{align}\label{eq:stable-motion}
    \ddot q_R(k_0,\tau) &+ |\omega^2(k_0)|~q_R(k_0,\tau) = 0 \nonumber \\
    \ddot q_I(k_0,\tau) &+ |\omega^2(k_0)|~q_I(k_0,\tau) = 0.
\end{align}
It can be immediately noticed that the equations of motion for the
real and the imaginary part of the field are identical.
This is a consequence of the constraint in
Eq.~\eqref{eq:stable-surface} needed to keep the field on a bounded
manifold. This is perfectly consistent with the fact that the
constrained symplectic dynamics evolves along a one-dimensional
manifold, analogous to the choice of convergent contours in the
Feynman path integral: the analytic prolongation of the fields to
$\mathbb{C}$ does not introduce extra degrees of freedom. In
particular, from the constraints written in
Eq.~\eqref{eq:stable-surface0} it is evident that it is perfectly
sufficient to solve the dynamics for $q_R(k_0,\tau)$, from which also
$q_I(k_0,\tau)$ is then univocally determined. In summary: despite the
need to consider the analytical prolongation to $\mathbb{C}$ of all
fields, yet, in order to guarantee consistency, the problem is reduced to study the dynamics of a single
real field, either the real part $q_R(k_0,\tau)$ or the imaginary one $q_I(k_0,\tau)$, since the two are not independent as long as one wants to preserve convergence.

At this stage we need a precise correspondence between expectation values
computed in the standard (Feynman) path-integral approach and those obtained
from the constrained symplectic quantization dynamics.
While a formal equivalence between the microcanonical partition function
and the Feynman path integral has been established in
Eq.~\eqref{eq:equivalence-sq-fey}, this equivalence remains incomplete until
the relation between (i) the stable manifold selected by the constraints
in Eq.~\eqref{eq:stable-surface} and (ii) the complex integration contours
$\boldsymbol{\Gamma}(x_0)$ appearing in Eq.~\eqref{eq:micro-part-func} is made
explicit.
To introduce the issue in a concrete setting, consider the endpoint--conditioned
matrix element with an operator insertion,
\begin{align}
  & \langle q_f,t_f| \hat{q}^2(t') | q_i,t_i\rangle \nonumber \\
  & = \frac{1}{\langle q_f,t_f| q_i,t_i\rangle}
  \int_{q(t_i)=q_i}^{q(t_f)=q_f} \mD q ~q^2(t')~\exp\left\lbrace \frac{i}{\hbar}\int_{t_i}^{t_f} dt \left[\frac{m}{2} \left( \frac{dq(t)}{dt} \right)^2 -  \frac{1}{2} m \Omega^2 q^2(t) \right] \right\rbrace,  
\end{align}
where $t_i\leq t' \leq t_f$ and which, for clarity, we rewrite using the notation introduced in this
section:
\begin{align}
  & \langle q_f,x_0^{\text{f}}| \hat{q}^2(x_0') | q_i,x_0^{\text{i}}\rangle  \nonumber \\
  & = \frac{1}{\langle q_f,x_0^{\text{f}}| q_i,x_0^{\text{i}}\rangle}
  \int_{q(x_0^{\text{i}})=q_i}^{q(x_0^{\text{f}})=q_f} \mD q ~q^2(x_0')~\exp\left\lbrace \frac{i}{\hbar} \int_{x_0^{\text{i}}}^{x_0^{\text{f}}} dx_0 \left[\frac{m}{2} \left( \frac{dq(x_0)}{dx_0} \right)^2 -  \frac{1}{2} m \Omega^2 q^2(x_0) \right]\right\rbrace.  
\label{eq:qm-expectq2}
\end{align}
In the present symplectic quantization framework, the fundamental dynamical object is the analytically
continued field $q(x_0,\tau)$, evolving in the fictitious time $\tau$ according
to the constrained Hamiltonian flow.
To reproduce the endpoint--conditioned amplitude in
Eq.~\eqref{eq:qm-expectq2}, one considers the dynamics on the interval
$[x_0^{\text{i}},x_0^{\text{f}}]$ with fixed boundary data \emph{for all} $\tau$,

\begin{align}
  q(x_0^{\text{i}},\tau) &=q_i ~~~\forall~\tau \nonumber \\
  q(x_0^{\text{f}},\tau) &=q_f ~~~\forall~\tau, 
\end{align}
and, keeping such boundary condition,
compute the expectation value over quantum fluctuations as
\begin{align}
  \langle q^2(x_0') \rangle|_{q_i(x_0^{\text{i}}), q_f(x_0^{\text{f}})}=
  \lim_{\Delta\tau\rightarrow\infty} \frac{1}{\Delta\tau} \int_{\tau_0}^{\tau_0+\Delta\tau}~d\tau~q_\tau^2(x_0')|_{ q_i(x_0^{\text{i}}), q_f(x_0^{\text{f}})},
\label{eq:sq-average}
\end{align}
where $q_\tau^2(x_0')|_{q_i(x_0^{\text{i}}), q_f(x_0^{\text{f}})}$ denotes the solution
of the symplectic dynamics with fixed boundaries along the $x_0$
axis. In order to connect the expression in Eq.~\eqref{eq:sq-average}
to the Feynman path integral representation of the expectation value
of $\hat{q}(x_0')$ in Eq.~\eqref{eq:qm-expectq2} it is necessary to
assume the equivalence between dynamical and thermodynamic averages in
the symplectic quantization approach:
\begin{align}
\lim_{\Delta\tau\to\infty}\frac{1}{\Delta\tau}
\int_{\tau_0}^{\tau_0+\Delta\tau}\!d\tau\;
q^2(x_0',\tau)
\;\cong\;
\frac{1}{\Omega_\infty[\hbar]}
\int_{\boldsymbol{\Gamma}(x_0)}
\mathcal{D}\bar{\pi}\,\mathcal{D}\pi\,\mathcal{D}\bar{q}\,\mathcal{D}q\;
q^2(x_0')\;
\delta\!\left(\mathcal{A}-\mathbb{H}[q,\bar{q},\pi,\bar{\pi}]\right),
\label{eq:ergodicity-sq}
\end{align}
where $\boldsymbol{\Gamma}(x_0)$ denotes the contour(s) in the complexified
phase space compatible with the constraints.
We can now state the key point.

In our setting the variables $(q,\pi)$ are generically complex along the constrained
Hamiltonian flow, and therefore the instantaneous value $q^2(x_0',\tau)$ is
in general a complex number.
This \emph{does not} imply that symplectic quantization predicts a complex value for the physical
matrix element in Eq.~\eqref{eq:qm-expectq2}.
Rather, it reflects the fact that we are evaluating the microcanonical functional
integral on a nontrivial contour $\boldsymbol{\Gamma}(x_0)$ in the complexified
space of fields, determined by the same constraints that define the stable
manifold of the dynamics.
By contrast, the standard Feynman representation in
Eq.~\eqref{eq:qm-expectq2} is written on the usual ``real'' configuration-space
contour for the oscillator variable.
A precise identification of $\boldsymbol{\Gamma}(x_0)$ therefore provides the
missing bridge between the two formulations: once the contour correspondence is
made explicit, the SQ microcanonical average reproduces the conventional
Feynman result for operator insertions. In summary, computing expectation values over quantum fluctuations by means of
the symplectic quantization dynamics requires an intermediate step with respect to the standard
quantum--mechanical formulae: symplectic quantization samples a microcanonical integral over
analytically continued fields on contours $\boldsymbol{\Gamma}(x_0)$ that are
not, in general, the standard ones.
The simplest way to illustrate how symplectic quantization simulations are compared with the usual
quantum--mechanical expressions is to work through an explicit example:
let us consider the expectation value of
the squared position operator for the quantum harmonic oscillator,
$\langle q_f,x_0^{\text{f}}| \hat{q}^2(x_0')| q_i,x_0^{\text{i}}\rangle$, and let us
assume for simplicity periodic boundary conditions in time. In
this case the expectation over quantum fluctuations can be written as
\begin{align}
  \langle \hat{q}^2(x_0') \rangle_{\text{P.B.}} =
  \int_{-\infty}^\infty dq_i~ \langle q_i,x_0^{\text{f}}| \hat{q}^2(x_0')| q_i,x_0^{\text{i}}\rangle =
  \frac{\int_{q(x_0^{\text{f}})=q(x_0^{\text{i}})} \mD q~q^2(x_0')~e^{\frac{i}{\hbar}S[q]}}{\int_{q(x_0^{\text{f}})=q(x_0^{\text{i}})} \mD q~e^{\frac{i}{\hbar}S[q]}},
\label{eq:q2-quant-exp}
\end{align}
where the two boundary conditions have been identified, $q(x_0^{\text{f}})=q(x_0^{\text{i}})$, and integrated over. To lighten the notation, in the following we assume that, in the absence of further specifications, all path-integral expressions are understood with periodic boundary conditions in $x_0$:
\begin{align}
  \Tr[e^{-i(x_0^{\text{f}}-x_0^{\text{i}})\hat{H}/\hbar}] = \int \mD q~e^{\frac{i}{\hbar}S[q]} =
  \int_{q(x_0^{\text{f}})=q(x_0^{\text{i}})} \mD q~e^{\frac{i}{\hbar}S[q]}.
\end{align}
The whole point is that in Eq.~\eqref{eq:q2-quant-exp} standard quantum-mechanical formulae assume that, for each coordinate time $x_0$, the position field is integrated over the real axis, namely $\int \mD q = \prod_{x_0} \int_{-\infty}^\infty dq(x_0)$. This differs from the integration paths selected by the microscopic dynamics of symplectic quantization. The crucial observation connecting standard quantum-mechanical amplitudes with the microcanonical approach of symplectic quantization is that expectation values are invariant under deformations of the integration contours in the complex plane. Therefore, we can elaborate on the fact that
\begin{align}
  \langle \hat{q}^2(x'_0) \rangle_{\text{P.B.}}
  & = \frac{\int_{-\infty}^\infty\prod_{x_0} dq(x_0)~q^2(x'_0)~e^{\frac{i}{\hbar}S[q]}}{\int_{-\infty}^\infty\prod_{x_0} dq(x_0)~e^{\frac{i}{\hbar}S[q]}}  \label{eq:equivalence-1} \\
  & = \frac{\int \prod_{\gamma_{q(x_0)}} dq(x_0)~q^2(x'_0)~e^{\frac{i}{\hbar}S[q]}}{\int\prod_{\gamma_{q(x_0)}} dq(x_0)~~e^{\frac{i}{\hbar}S[q]}} \label{eq:equivalence-2}\\
  & = \frac{\int \prod_{\gamma_{q(x_0)}} dq_R(x_0)\,dq_I(x_0)~[q_R(x'_0)+i\,q_I(x'_0)]^2~e^{\frac{i}{\hbar}S[q_R+i q_I]}}{\int\prod_{\gamma_{q(x_0)}}dq_R(x_0)\,dq_I(x_0)~e^{\frac{i}{\hbar}S[q_R+i q_I]}} \, .
  \label{eq:equivalence-3}
\end{align}
The crucial observation for what follows is that the contours $\gamma_{q(x_0)}$ must be paths in the complex plane that correspond to the constraints imposed on the symplectic dynamics in Eq.~\eqref{eq:stable-surface0}. To this purpose, taking advantage of the periodic boundary conditions in $x_0$, it is convenient to introduce the Fourier modes of the position field,
\begin{align}
  q(x'_0) &= \frac{1}{T\sqrt{2\pi}} \sum_{k'_0=-\infty}^\infty e^{-i k'_0 x'_0}~\tilde{q}(k'_0),  \nonumber \\
  \tilde{q}(k'_0) &=  \frac{1}{\sqrt{2\pi}}\int_{0}^{T} dx_0~e^{i k'_0 x'_0}~q(x'_0),
\end{align}
where $T = x_0^{\text{f}}-x_0^{\text{i}}$, and to write the expectation value in Fourier representation:
\begin{align}
  \langle \hat{q}^2(x'_0) \rangle_{\text{P.B.}}
  & = \frac{1}{2\pi T^2} \sum_{k'_0,p'_0=-\infty}^\infty e^{-i x'_0(k'_0+p'_0)} \left\langle \tilde{q}(k'_0)\tilde{q}(p'_0)\right\rangle_{\text{P.B.}}\nonumber \\
  & = \frac{1}{2\pi T^2} \sum_{k'_0=-\infty}^\infty e^{-i 2 x'_0k'_0} \left\langle \tilde{q}(k'_0)\tilde{q}(-k'_0)\right\rangle_{\text{P.B.}} \, ,
\end{align}
where in the last step we used translation invariance (for periodic boundary conditions) to enforce $p'_0=-k'_0$. We now consider the situation in which $q(x_0')\in\mathbb{C}$, so that the integration contours, also in Fourier space, are paths in the complex plane:
\begin{align}
  \left\langle \tilde{q}(k'_0)\tilde{q}(-k'_0)\right\rangle_{\text{P.B.}}
  =
  \frac{\int \prod_{\gamma_{q(k_0)}} dq_R(k_0)\,dq_I(k_0)~q(k'_0)\,q(-k'_0)~e^{\frac{i}{\hbar}S[q(k_0)]}}{\int\prod_{\gamma_{q(k_0)}}dq_R(k_0)\,dq_I(k_0)~e^{\frac{i}{\hbar}S[q(k_0)]}} \, .
  \label{eq:two-point-path-fourier}
\end{align}
It is only at this step that we can characterize the integration paths $\gamma_{q}$ in terms of the constraints written in Eq.~\eqref{eq:stable-surface0}. The first step is to manipulate the action in a form suitable to impose the constraints:
\begin{align}
  S[q]
  &= \int_{x_0^{\text{i}}}^{x_0^{\text{f}}} dx_0 \left[\frac{m}{2} \left( \frac{dq(x_0)}{dx_0} \right)^2 -  \frac{1}{2} m \Omega^2 q^2(x_0) \right] \nonumber \\
  &= \frac{m}{2} q(x_0)\frac{dq(x_0)}{dx_0}\Bigg|_{x_0^{\text{i}}}^{x_0^{\text{f}}}
  - \int_{x_0^{\text{i}}}^{x_0^{\text{f}}} dx_0~q(x_0)\left[\frac{m}{2}\frac{d^2}{dx_0^2} + \frac{1}{2}m\Omega^2\right]q(x_0) \nonumber \\
  &= - \frac{m}{2} \sum_{k_0=-\infty}^\infty \tilde{q}(k_0)\left[-k_0^2+\Omega^2\right]\tilde{q}(-k_0) \nonumber \\
  &= -\frac{1}{2}\sum_{k_0=-\infty}^\infty \omega^2(k_0)\,\tilde{q}(k_0)\tilde{q}(-k_0)\, ,
\end{align}
where $\omega^2(k_0)\equiv m(\Omega^2-k_0^2)$. The standard textbook step would now be to write $\tilde{q}(k_0)\tilde{q}(-k_0)=\tilde{q}(k_0)\tilde{q}^{\,*}(k_0)$, which is only correct when $q(x_0)\in\mathbb{R}$. In the present case, where $q(x_0)\in\mathbb{C}$, we write the Fourier modes in terms of their real and imaginary parts, so that the constraints can be exploited directly:
\begin{align}\label{eq:integrationpathfeynmann}
  S[q]
  &= - \frac{1}{2} \sum_{k_0=-\infty}^\infty \omega^2(k_0)\Big[ q_R(k_0)q_R(-k_0)-q_I(k_0)q_I(-k_0)
  + i\,q_R(k_0)q_I(-k_0)+ i\,q_I(k_0)q_R(-k_0)\Big] \nonumber \\
  &= i \sum_{\{k_0\,|\,\omega^2(k_0)\ge 0\}} \omega^2(k_0)\,q_R(k_0)q_R(-k_0)
     - i \sum_{\{k_0\,|\,\omega^2(k_0)< 0\}} \omega^2(k_0)\,q_R(k_0)q_R(-k_0)\, .
\end{align}
The second line of Eq.~\eqref{eq:integrationpathfeynmann} follows by imposing, in the functional integration of Eq.~\eqref{eq:equivalence-3}, the constraints $q_R(k_0)=-q_I(k_0)$ for $\omega^2(k_0)>0$ and $q_R(k_0)=q_I(k_0)$ for $\omega^2(k_0)<0$, which are equivalent to rotating the integration contour of each mode by an angle $\pi/4$ clockwise for $\omega^2(k_0)>0$ and counterclockwise for $\omega^2(k_0)<0$, as shown in Fig.~\ref{fig:contours}. 
\begin{figure}[t]
\centering
\begin{tikzpicture}[scale=1.2]
    \draw[-{Latex}] (-3.5,0) -- (3.5,0)
        node[below left] {$\operatorname{Re} q(k_0)$};
    \draw[-{Latex}] (0,-3.5) -- (0,3.5)
        node[below left] {$\operatorname{Im} q(k_0)$};
    \node at (0,0) [below right] {$0$};

    \draw[blue, thick, dashed, -{Stealth}] (-3, 3) -- (3, -3);
    \node[blue, rotate=-45, above, xshift=-3cm, yshift=0.1cm]
        {\small contour (A): $k_0^2 < \Omega^2$};
    \node[blue, rotate=-45, below, xshift=-3cm, yshift=-0.1cm]
        {$q(k_0)=e^{-i\pi/4}\tilde q(k_0)$};

    \draw[red, thick, -{Stealth}] (-3, -3) -- (3, 3);
    \node[red, rotate=45, above, xshift=-2.9cm, yshift=0.1cm]
        {\small contour (B): $k_0^2 > \Omega^2$};
    \node[red, rotate=45, below, xshift=-2.9cm, yshift=-0.1cm]
        {$q(k_0)=e^{+i\pi/4}\tilde q(k_0)$};

    \draw[->, thick] (2,0) arc[start angle=0, end angle=45, radius=2];
    \node at (2.2,1.2) {\scriptsize $e^{+i\pi/4}$};

    \draw[->, thick] (2,0) arc[start angle=0, end angle=-45, radius=2];
    \node at (2.2,-1.2) {\scriptsize $e^{-i\pi/4}$};
\end{tikzpicture}
\caption{Integration contours in the complex $q(k_0)$ plane. For modes with
$k_0^2<\Omega^2$, the contour is rotated by $-\pi/4$, corresponding to
$q(k_0)=e^{-i\pi/4}\tilde q(k_0)$. For modes with $k_0^2>\Omega^2$, the contour
is rotated by $+\pi/4$, corresponding to
$q(k_0)=e^{+i\pi/4}\tilde q(k_0)$. With this choice, the phase of the quadratic
holomorphic action is rotated into a real damping direction.}
\label{fig:contours}
\end{figure}

Recalling that $q_R(x_0)$ and $q_I(x_0)$ are real functions, their Fourier components satisfy $q_R(-k_0)=q_R^{*}(k_0)$ and $q_I(-k_0)=q_I^{*}(k_0)$, and we obtain the following expression for the measure:
\begin{align}
  \exp\!\left( \frac{i}{\hbar}S[q]\right)
  = \exp\!\left(- \frac{1}{\hbar}\sum_{\{k_0\,|\,\omega^2(k_0)\ge 0\}}\omega^2(k_0)\,|q_R(k_0)|^2
  + \frac{1}{\hbar}\sum_{\{k_0\,|\,\omega^2(k_0)<0\}}\omega^2(k_0)\,|q_R(k_0)|^2 \right).
  \label{eq:sq-factor-measure-k}
\end{align}
The kernel in Fourier space of the two-point correlation function can then be written as
\begin{align}
  \left\langle \tilde{q}(k'_0)\tilde{q}(-k'_0)\right\rangle_{\text{P.B.}}
  &=
  \left\langle \tilde{q}_R(k'_0)\tilde{q}_R(-k'_0)\right\rangle
  -\left\langle \tilde{q}_I(k'_0)\tilde{q}_I(-k'_0)\right\rangle \nonumber\\
  &\hspace{1cm}+\, i\left[\left\langle \tilde{q}_R(k'_0)\tilde{q}_I(-k'_0)\right\rangle
  +\left\langle \tilde{q}_I(k'_0)\tilde{q}_R(-k'_0)\right\rangle\right].
  \label{eq:two-point-complete}
\end{align}
This expression can be simplified by using the constraints in the functional integration. Exploiting the Gaussian measure in Eq.~\eqref{eq:sq-factor-measure-k}, one finds
\begin{align}
  \omega^2(k'_0) < 0 ~~~&\Longrightarrow~~~\left\langle \tilde{q}(k'_0)\tilde{q}(-k'_0)\right\rangle_{\text{P.B.}} = \frac{2i}{\omega^2(k_0)} = - 2i \left\langle \tilde{q}_R(k_0)\tilde{q}_R(-k'_0)\right\rangle,  \nonumber\\
  \omega^2(k'_0) > 0 ~~~&\Longrightarrow~~~\left\langle \tilde{q}(k'_0)\tilde{q}(-k'_0)\right\rangle_{\text{P.B.}} = \frac{2i}{\omega^2(k'_0)} = 2i \left\langle \tilde{q}_R(k'_0)\tilde{q}_R(-k'_0)\right\rangle .
  \label{eq:corresp-oqm-sq}
\end{align}
The quantities on the left-hand side of Eq.~\eqref{eq:corresp-oqm-sq} are the expectation values from ordinary quantum mechanics, whereas the quantities on the right-hand side are those that can be sampled by means of the constrained symplectic-quantization dynamics, by computing time averages along the intrinsic time:
\begin{align}
  \left\langle \tilde{q}_R(k'_0)\tilde{q}_R(-k'_0)\right\rangle
  =
  \lim_{\Delta\tau\rightarrow\infty}\frac{1}{\Delta\tau}\int_{\tau_0}^{\tau_0+\Delta\tau} d\tau~\tilde{q}_R(k'_0,\tau)\tilde{q}_R(-k'_0,\tau),
\label{eq:ensemble-dynamical-av}
\end{align}
where the fields on the right-hand side of Eq.~\eqref{eq:ensemble-dynamical-av} denote the solutions of the symplectic-quantization Hamiltonian dynamics. Following the same logic it is possible to identify which correlation functions in the symplectic-quantization approach correspond to standard observables in quantum mechanics such as $\langle q(x_0)\rangle$, $\langle q^n(x_0)\rangle$, and the two-point correlator $\langle q(x_0)q(x_0')\rangle$, with periodic boundary conditions in coordinate time or with other boundary conditions, for example Dirichlet. To lighten the discussion, the derivation of the corresponding expressions is left to the appendices. In what follows we will limit ourselves to reporting the results, i.e. the expression of quantum-mechanical correlators in terms of the real and imaginary parts of analytically continued fields in the symplectic quantization approach. At this stage, the only preliminary discussion needed before the presentation of results concerns how the numerical simulation of the symplectic-quantization dynamics is realized, and in particular how the constraints on the real and imaginary parts are implemented in practice.

\section{Constrained symplectic dynamics: numerical algorithm}
\label{sec:numerical-algorithm}

In the present section we present a detailed account of the numerical
approach to the symplectic quantization dynamics of the quantum
harmonic oscillator. There is a general part of the discussion, not
specific of the problem, which deals with the discretization on a
lattice of the equations of motion. What is on the contrary ``problem
dependent'' is the way to implement in the microscopic Hamiltonian
dynamics the constraints required to have stable solutions. This is a
technical point which, compared to the strategy of computing directly
the functional integral in quantum field theory by means of the
Lefschetz-Thimbles approach, plays a role analogous to the choice of
the most appropriate contours in the complex action plane along which
is more convenient to perform functional integration in order to
guarantee convergence. Let us then first present the discretized
version of our equations of motion and then explain how the
constraints are implemented in practice. We recall in first place the
shape of the effective separable Hamiltonian which generates the
symplectic dynamics:
\begin{align}
\mathbb{H}[q,\bar{q},\pi,\bar{\pi}] = \mathbb{K}[\pi,\bar{\pi}] + \mathbb{V}[q,\bar{q}],
\end{align}
where
\begin{align}
  \mathbb{K}[\pi,\bar{\pi}] &= \int dx_0~\pi(x_0)~\bar{\pi}(x_0) \nonumber \\
  \mathbb{V}[q,\bar{q}] &= 2~\text{Im} S[q,\bar{q}] = \frac{S[q]-\bar{S}[\bar{q}]}{i} \nonumber \\
  & = \frac{m}{2i} q(x_0)\frac{dq(x_0)}{dx_0}\Bigg|_{x^{\text{i}}_0}^{x^{\text{f}}_0} - \frac{1}{i}\int_{x_0^{\text{i}}}^{x_0^{\text{f}}} dx_0~q(x_0)\left[\frac{m}{2} \frac{d^2}{dx_0^2} +  \frac{1}{2} m \Omega^2 \right]q(x_0) - \nonumber \\
  &- \frac{m}{2i} \bar{q}(x_0)\frac{d\bar{q}(x_0)}{dx_0}\Bigg|_{x^{\text{i}}_0}^{x^{\text{f}}_0} + \frac{1}{i}\int_{x_0^{\text{i}}}^{x_0^{\text{f}}} dx_0~\bar{q}(x_0)\left[\frac{m}{2} \frac{d^2}{dx_0^2} +  \frac{1}{2} m \Omega^2 \right]\bar{q}(x_0) 
\end{align}
Let us now consider the case of periodic boundary conditions in the $x_0$-direction for the field $q(x_0,\tau)$. In this case the boundary terms generated by partial integration in the action do not affect the equations of motion, since periodicity implies
\begin{equation}
q(x_0^{\text{f}},\tau)=q(x_0^{\text{i}},\tau),\qquad 
\partial_{x_0}q(x_0^{\text{f}},\tau)=\partial_{x_0}q(x_0^{\text{i}},\tau)\qquad \forall~\tau,
\end{equation}
and therefore
\begin{equation}
q(x_0,\tau)\,\partial_{x_0}q(x_0,\tau)\Big|_{x_0^{\text{i}}}^{x_0^{\text{f}}}=0
\qquad\text{(and analogously for $\bar q$)}.
\end{equation}
As a consequence, the equations of motion are entirely determined by the bulk terms and read
\begin{align}
  \dot{q}(x_0,\tau) &= \bar{\pi}(x_0,\tau), \nonumber\\
  \dot{\bar{q}}(x_0,\tau) &= \pi(x_0,\tau), \nonumber\\
  \dot{\pi}(x_0,\tau) &= -\,i m \left(\frac{\partial^2}{\partial x_0^2} +\Omega^2\right)q(x_0,\tau), \nonumber\\
  \dot{\bar{\pi}}(x_0,\tau) &= +\,i m \left(\frac{\partial^2}{\partial x_0^2} +\Omega^2\right)\bar{q}(x_0,\tau),
  \label{eq:Hamiltonian-equations}
\end{align}
where the dot denotes $\partial_\tau$. Let us now comment on Dirichlet (fixed) boundary conditions, which are relevant for instance to transition amplitudes with fixed endpoints. For Dirichlet boundaries one imposes
\begin{equation}
q(x_0^{\text{i}},\tau)=q(x_0^{\text{f}},\tau)=0\qquad \forall~\tau,
\end{equation}
(and similarly for $\bar q$), so that the boundary term produced by partial integration vanishes identically:
\begin{equation}
q(x_0,\tau)\,\partial_{x_0}q(x_0,\tau)\Big|_{x_0^{\text{i}}}^{x_0^{\text{f}}}=0
\qquad\text{(and analogously for $\bar q$)}.
\end{equation}
Therefore the equations of motion \eqref{eq:Hamiltonian-equations} remain unchanged, while the boundary conditions select the admissible class of field configurations and, in the lattice discretization, determine how the $x_0$-Laplacian is implemented at the endpoints (i.e. the boundary sites are held fixed and are not updated by the $\tau$-evolution).


From the above Eq.~\eqref{eq:Hamiltonian-equations}, by writing
explicitly the equations in terms of the real and imaginary parts of
the fields, one gets in real space (coordinate time):
\begin{align}
    \ddot{q}_R(x_0,\tau) &= m\left(\Omega^2+\frac{\partial^2}{\partial x_0^2}\right)~q_I(x_0,\tau) \nonumber\\
    \ddot{q}_I(x_0,\tau) &= m\left(\Omega^2+\frac{\partial^2}{\partial x_0^2}\right)~q_R(x_0,\tau).
    \label{eq:coupled-real-imag-space}
\end{align}
corresponding in Fourier space to
\begin{align}
    \ddot{q}_R(k_0,\tau) &= \omega^2(k_0)~q_I(k_0,\tau) \nonumber\\
    \ddot{q}_I(k_0,\tau) &= \omega^2(k_0)~q_R(k_0,\tau),
    \label{eq:coupled-real-imag-Fourier}
\end{align}
where we recall the dispersion relation
\begin{align}
  \omega^2(k_0) = m (\Omega^2-k_0^2).
  \label{eq:dispersion}
\end{align}
We have seen in Sec.~\ref{sec:Harm-Osc-dynamics} that in the case of
the harmonic oscillator the constraints which
allow to avoid unbounded solutions are straightforward in Fourier
space, leading to a set of stable uncoupled equations for each one of
the modes $k_0$. One could take simply advange of this and draw a
straightforward correspondence with quantum mechanics by exactly
integrating the dynamics in Eq.~\eqref{eq:stable-motion} and comparing
the results obtained with the standard formalism of quantum mechanics
using the correspondence between analytically continued fields and
standard fields discussed in the previous section,
Sec.~\ref{sec:Harm-Osc-dynamics}. But here we aim at a more general
formalism, allowing to handle numerically also in presence of
interactions, for which it is in general computationally prohibitive
the attempt to solve the dynamics directly in Fourier space since the
standard interactions of quantum mechanics and quantum field theory,
which are local in space, becomes highly non local in Fourier
space. It is for this reason that, before discretizing the coordinate
time $x_0$ on a lattice and writing the corresponding equations, we
need to define a protocol that allows us to implement the constraints
between $q_R(k_0,\tau)$ and $q_I(k_0,\tau)$ also in coordinate space
equations. As clearly explained in Sec.~\eqref{sec:Harm-Osc-dynamics},
the implementation of the constraints on the dynamics allows to a
complete description of the system simply from the dynamics of the
real part of the position field
\begin{align}
  \omega^2(k_0) > 0 ~~~\Longrightarrow~~~\ddot{q}_R(k_0,\tau) &= -\omega^2(k_0)~q_R(k_0,\tau) ~~~\text{(odd modes)} \nonumber \\
  \omega^2(k_0) < 0 ~~~\Longrightarrow~~~\ddot{q}_R(k_0,\tau) &= +\omega^2(k_0)~q_R(k_0,\tau) ~~~\text{(even modes)} .
  \label{eq:stable-motion-2}
\end{align}
The above equation cannot be easily Fourier transformed back to coordinate
space because we would get highly nonlocal dynamical
equations. The most convenient choice to impose in coordinate space the
constraint embedded in Eq.~\eqref{eq:stable-motion-2} is to decompose
the position field into an \emph{even} and an \emph{odd} part,
corresponding respectively to the summation over modes $k_0$ for which
the constraint is $q_R(k_0,\tau)=q_I(k_0,\tau)$ (even),
\begin{align}\label{eq:qE-def}
q_R^E(x_0,\tau) = \sum_{\lbrace k_0 ~|~ {\omega^2(k_0)<0}\rbrace} dk_0 \, e^{i k_0 x_0} q_R(k_0,\tau),
\end{align}
and to the summation over modes $k_0$ for which
the constraint is $q_R(k_0,\tau)=-q_I(k_0,\tau)$ (odd),
\begin{equation}\label{eq:qo-def}
  q_R^O(x_0,\tau) = \sum_{\lbrace k_0 ~|~ \omega^2(k_0)>0\rbrace} dk_0 \, e^{i k_0 x_0} q_R(k_0,\tau).
\end{equation}
By then writing the equations of motion in terms of odd end even
components of the field one gets
\begin{align}\label{eq:even-odd-dynamics}
    \frac{d^2}{d\tau^2}q_R^O(x_0,\tau) &= - m \left(\Omega^2 + \frac{\partial^2}{\partial x_0^2}\right) q_R^O(x_0,\tau) \nonumber \\
    \frac{d^2}{d\tau^2}q_R^E(x_0,\tau) &= + m \left(\Omega^2 + \frac{\partial^2}{\partial x_0^2}\right) q_R^E(x_0,\tau).
\end{align}
The equations in Eq.~\eqref{eq:even-odd-dynamics} are eventually the
version which we discretized and integrated. Formally, they are both
stable, but numerical tests on the dynamics prove that they are
nevertheless plagued by an instability problem due to
discretization. What happens is the following. Let us for instance
consider the field $q_R^O(x_0,\tau)$, which should be at all times the
superposition only of modes $q_R(k_0,\tau)$ such that
$\omega^2(k_0)>0$: it happens that at a certain point along the
numerical integration of the equations of motion, it starts to get
also components from modes $k_0$ with $\omega^2(k_0)<0$ so that the
amplitude of $q_R(k_0,\tau)$ eventually starts to grow exponentially.
We solved this problem by putting by hand to zero the {\it even}
components of the field $q_R^O(x_0,\tau)$ periodically along the
numerical integration and doing the viceversa for $q_R^E(x_0,\tau)$.\\\\

We discretize the coordinate time direction using exactly $M$ lattice sites
\begin{equation}
x_0^{(\ell)} = x_0^{\mathrm{i}}+\ell a,\qquad \ell=0,1,\dots,M-1.
\end{equation}
For periodic boundary conditions (P.B.) the extent is $T=Ma$ and fields satisfy
$q(\ell+M,\tau)=q(\ell,\tau)$, so that $x_0^{\mathrm{f}}\equiv x_0^{\mathrm{i}}+T$.
For Dirichlet boundary conditions (D.B.) the endpoints are distinct and we set
$T=(M-1)a$, with $q(0,\tau)=q_i$ and $q(M-1,\tau)=q_f$ for all $\tau$. In the case of periodic boundary conditions in $x_0$ the field admits the standard Fourier representation
\begin{align}
  q_R(x_0^{(\ell)},\tau) = \frac{1}{\sqrt{M}}\sum_{n=-M/2}^{M/2-1} 
  e^{i k_0^{(n)} x_0^{(\ell)}}~ q_R(k_0^{(n)},\tau),
\end{align}
where the lattice momenta are $k_0^{(n)}=\frac{2\pi n}{T}$, with
$T = x_0^{\text{f}}-x_0^{\text{i}} = Ma$. For boundary conditions
different from periodic ones, the structure of the mode expansion is
modified accordingly; the
explicit mode decomposition in the Dirichlet case is reported in the
appendices \ref{app:dirichlet-lattice}. The dispersion relation on the
lattice reads as
\begin{align}\label{eq:omega_periodic}
  \Omega^2(n) = m \left[ \Omega^2 - \frac{2}{a^2} \left(1-\cos\left(k_0^{(n)} a \right)\right) \right].
\end{align}
In order to lighten the notation we will denote the position field and
its Fourier transform respectivel as
\begin{align}
  q_R(x_0^{(\ell)},\tau) &= q_R(\ell,\tau) \nonumber \\
  q_R(k_0^{(n)},\tau) &= \hat{q}_R(n,\tau),
\end{align}
so that the Fourier representation of the field can be rewritten as
\begin{align}
q_R(\ell,\tau) = \frac{1}{\sqrt{M}}\sum_{n=-M/2}^{M/2-1} e^{i k_0^{(n)} x_0^{(\ell)}}~\hat{q}_R(n,\tau).
\end{align}
Analogously to the continuum, the even and odd component of the
position field are defined by separating the modes according to the
sign of $\Omega^2(n)$, as defined in Eq.~\ref{eq:omega_periodic};
\begin{align}
  q_R^O(\ell,\tau) &= \sum_{\lbrace n ~|~ \omega^2(n)>0\rbrace}  e^{i k_0^{(n)} x_0^{(\ell)}} \hat{q}_R(n,\tau), \nonumber \\
  q_R^E(\ell,\tau) &= \sum_{\lbrace n ~|~ \omega^2(n)<0\rbrace}  e^{i k_0^{(n)} x_0^{(\ell)}} \hat{q}_R(n,\tau).
  \label{eq:lattice-odd-even}
\end{align}
The two lattice fields just defined in Eq.~\eqref{eq:lattice-odd-even}
obey the following equations
\begin{align}
  \frac{d^2}{d\tau^2} q_R^O(\ell,\tau)~&=~ -m \left[ \Delta q_R^O(\ell,\tau) + \Omega^2 q_R^O(\ell,\tau)\right] \label{eq:lattice-odd-coordinate}\\
  \frac{d^2}{d\tau^2} q_R^E(\ell,\tau)~&=~ ~~m \left[ \Delta q_R^E(\ell,\tau) + \Omega^2 q_R^E(\ell,\tau)\right] \label{eq:lattice-even-coordinate}\\
\end{align}
where the symbol $\Delta$ denotes the standard lattice Laplacian
\begin{align}
   \Delta q_R^O(\ell,\tau) = \frac{q_R^O(\ell+1,\tau)-2q_R^O(\ell,\tau)+q_R^O(\ell-1,\tau)}{a^2}
\end{align}
The dynamics in the intrinsic time $\tau$ of
Eq.~\eqref{eq:lattice-odd-coordinate} and
Eq.~\eqref{eq:lattice-even-coordinate} is then integrated numerically
using the leapfrog algorithm, which is a symplectic
integrator. Crucially, we have guaranteed that the dynamics of the two
fields $q_R^O(\ell,\tau)$ and $q_R^E(\ell,\tau)$ remains on the stable
manifold by means of the following protocol
\begin{enumerate}
\item[{\bf 1)}] Set to zero in the initial condition, i.e., at
  $\tau=0$, the amplitude of all modes with $\omega^2(n) > 0 $ for
  $q_R^O(\ell,\tau)$ and all modes with $\omega^2(n) < 0 $ for
  $q_R^E(\ell,\tau)$:
  \begin{align}
    \hat{q}_R^O(n,\tau) = 0 ~~~\forall~n~|~\omega^2(n) < 0 \nonumber \\
    \hat{q}_R^E(n,\tau) = 0 ~~~\forall~n~|~\omega^2(n) > 0  
  \end{align}
\item[{\bf 2)}] Transform back from Fourier to real space and
  numerically integrate on the lattice the dynamics for the fields
  $q_R^O(\ell,\tau)$ and $q_R^E(\ell,\tau)$.
\item[{\bf 3)}] After the numerical integration step compute again the
  Fourier components of the fields $q_R^O(\ell,\tau+\delta\tau)$ and
  $q_R^E(\ell,\tau+\delta\tau)$ and set again by hand to zero the amplitude
  of the Fourier components which should not contribute, namely:
  \begin{align}
    \hat{q}_R^O(n,\tau+\delta\tau) = 0 ~~~\forall~n~|~\omega^2(n) < 0 \nonumber \\
    \hat{q}_R^E(n,\tau+\delta\tau) = 0 ~~~\forall~n~|~\omega^2(n) > 0  
  \end{align}
  Repeat from point {\bf 2)} above.
\end{enumerate}
Let us stress that the solution of the dynamics for
$q_R(\ell,\tau)=q_R^O(\ell,\tau)+q_R^E(\ell,\tau)$ is perfectly
sufficient to know the behaviour of $q(\ell,\tau) = q_R(\ell,\tau) + i
q_I(\ell,\tau)$, since at any time from the knowledge of the
Fourier modes $\hat{q}_R(n,\tau)$ we also have $\hat{q}_I(n,\tau)$. They are
in fact uniquely determined from the constraint in
Eq.~\eqref{eq:stable-surface0}. The leap-from algorithm 
keeps track of both $q(\ell,\tau)$ and
$\pi(\ell,\tau)$, so we can compute at any time the value of the
generalized Hamiltonian
\begin{align}
  \mathbb{H}_\tau[\pi,\bar{\pi},q,\bar{q}] &= \mathbb{K}_\tau[\pi,\bar{\pi}] + \mathbb{V}_\tau[q,\bar{q}]
\end{align}
where
\begin{align}
  \mathbb{K}_\tau[\pi,\bar{\pi}] & = \frac{1}{2} \sum_{\ell=1}^M \left[ \pi_R^2(\ell,\tau)+\pi_I^2(\ell,\tau)\right] \nonumber \\
  \mathbb{V}_\tau[q,\bar{q}] &= - \frac{m}{2} \sum_{\ell=1}^M \left[ q_R(\ell,\tau) \Delta q_I(\ell,\tau)+ q_I(\ell,\tau)\Delta q_R(\ell,\tau)
  + \Omega^2~ q_R(\ell,\tau) q_I(\ell,\tau)\right].
  \label{eq:discrete-Hamiltonian}
\end{align}
\begin{figure}[H]
    \centering
    \includegraphics[width=0.75\linewidth]{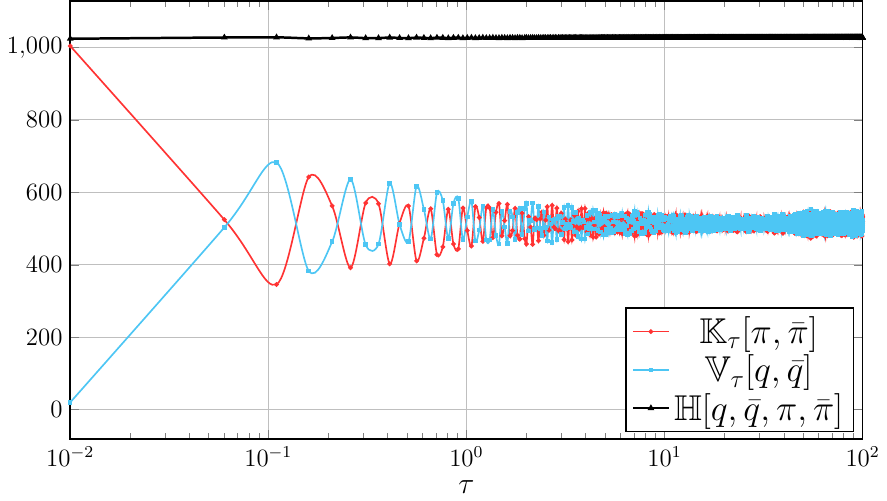}
    \caption{Time evolution of the kinetic ($E_{kin}=\mathbb{K}[\pi]$) and potential ($E_{pot}=\mathbb{V}[q]$) energy for $m=1.0$, $\Omega=2.5$, $a=0.1$, with integration parameters $d\tau=0.01$, and total simulation time $\Delta \tau=100$. The system size is $M=1024$, and we use periodic boundaries.}
    \label{fig:equipartition}
\end{figure}

By monitoring the behaviour of the generalized Hamiltonian in
Eq.~\eqref{eq:discrete-Hamiltonian} we were able to test both the
conservation of the total generalized energy
$E=\mathbb{H}_\tau[\pi,\bar{\pi},q,\bar{q}]$, show in
Fig.~\ref{fig:equipartition}, which is the first diagnostic of the
good functioning of the symplectic integrator, and the fast reaching
of equipartition, i.e., the relaxation to a stationary state such that
\begin{align}
\langle \mathbb{K}_\tau[\pi,\bar{\pi}] \rangle = \langle \mathbb{V}_\tau[\phi,\bar{\phi}] \rangle,
\end{align}
where the angular brackets denote in this case the average over
dynamical fluctuations.\\ As thoroughly discussed a recent review on
the foundations of statistical mechanics~\cite{Baldovin:2024dpo}, the
only necessary condition for the thermalization of {\it almost all}
the physically relevant observables is to have a large number of
degrees of freedom, so that we do not have to bother about the lack of
non-linear terms in $\mathbb{H}_\tau[\pi,\bar{\pi},q,\bar{q}]$ for
what concerns thermalization.\\ \\

The stage is now completely set for the presentation of our numerical
results, which will be done in the next sections.

\section{Constrained symplectic dynamics: numerical results}
\label{sec:numerical-results}

In this section, we present numerical results for the constrained
symplectic-quantization dynamics of the quantum harmonic oscillator
discussed in the previous sections. First, in Sec.~\ref{ss:average} we
analyze the one-point function $\langle \hat
q(x_0)\rangle_{\mathrm{P.B.}}$ with periodic boundary conditions
(hence the subscript ``P.B.''), using it as a diagnostic for finite
intrinsic-time effects and exploiting its study as a benchmark to test
the convergence of dynamical averages. Second, in
Sec.~\ref{subsec:two-point-periodic} we present the results for the
two-point correlation function $\langle \hat q(x_0)\hat
q(x_0')\rangle_{\mathrm{P.B.}}$ on a Minkowskian time lattice, finding
a perfect agreement between numerical simulation and analytical
prediction. We then present in Sec.~\ref{sec:lat-commutator} the
calculation on a Minkowskian time lattice of the expectation value
$\langle [\hat{q},\hat{p}] \rangle$, again using periodic boundary
conditions in time, showing that, in perfect agreement with the
theory, the simulations yield $\textrm{Im}(\langle [\hat{q},\hat{p}]
\rangle) = \hbar$ for different values of $\hbar$, which enters as a
parameter of the simulation. Finally, in
Sec.~\ref{subsec:q2k-dirichlet-spectrum} we study the expectation
values of even powers of the operator position with fixed-end
(Dirichlet) boundary conditions in Minkowskian time (referred to with
the subscript ``D.B.''), i.e., $\langle
q^{2k}(x_0)\rangle_{\mathrm{D.B.}(T)}$ with $q(x_0^f)=q(x_0^i)$,
showing that the power spectrum of this observable has a discrete
structure of peaks with a spacing in frequency which precisely
corresponds to the structure of harmonic oscillator excited states.

\subsection{Average position: \texorpdfstring{$\langle \hat{q}(x_0) \rangle$}: finite intrinsic time effects (periodic boundaries)}
\label{ss:average}

\begin{figure}[H]
    \centering
    \includegraphics[width=1\linewidth]{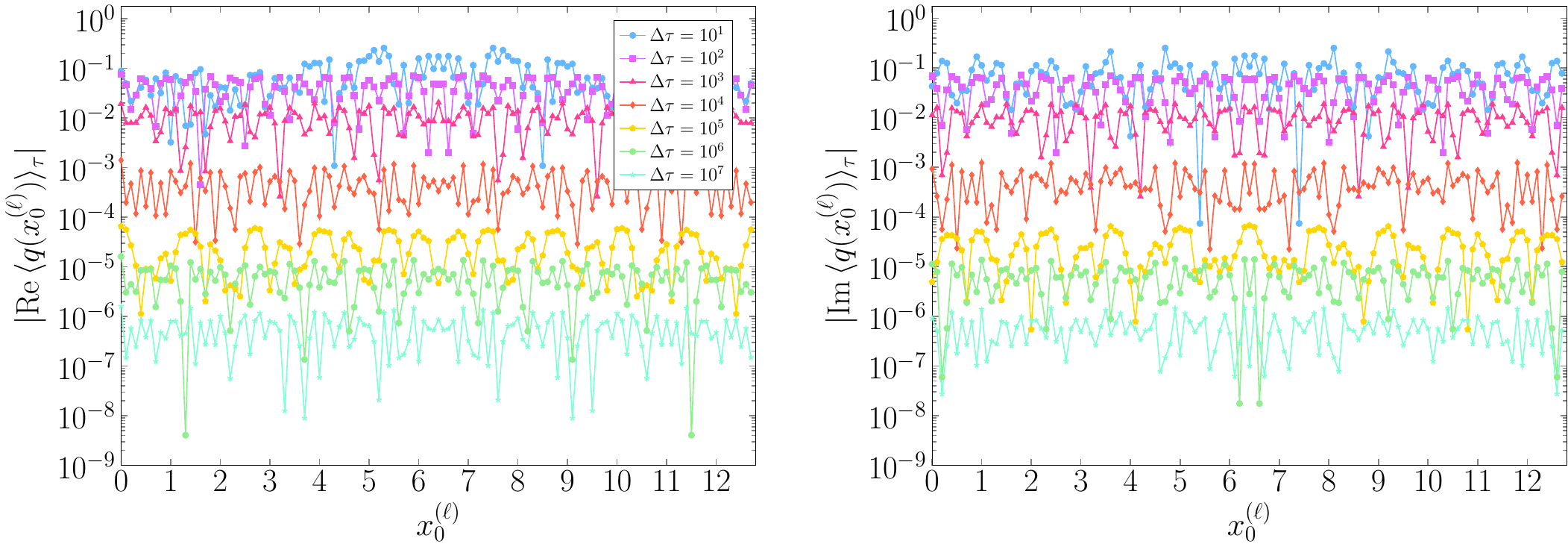}
    \caption{Expectation value of $q(x_0^{(\ell)})$ with periodic
      boundary conditions and for different values of the time window
      $\tau$ across which sampling is realized. Simulation parameters:
      $M=128$, $a=0.1$, $\Omega=2.5$ and $m=1$.}
    \label{fig:means_boundary_comparison}
\end{figure}

Here we show the symplectic quantization results for the position
expectation value for the quantum harmonic oscillator with periodic
boundary conditions in time. This is an insightful example to
illustrate how the symplectic quantization approach works, in
particular how the finite intrinsic time effects might affect the
comparison between observables sampled from the symplectic
quantization dynamics and standard quantum mechanical expectation
values. First, let us recall the expression of the average
value of the position operator with periodic boundary conditions in
time:
\begin{align}
  \langle \hat{q}(x_0) \rangle_{\text{P.B.}} =
  \frac{\Tr[e^{-\frac{i}{\hbar} \hat{H}(x_0^{\text{f}}-x_0^{\text{i}})} \hat{q}(x_0)]}{\Tr[e^{-\frac{i}{\hbar} \hat{H}(x_0^{\text{f}}-x_0^{\text{i}})}]} =
  \frac{\int_{\text{P.B.}}\mD q~q(x_0)~e^{\frac{i}{\hbar}S[q]}}{\int_{\text{P.B.}}\mD q~e^{\frac{i}{\hbar}S[q]}} = 0.
  \label{eq:qav-qm}
\end{align}
It is clear that if we assume for the expression in
Eq.~\eqref{eq:qav-qm} the action $S(q)$ of the harmonic oscillator
written in Eq.~\eqref{eq:action}, then we necessarily have $\langle
\hat{q}(x_0) \rangle =0$. Let us now explain why finite intrinsic time
effects occur when the symplectic quantization protocol is implemented
on a discrete Minkowskian time lattice, discretized into $M$ points,
with the discrete index $\ell=0,\dots,M-1$ replacing $x_0$, according
to the logic $x_0 \rightarrow x_0^{(\ell)} \rightarrow \ell$. Clearly,
according to the whole previous introduction, in the symplectic
quantization approach ooperators are represented by fields, as in functional formulations. For the position field on a periodic time lattice
we can therefore exploit the following representation
\begin{align}
  q{(\ell)} = \frac{1}{\sqrt{M}}\sum_{n=-M/2}^{M/2-1}  e^{i k_0^{(n)} x_0^{(\ell)}}~\tilde{q}{(n)},
  \label{eq:latt-fou}
\end{align}
with
\begin{equation}
k_0^{(n)}=\frac{2\pi n}{T},\qquad T=Ma,
\qquad
n=-\frac{M}{2},\dots,\frac{M}{2}-1
\end{equation}
and where we have set
\begin{equation}
q(\ell)=q(x_0^{(\ell)}) \qquad x_0^{(\ell)} = x_0^{\mathrm{i}}+\ell a,
\qquad
\ell=0,1,\dots,M-1.
\qquad
T = Ma,
\end{equation}
The periodic boundary condition $q(x_0^{\text{f}})=q(x_0^{\text{i}})$
is assumed. The expectation value $\langle q{(\ell)}
\rangle_{\text{P.B.}}$ can be accordingly written as
\begin{align}
  \langle q{(\ell)} \rangle_{\text{P.B.}}
  = \frac{1}{\sqrt{M}}\sum_{n=-M/2}^{M/2-1}
  e^{i k_0^{(n)} x_0^{(\ell)}}~\langle \tilde{q}{(n)} \rangle_{\text{P.B.}},
  \label{eq:qav-fou}
\end{align}
so that a sufficient condition for having $\langle q{(\ell)}
\rangle_{\text{P.B.}}$ is $\langle \tilde
q{(n)}\rangle_{\mathrm{P.B.}}=0$ for all modes. Let us now argue why this is
precisely so in the symplectic quantization approach. Let us recall
that according to symplectic quantization, as thoroughly explained in
Sec.~\eqref{sec:Harm-Osc-dynamics}, the Fourier components of the position field take values in the complex plane,
$\tilde q(n)\in\mathbb C$. For the harmonic oscillator, the corresponding
constrained measure factorizes into real and imaginary parts, see
Eq.~\eqref{eq:sq-factor-measure-k}, yielding:
\begin{align}
\langle \tilde{q}{(n)} \rangle_{\text{P.B.}}
=
\langle \tilde{q}_R{(n)} \rangle_{\text{P.B.}}
+ i \langle \tilde{q}_I{(n)} \rangle_{\text{P.B.}}.
\end{align}
Due to the symmetry between real and imaginary part which must be
considered in order to have a well-defined analytic continuation of
the position field to complex value, about which we presented a
detailed discussion in Sec.~\eqref{sec:Harm-Osc-dynamics}, we have that
the above expression, depending on the frequency, simplifies to
\begin{align}
  \omega^2(k_0^{(n)}) > 0 ~~~&\Longrightarrow~~~\langle \tilde{q}{(n)} \rangle_{\text{P.B.}} = (1-i)~\langle \tilde{q}_R{(n)} \rangle_{\text{P.B.}} = 0 \nonumber \\
  \omega^2(k_0^{(n)}) < 0 ~~~&\Longrightarrow~~~\langle \tilde{q}{(n)} \rangle_{\text{P.B.}} = (1+i)~\langle \tilde{q}_R{(n)} \rangle_{\text{P.B.}} = 0,
\end{align}
where the vanishing of the last term on the right of the two
expressions is due to the Gaussian nature of the measure:
\begin{align}
  \omega^2(k_0^{(n)}) > 0 ~~~&\Longrightarrow~~~\langle \tilde{q}_R{(n)} \rangle_{\text{P.B.}} \propto \int d\tilde{q}_R{(n)}~\tilde{q}_R{(n)}~e^{-\omega^2(n)\left(\tilde{q}_R{(n)}\right)^2} \nonumber \\
  \omega^2(k_0^{(n)}) < 0 ~~~&\Longrightarrow~~~\langle \tilde{q}_R{(n)} \rangle_{\text{P.B.}} \propto \int d\tilde{q}_R{(n)}~\tilde{q}_R{(n)}~e^{\omega^2(n)\left(\tilde{q}_R{(n)}\right)^2}.
  \label{eq:qk-ave-explicit}
\end{align}
A crucial effect of the symplectic quantization numerical approach is
that, by estimating statistical expectations values as the ones in
Eq.~\eqref{eq:qk-ave-explicit} by means of dynamical averages along
intrinsic-time dynamics, non-zero values of $\langle q(\ell)
\rangle_{\text{P.B.}}$ appear due to the finite duration of
simulations. Indeed, in principle the equivalence between dynamical and
thermodynamic averages 
\begin{align}
  \frac{1}{\tau} \int_{\tau_0}^{\tau_0+\tau} d\tau~q(\ell,\tau) \approx
  \langle q(\ell) \rangle_{\text{P.B.}},
  \label{eq:qx0-av-finite-T}
\end{align}
is guaranteed only asymptotically, namely for $\tau \gg 1$. In our
simulations the finite intrinsic time effect is manifest as a
deviation from zero of the expectation $\langle q(\ell)
\rangle_{\text{P.B.}}$, as is shown in
Fig.~\ref{fig:means_boundary_comparison}, where the estimate of
$|\langle q(\ell)\rangle_{\text{P.B.}}|$ sampled along the dynamics is
shown for different values of $\tau$. The presence of a non zero value
for the expectation value of the position in the quantum harmonic
oscillator can be easily rationalized within the symplectic
quantization approach by arguing that at too short simulation times
the sampling of the Gaussian probability distribution which guarantees
the vanishing of $\langle q(\ell) \rangle_{\text{P.B.}}$ is not
efficient, so that the average is not precisely zero.
To completely clarify the theoretical understanding of the
finite-$\tau$ effects let us draw the same argument from the
perspective of quantum-mechanical formulae. First of all, let us
recall that, according to the quantum-mechanical expression for the
expectation value of the position operator reported in
Eq.~\eqref{eq:qav-qm}, its vanishing relies on the vanishing of the
following trace:
\begin{equation}
  \Tr[e^{\frac{i}{\hbar}\hat{H}T} \hat{q}(x_0)] = \sum_{n,m}
  e^{\frac{i}{2\hbar}(E_m + E_n)T} \langle \psi_n | \hat{q}(x_0) |
  \psi_m \rangle \int dq\, \langle q | \psi_n \rangle \langle \psi_m |
  q \rangle,
  \label{eq:qav-trace}
\end{equation}
where $T = x_0^{\text{f}}-x_0^{\text{i}}$. The expression of the
position operator in terms of ladder operators is
\begin{equation}
\hat{q}(x_0) = \sqrt{\frac{\hbar}{2 m \Omega}} \left(e^{-i \Omega x_0}~\hat{a} + e^{i \Omega x_0}~\hat{a}^\dagger~\right),
\end{equation}
so that, by recalling that $\hat{a} ~|\psi_m\rangle =
\sqrt{m}~|\psi_{m-1}\rangle$ and that $\hat{a}^\dagger |\psi_m\rangle
= \sqrt{m+1}~|\psi_{m+1}\rangle$, one gets
\begin{equation}
  \langle \psi_n | \hat{q}(x_0) | \psi_m \rangle = \sqrt{\frac{\hbar}{2 m \Omega}} \left( \sqrt{m}~e^{-i\Omega x_0}~\delta_{n,m-1} + \sqrt{m+1}~e^{i\Omega x_0}~\delta_{n,m+1} \right).
  \label{eq:qav-ladder}
\end{equation}
By plugging the expression in Eq.~\eqref{eq:qav-ladder} into the trace
of Eq.~\eqref{eq:qav-trace} it becomes clear that the vanishing of the
whole expression is guaranteed by the identity
\begin{align}
\int_{-\infty}^\infty dq \, \langle q | \psi_n \rangle \langle \psi_m
| q \rangle = \delta_{nm}.
\label{eq:diracdeltaR}
\end{align}
It is precisely the lack of a uniform enough sampling of $q(x_0)$
values along the symplectic quantization dynamics which prevents to
use the identification of the integration over $q(x_0)$ with the
Kronecker delta, as written in Eq.~\eqref{eq:diracdeltaR}, thus also
preventing to argue the vanishing of the trace in
Eq.~\eqref{eq:qav-trace}.
What is shown in Fig.~\ref{fig:means_boundary_comparison} is that the finite-time sampling leads to deviations from zero in the average
value of $q(x_0)$. Consistently with the explanation given above, these
deviations decrease as the width of the sampling window increases.


\subsection{Two-point correlation function (periodic boundaries)}
\label{subsec:two-point-periodic}

\begin{figure}[H]
  \centering
  \includegraphics[width=1\linewidth]{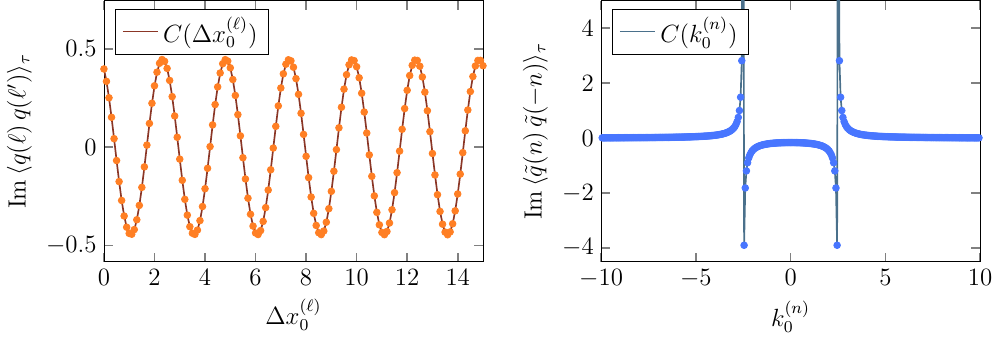}
  \caption{
    Harmonic-oscillator two-point function on the lattice with periodic boundary conditions.
    \textbf{Left:} coordinate-space correlator $\Im\,\langle q(\ell)\,q(\ell')\rangle_{\tau}$
    compared with the analytic lattice prediction~\eqref{eq:2pt-ho-lattice-PBC}.
    \textbf{Right:} Fourier-spectrum correlator $\Im\,\langle \tilde q(n)\,\tilde q(-n)\rangle_{\tau}$
    compared with the prediction~\eqref{eq:2pt-ho-momentum-num-conv}.
    Simulation parameters: $M=1024$, $a=0.1$, $m=1$, $\Omega=2.5$, $dt=0.01$, periodic boundaries.
  }
  \label{fig:correlation-lattice}
\end{figure}

The first numerical piece of evidence in favour of the symplectic
quantization approach we wish to present is the study of the two point
correlation function $\langle q(x_0)q(x_0')\rangle_{\text{P.B.}}$ with
periodic boundary conditions in Minkowskian time. Its expression,
as detailed in App.~\ref{app:periodic-twopoint}, can be explicitly
computed from standard quantum-mechanical formulae and reads as
\begin{align}
  \langle q(x_0) q(x_0') \rangle_{\text{P.B.}}
  = -\frac{i}{2m\Omega}\,
  \frac{\sin\!\big[\Omega\,\Delta x_0\big]
        + \sin\!\big[\Omega\,(T-\Delta x_0)\big]}
       {1-\cos(\Omega T)},
  \label{eq:2pt-ho-continuum-PBC}
\end{align}
namely is purely imaginary, with $T$ the total length of the time
interval and $\Omega$ the frequency of the harmonic oscillator. Let
us explain what we actually measured in terms of the analytically
continued field $q(x_0) \in \mathbb{C}$ which enters the symplectic
quantization approach in order to extract the two-point correlator
written in Eq.~\eqref{eq:2pt-ho-continuum-PBC}. By writing the
two-point correlation function in terms of the constrained averages
over the analytically continued field $q(x_0) = q_R(x_0)+i q_I(x_0)$
introduced in Sec.~\ref{sec:numerical-algorithm} one gets:
\begin{align}
  \langle q(x_0)q(x_0')\rangle_{\text{P.B.}}
  &= \langle q_R(x_0)q_R(x_0')\rangle_{\text{P.B.}}\nonumber
   - \langle q_I(x_0)q_I(x_0')\rangle_{\text{P.B.}}	\\
   &+ i \langle q_R(x_0)q_I(x_0')\rangle_{\text{P.B.}}
   + i \langle q_I(x_0)q_R(x_0')\rangle_{\text{P.B.}}.
  \label{eq:two-point-qi-qr}
\end{align}
The expression in Eq.~\eqref{eq:two-point-qi-qr} can be simplified by
taking into account the constraints which guarantee the convergence of
functional integrals within the symplectic quantization approach. Such
constraints are implemented in our numerical protocol by decomposing
the two components of the position field in even and odd
parts~\eqref{eq:qE-def} and \eqref{eq:qo-def}:
\begin{align}
  q_R(x_0) &=  q_R^E(x_0) +  q_R^O(x_0) ,
  \label{eq:qr-qi-eo-a} \\
  q_I(x_0) &=  q_I^E(x_0) +  q_I^O(x_0).
  \label{eq:qr-qi-eo-b}
\end{align}
Due to the constraints connecting real and imaginary part, see
Sec.~\ref{sec:Harm-Osc-dynamics}, the two-point correlation function
can be then rewritten as
\begin{align}
  \langle q(x_0)q(x_0')\rangle_{\text{P.B.}}
  &= 2 \Big[ \langle q_R^O(x_0) q_R^E(x_0')\rangle_{\text{P.B.}}
    + \langle q_R^E(x_0) q_R^O(x_0')\rangle_{\text{P.B.}}\\
    &+ i \langle q_R^E(x_0) 
    q_R^E(x_0')\rangle_{\text{P.B.}}
    - i \langle q_R^O(x_0) q_R^O(x_0')\rangle_{\text{P.B.}} \Big] ,
  \label{eq:two-point-qi-qr-simply}
\end{align}
an expression which can be further simplified by taking into account
the constrained integration measure
Eq.~\eqref{eq:sq-factor-measure-k}, which in terms of $q_R^E(x_0)$ and
$q_R^O(x_0)$ reads as
\begin{align}
  & \exp\left( \frac{i}{\hbar} S[q_R^O,q_R^E]\right) = \nonumber \\
  & =\exp\left\{
  \frac{m}{2\hbar} \int dx_0~q_R^O(x_0)
  \left(\Omega^2+\frac{\partial^2}{\partial x_0^2}\right)q_R^O(x_0)
  -\frac{m}{2\hbar} \int dx_0~q_R^E(x_0)
  \left(\Omega^2+\frac{\partial^2}{\partial x_0^2}\right)q_R^E(x_0).
  \right\} 
\end{align}
From the above functional probability density one immediately finds
\begin{align}
  \langle q_R^O(x_0) q_R^E(x_0')\rangle_{\text{P.B.}}
  = \langle q_R^E(x_0) q_R^O(x_0')\rangle_{\text{P.B.}} = 0 .
\end{align}
Thus the two-point function of the quantum harmonic oscillator,
written in terms of the analytically continued field
$q(x_0)\in\mathbb{C}$, simplifies to
\begin{align}
  \langle q(x_0)q(x_0')\rangle_{\text{P.B.}}
  = 2i \Big[ \langle q_R^E(x_0) q_R^E(x_0')\rangle_{\text{P.B.}}
    - \langle q_R^O(x_0) q_R^O(x_0')\rangle_{\text{P.B.}} \Big]
  \label{eq:two-point-qi-qr-simply-2}
\end{align}
and it is purely imaginary, as it should in agreement with the
expression of $\langle q(x_0)q(x_0')\rangle_{\text{P.B.}}$ in
Eq.~\eqref{eq:2pt-ho-continuum-PBC}. Clearly, since the simulation is
carried out on a discretize time lattice what we actually sample
numerically is the following difference of the two point real
correlation functions
\begin{align}
2i\left(\langle q_R^E(\ell) q_R^E(\ell')\rangle_{\text{P.B.}}
  - \langle q_R^O(\ell) q_R^O(\ell')\rangle_{\text{P.B.}}\right) =   \langle q(\ell)q(\ell')\rangle_{\text{P.B.}}
\label{eq:two-point-corr}
\end{align}
In the left panel of Fig.~\ref{fig:correlation-lattice} the two-point
correlation function that we obtained numerically from the correlator
of fields $q_R^O(\ell)$ and $q_R^E(\ell)$ using
Eq.~\eqref{eq:two-point-corr} is compared with the analytical
expression of the two-point correlation function on the lattice
$\langle q(\ell)q(\ell')\rangle_{\text{P.B.}}$, showing a perfect
agreement. The continuous line in Fig.~\ref{fig:correlation-lattice} represents
the analytical formula
\begin{align}
  \langle q(\ell) q(\ell') \rangle_{\text{P.B.}}^{\text{(lat)}}
  = -\frac{i\,a}{2m\sin(\kappa a)}\,
  \frac{\sin\!\big[\kappa\,a(\ell-\ell')\big]
        + \sin\!\big[\kappa\,(T-a(\ell-\ell'))\big]}
       {1-\cos(\kappa T)} ,
  \label{eq:2pt-ho-lattice-PBC}
\end{align}
where $\ell$ is an integer index, $\ell=0,\ldots,M-1$, and $\kappa$ is defined
by the dispersion relation
\begin{align}
  \cos(\kappa a) = 1 - \frac{1}{2} a^2 \Omega^2 .
  \label{eq:kappa-dispersion}
\end{align}
All details of the derivation are shown in
App.~\ref{app:periodic-twopoint}. The expression of the two-point
correlator shown in Eq.~\eqref{eq:2pt-ho-continuum-PBC} can be
obtained by Eq.~\eqref{eq:2pt-ho-lattice-PBC} by keeping $T$ fixed and
taking simultaneously $a\to 0$ and $M\rightarrow\infty$. Doing so, one
has
\begin{align}
a (\ell -\ell') &= x_0 -x'_0 = \Delta x_0 \nonumber \\
\sin(\kappa a) &\approx \kappa a \nonumber \\
\cos(\kappa a ) &\approx 1-\frac{1}{2} \kappa^2 a^2,
\end{align}
from which we can identify $\kappa = \Omega$. The oscillatory behaviour reported in
Eqns.~\eqref{eq:2pt-ho-lattice-PBC},~\eqref{eq:2pt-ho-continuum-PBC}
is perfectly reproduced by the numerical data shown in
Fig.~\ref{fig:correlation-lattice}. As shown in detail in
App.~\ref{app:periodic-twopoint}, the oscillatory behaviour of the
two-point correlation function is the genuine landmark of the
real-time quantum dynamics and of the concurrence of the harmonic
oscillator excited states to correlation functions, two facts excluded
by definition within the Euclidean quantum mechanics formulation, the
one usually exploited for numerical simulations.


In the right panel of Fig.~\ref{fig:correlation-lattice} we also
compare the numerical data for the two-point correlation function in
frequency space with the corresponding analytical prediction. By
introducing the discrete Fourier modes
\begin{align}
  \tilde q\!\left(n\right)
  \;\equiv\;
  \frac{1}{\sqrt{M}}\sum_{\ell=0}^{M-1} e^{-i k_0^{(n)} x_0^{(\ell)}}\, q\!\left(\ell\right),
  \label{eq:fourier-def-lat}
\end{align}
where $k_0^{(n)}=2\pi n/T$, $T = x_0^f - x_0^i$ and
$n=-\frac{M}{2},\dots,\frac{M}{2}-1$, the analytical formula for the
propagator in Fourier space reads 
\begin{align}
  \Big\langle \tilde q\!\left(n\right)\,\tilde q\!\left(-n\right)\Big\rangle_{\text{P.B.}}^{\text{(lat)}}
  = \frac{i}{m}\,
  \frac{1}{\frac{4}{a^2}\sin^2\!\left(k_0^{(n)} a/2\right)-\Omega^2}\,.
  \label{eq:2pt-ho-momentum-lattice}
\end{align}
In our simulations we consider the analytically continued complex
field on the lattice $\tilde q(n)$:
\begin{equation}
  \tilde q(n)
  \equiv \tilde q_R\!\left(n\right)
  + i\,\tilde q_I\!\left(n\right),
  \qquad
  \tilde q_{R,I}\!\left(n\right)\in\mathbb{R},
\end{equation}
so that in terms of real and imaginary parts of $\tilde q(n) \in
\mathbb{C}$ the original correlator reads as
\begin{align}
  \left\langle \tilde{q}(n)\tilde{q}(-n)\right\rangle
  &=
  \Big\langle
  \tilde q_R\!\left(n\right)\tilde q_R\!\left(-n\right)
  - \tilde q_I\!\left(n\right)\tilde q_I\!\left(-n\right)
  \Big\rangle
  \nonumber\\
  &\hspace{1.2cm}
  +\, i\Big\langle
  \tilde q_R\!\left(n\right)\tilde q_I\!\left(-n\right)
  + \tilde q_I\!\left(n\right)\tilde q_R\!\left(-n\right)
  \Big\rangle,
  \label{eq:2pt-ho-momentum-num}
\end{align}
where the real part of the expression on the right of
Eq.~\eqref{eq:2pt-ho-momentum-num} vanishes due to the symmetry
characterizing the constrained averages of symplectic quantization (see
Sec.~\ref{sec:Harm-Osc-dynamics}), yielding
\begin{align}
  C(k_0^{(n)}) = \left\langle \tilde{q}(n)\tilde{q}(-n)\right\rangle =
  i [\left\langle \tilde q_R(n)\tilde q_I(-n) \right\rangle  
  + \left\langle \tilde q_I(n)\tilde q_R(-n)\right\rangle]
  \label{eq:2pt-ho-momentum-num-conv}
\end{align}

In the right panel of Fig.~\ref{fig:correlation-lattice} we compare
the numerical data for $\Im\,\langle \tilde q(n)\,\tilde q(-n)\rangle$
as obtained from the correlators on the right-hand side of
Eq.~\eqref{eq:2pt-ho-momentum-num-conv} with the analytical prediction
in Eq.~\eqref{eq:2pt-ho-momentum-num}, finding perfect agreement.

\subsection{Uncertainty principle}
\label{sec:lat-commutator}

\begin{figure}[H]
  \centering
  \includegraphics[width=0.70\textwidth]{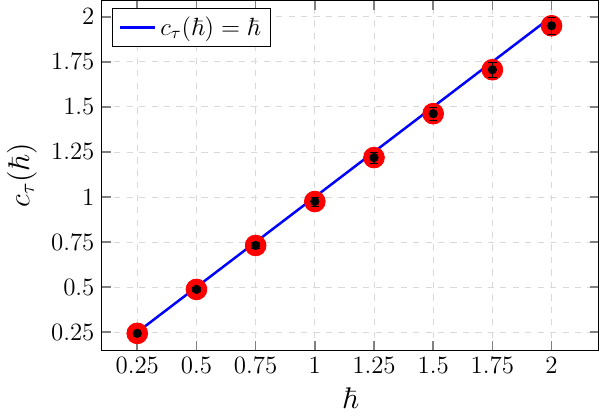}
  \caption{ Numerical determination of the equal--time commutator from
    the constrained symplectic dynamics. The solid line is the theoretical prediction,
$\mathcal F_\hbar=\hbar$, namely a straight line passing through the
origin with unit slope, while points represent the functional
    $\mathcal{F}_\hbar[q_R^E(\ell),q_R^O(\ell)]$, see
    Eq.~\eqref{eq:c-hbar-qREqRO-final} in the text, measured in
    numerical simulations.}
  \label{fig:comm-vs-hbar}
\end{figure}

A fundamental step to assess the validity of symplectic quantization
is to test its consistency with the uncertainty principle of quantum
mechanics, which, in its operatorial form, is equivalent to the equal
time commutation relation
\begin{equation}
  [\hat q(x_0),\hat p(x_0)] = i\hbar.
  \label{eq:canonical-comm-rel}
\end{equation}
If one then considers quantum mechanics in periodic time the operatorial
identity of Eq.~\eqref{eq:canonical-comm-rel} is equivalent to the following
expectation over quantum fluctuations, where the trace is related to the periodicity
in time
\begin{align}
 i \hbar \equiv \frac{1}{\mZ(\hbar)}\,\Tr\!\left([\hat{q}(x_0),\hat{p}(x_0)]\right),
\label{eq:trace-hbar}
\end{align}
In order to present the numerical protocol to sample the expectation
value in Eq.~\eqref{eq:trace-hbar} by means of the symplectic
quantization dynamics a necessary preliminary step is to recall how an
equal-time commutation relation between operators translates into a
functional formalism. A concise and insightful explanation of how this
can be done in the path-integral formalism can be found for instance
in~\cite{schulman2005techniques}, part of which we have reported for
the ease of the reader in
App.~\ref{app:commutator-derivation}. Basically, when operators are
converted into fields, a different ordering of equal time operators
can be simply represented by first computing a path-integral where the
rightmost operator corresponds to a field acting slightly before the
one to its left: We in fact consider $x_0^+ \geq x_0$ in the following
expression, and then we take the limit of equal times at the end:
\begin{align}
  \Tr\!\left([\hat{q}(x_0),\hat{p}(x_0)]\right) = \lim_{x_0^{+}\rightarrow x_0}
  \int \mathcal{D}q(x_0)~e^{\frac{i}{\hbar}S[q]}~\Bigl[q(x_0^{+})p(x_0)-p(x_0^{+})q(x_0)\Bigr].
  \label{eq:comm-path-int-1}
\end{align}
When discretizing the path integral the momentum itself reads
as a function of the position, therefore it is more convenient to recast
Eq.~\eqref{eq:comm-path-int-1} into the following equivalent
expression, where the position field $q(x_0)$ is computed at equal
times:
\begin{align}
 \Tr\!\left([\hat{q}(x_0),\hat{p}(x_0)]\right) =\lim_{\substack{x_0^{+}\rightarrow x_0 \\ x_0^{-}\rightarrow x_0}}\int
 \mathcal{D}q(x_0)~e^{\frac{i}{\hbar}S[q]}~\Bigl[q(x_0)p(x_0^{-})-p(x_0^{+})q(x_0)\Bigr],
 \label{eq:comm-path-int-2}
\end{align}
where $x_0^+ \geq x_0$ and $x_0^- \leq x_0$. When discretizing the
path integral of Eq.~\eqref{eq:comm-path-int-2} on a time-lattice with
spacing $a=T/M$, where $T = x_0^{\mathrm{f}} - x_0^{\mathrm{i}}$ is
the length of the time interval over which periodic boundary
conditions are taken and $M$ is the number of points in the
discretization, if a generic point on the lattice is denoted as
$x_0^{(\ell)} = x_0^{\mathrm{i}}+\ell a$, the following convention for
the calculation of momenta at a time slightly preceding $x_0$, $x_0^-
\leq x_0$, and a time slightly following $x_0$, $x_0^+ \geq x_0$, can
be adopted:
\begin{align}
p(x_0^{-}) = p\!\left(x_0^{\ell-\frac12}\right) = \frac{m}{a}\Bigl[q(\ell)-q(\ell-1)\Bigr] \nonumber \\
p(x_0^{+}) = p\!\left(x_0^{\ell+\frac12}\right) = \frac{m}{a}\Bigl[q(\ell+1)-q(\ell)\Bigr].
\label{eq:lattice-momenta-def}
\end{align}
From the definition of momenta in Eq.~\eqref{eq:lattice-momenta-def}
the discretized path-integral of the commutator reads as
\begin{align}
\frac{1}{\mZ(\hbar)} \Tr \left([\hat{q}(x_0),\hat{p}(x_0)]\right) &= \frac{1}{\mZ(\hbar)}\int \prod_{\ell=0}^{M-1} dq(\ell)~e^{\frac{i}{\hbar}S[q]}~
\Bigl[q(\ell)p\!\left(x_0^{\ell-\frac12}\right)-q(\ell)p\!\left(x_0^{\ell+\frac12}\right)\Bigr] \nonumber \\
&= \frac{m}{a}\Bigl[
2 \langle  q^2(\ell)\rangle_{\text{P.B.}} - \langle q(\ell)q(\ell-1)\rangle_{\text{P.B.}} -
\langle q(\ell)q(\ell+1)\rangle_{\text{P.B.}}
\Bigr].
\label{eq:comm-correlators}
\end{align}
In order to cope with the symplectic quantization formalism the
expression in Eq.~\eqref{eq:comm-correlators} needs now to be
rewritten in terms of the analytically continued complex field
$q(\ell)=q_R(\ell) + i q_I(\ell)$. The constraints which characterize
the symplectic quantization measure can be then consistently
implemented by further decomposing the real and imaginary parts in odd
and even components with respect to the symmetry which define the
constraint, see Sec.~\ref{sec:Harm-Osc-dynamics}:
\begin{align}
q_R(\ell) &= q_R^E(\ell) + q_R^O(\ell) \nonumber \\
q_I(\ell) &= q_I^E(\ell) + q_I^O(\ell) . \nonumber
\end{align}
Using the constraints on the functional measure,
Eqs.~\eqref{eq:stable-surface0} from Sec.~\ref{sec:Harm-Osc-dynamics}
and Eq.\eqref{eq:lattice-odd-even} from
Sec.~\ref{sec:numerical-algorithm}, the imaginary part of any
two-point correlation function can be then rewritten directly in terms
of the even and odd components of the real part of the field as
\begin{equation}
  \Im\bigl\langle q(\ell)\,q(\ell')\bigr\rangle_{\text{P.B.}}
  =
  2\Bigl[
    \bigl\langle q_R^E(\ell)\,q_R^E(\ell')\bigr\rangle_{\text{P.B.}}
    -
    \bigl\langle q_R^O(\ell)\,q_R^O(\ell')\bigr\rangle_{\text{P.B.}}
  \Bigr] .
  \label{eq:Imqq-qREqRO}
\end{equation}
Let us now recall a result already mentioned in Sec.~\ref{subsec:two-point-periodic}:
any two-point correlation function for the quantum harmonic oscillator in periodic
time is purely imaginary, i.e.,
\begin{align}
\langle q(\ell)\,q(\ell')\bigr\rangle_{\text{P.B.}} =  i~\textrm{Im}\langle q(\ell)\,q(\ell')\bigr\rangle_{\text{P.B.}}
\label{eq:two-points-tper-im}
\end{align}
By exploiting the identity of Eq.~\eqref{eq:two-points-tper-im} we can
write the expectation over quantum fluctuations of the commutator as:
\begin{align}
i \hbar~\frac{1}{\mZ(\hbar)} \Tr \left([\hat{q}(x_0),\hat{p}(x_0)]\right)  = \frac{m}{a}~\Bigl[
  2\,\Im\bigl\langle q^2(\ell)\bigr\rangle_{\text{P.B.}}
  -\Im\bigl\langle q(\ell)\,q(\ell-1)\bigr\rangle_{\text{P.B.}}
  -\Im\bigl\langle q(\ell)\,q(\ell+1)\bigr\rangle_{\text{P.B.}}\Bigr],
\end{align}
so that, using then the identity of Eq.~\eqref{eq:Imqq-qREqRO}, we can
read off $\hbar$ from the following combination of two-point
correlation functions measured along the symplectic quantization
dynamics
\begin{align}
  \hbar &= \mathcal{F}_\hbar[q_R^E(\ell),q_R^O(\ell)] = \nonumber \\
  &= ~\frac{m}{a} ~ \Bigl\{
      4\Bigl[
        \bigl\langle q_R^E(\ell)\,q_R^E(\ell)\bigr\rangle_{\text{P.B.}}
        - \bigl\langle q_R^O(\ell)\,q_R^O(\ell)\bigr\rangle_{\text{P.B.}}
      \Bigr] -2\Bigl[
        \bigl\langle q_R^E(\ell)\,q_R^E(\ell-1)\bigr\rangle_{\text{P.B.}} -
        \bigl\langle q_R^O(\ell)\,q_R^O(\ell-1)\bigr\rangle_{\text{P.B.}}
      \Bigr] \nonumber \\
  & -2\Bigl[ \bigl\langle q_R^E(\ell)\,q_R^E(\ell+1)\bigr\rangle_{\text{P.B.}}
        - \bigl\langle q_R^O(\ell)\,q_R^O(\ell+1)\bigr\rangle_{\text{P.B.}}
      \Bigr]
    \Bigr\}.
\label{eq:c-hbar-qREqRO-final}
\end{align}
The functional $\mathcal{F}_\hbar[q_R^E(\ell),q_R^O(\ell)]$, which is
defined in Eq.~\eqref{eq:c-hbar-qREqRO-final}, finally yields an
explicit account of the observable that we consider in our constrained
symplectic dynamics simulations as estimate of the commutator. As
usual, the statistical expectations appearing in rightmost term of
Eq.~\eqref{eq:c-hbar-qREqRO-final} are obtained as dynamical
averages. The subscript $\hbar$ in $\mathcal{F}_\hbar$ denotes the
parametric dependence on $\hbar$ of the symplectic quantization
dynamics (see Sec.~\ref{sec:numerical-algorithm}). In order to provide
a more robust evidence of the equal-time commutation relation $i\hbar
= \langle [\hat{q},\hat{p}] \rangle$ we performed different
simulations varying $\hbar$ and testing that we always get
\begin{align}
\mathcal{F}_\hbar[q_R^E(\ell),q_R^O(\ell)] = \hbar.
\end{align}
The result, which provide a beautiful evidence of how the commutation
relation between position and momentum operator is reproduced by
constrained symplectic quantization, is represented in
Fig.~\ref{fig:comm-vs-hbar}, where we have plotted the measured
$\mathcal{F}_\hbar[q_R^E(\ell),q_R^O(\ell)]$ as a function of $\hbar$,
showing that the resulting values perfectly fall, without any adjusted
parameter, on a straight line passing from the origin and with unit
slope. The error bars in Fig.~\ref{fig:comm-vs-hbar} are computed as
the standard deviation corresponding to site to site fluctuations on
the time lattice of the quantity $\langle
[\hat{q}(\ell),\hat{p}(\ell)]\rangle$.

\subsection{Discrete spectrum of $\langle \hat{q}^{2k}\rangle$: evidence of excited eigenstates}
\label{subsec:q2k-dirichlet-spectrum}

In this section we present the numerical results for the expectation
value of even powers of the position operator with Dirichlet fixed
boundary conditions in time, which we refer to as $\langle
\hat{q}^{2k}(x_0)\big\rangle_{\text{D.B.}}$. Without any loss of
generality, we will consider for the fixed boundaries the case
$q(x_0^{\mathrm{i}})=q(x_0^{\mathrm{f}})$, with $x_0^{\mathrm{i}} \leq
x_0 \leq x_0^{\mathrm{f}}$. We have chosen the study of $\langle
\hat{q}^{2k}(x_0)\big\rangle_{\text{D.B.}}$ because it yields, in
particular from the observation of its Fourier spectrum, clear
evidence of the role played in higher order real-time correlation
functions by all the excited eigenstates of the harmonic oscillator,
which are usually suppressed for a large enough time span $T =
x_0^{\mathrm{f}}-x_0^{\mathrm{i}}$ within the Euclidean approach. The
general expression of $\langle
\hat{q}^{2k}(x_0)\big\rangle_{\text{D.B.}}$ with Dirichlet boundary
conditions
$q(x_0^{\mathrm{i}})=q(x_0^{\mathrm{f}})=\tilde{q}$ reads as
\begin{align}
  \langle \hat{q}^{2k}(x_0)\big\rangle_{\text{D.B.}} = \frac{\langle \tilde{q},x_0^{\mathrm{f}} \,|\, \hat{q}^{2k}(x_0)\,|\, \tilde{q},x_0^{\mathrm{i}} \rangle}{\langle \tilde{q},x_0^{\mathrm{f}} \,|\, \tilde{q},x_0^{\mathrm{i}} \rangle},
\end{align}
which, for the particular but not less general choice $\tilde{q}=0$ and for
a time span of length $T$, reads as
\begin{align}
  \big\langle \hat{q}^{2k}(x_0)\big\rangle_{\text{D.B.}}=
  \frac{\langle 0,x_0^{\mathrm{f}} | \hat{q}^{2k}(x_0) | x_0^{\mathrm{i}},0 \rangle}{\langle 0, x_0^{\mathrm{f}} | x_0^{\mathrm{i}},0 \rangle}.
       \label{eq:DB-def-normalised}
\end{align}
Before presenting the constrained symplectic quantization numerical
results let us elucidate why excited states of the harmonic oscillator
crucially concur to the functional form of $\big\langle
\hat{q}^{2k}(x_0)\big\rangle_{\text{D.B.}}$, a task which is most
easily accomplished exploiting standard quantum mechanical formulae
with the insertion of eigenstates completeness inside the expression
of the expectation value. By inserting two completeness of harmonic
oscillator eigenstates, respectively into the numerator and into the
denominator of Eq.~\eqref{eq:DB-def-normalised} one gets

\begin{align}
  \langle 0,x_0^{\mathrm{f}}| \hat{q}^{2k}(x_0) | 0,x_0^{\mathrm{i}}\rangle
  &=
  \sum_{m,n\ge 0}
  \langle 0,x_0^{\mathrm{f}}|\psi_m\rangle\,
  \langle \psi_m|\hat{q}^{2k}(x_0)|\psi_n\rangle\,
  \langle \psi_n|0,x_0^{\mathrm{i}}\rangle
  \nonumber\\
  &=
  \sum_{m,n\ge 0}
  \psi_m(0)\,\psi_n(0)\,
  \langle \psi_m|\hat{q}^{2k}|\psi_n\rangle\,
  e^{-\,\frac{i}{\hbar}E_m(x_0^{\mathrm{f}}-x_0)}\,
  e^{-\,\frac{i}{\hbar}E_n(x_0-x_0^{\mathrm{i}})}\nonumber \\
  \langle 0,x_0^{\mathrm{f}}|0,x_0^{\mathrm{i}}\rangle
  &= \sum_{n\ge 0}|\psi_n(0)|^2 e^{-\,\frac{i}{\hbar}E_nT} = \Delta(T)
  \label{eq:DB-numerator-spectral}
\end{align}
By exploiting the definition of the quantum harmonic oscillator excitation
spectrum
\begin{align}
E_n = \hbar\Omega \left( n + \frac{1}{2} \right)
\end{align}
and the fact that all odd eigenfunctions have a node at $q=0$ the
average value of Eq.~\eqref{eq:DB-def-normalised} can be conveniently
rewritten as
\begin{align}
  \big\langle \hat{q}^{2k}(x_0)\big\rangle_{\text{D.B.}}
  &=
  \frac{1}{\Delta(T)}
  \sum_{m,n\ge 0}
  \psi_m(0)\,\psi_n(0)\,
  \langle \psi_m|\hat{q}^{2k}|\psi_n\rangle\,
  e^{-\,\frac{i}{\hbar}(E_n-E_m)(x_0-x_0^{\mathrm{i}})}\,
  e^{-\,\frac{i}{\hbar}E_mT} \nonumber \\
  & = \frac{1}{\Delta(T)}
  \sum_{m,n\ge 0}
  \psi_{2m}(0)\,\psi_{2n}(0)\,
  \langle \psi_{2m}|\hat{q}^{2k}|\psi_{2n}\rangle\,
  e^{-2i\Omega(n-m)(x_0-x_0^{\mathrm{i}})}\,
  e^{-\frac{i}{\hbar} E_{2m} T}, \nonumber \\
  \label{eq:DB-spectral-general}
\end{align}
since all frequency differences in the argument of the exponential are
\begin{align}
  E_{2n}-E_{2m}=2(n-m)\hbar\Omega.
\end{align}
For the sake of simplicity let us specialize the discussion here in
the main text for $k=1$, i.e., to the case of the expectation value
$\langle \hat{q}^2(x_0)\rangle_{\text{D.B.}}$, leaving the derivation
for higher powers of $\hat{q}^2$ App.~\ref{app:ho-q4-q6}. At this
step, in order to further simplify the expression in
Eq.~\eqref{eq:DB-spectral-general} it is necessary to write the
position operator in terms of the creation and destruction operators
$\hat q=\sqrt{\frac{\hbar}{2m\Omega}}(a+a^\dagger)$, from which, using
also the commutation relation $[a,a^\dagger]=1$, we have for $\hat{q}^2$ the following
expression
\begin{align}
  \hat{q}^{2}
  = \frac{\hbar}{2m\Omega}\Bigl(a a + a^\dagger a^\dagger + 2 \hat{N} + 1\Bigr),
  \label{eq:q2-ladder}
\end{align}
where $\hat{N}$ is the number operator, so that
\begin{align}
  \langle \psi_{2m}|\hat{q}^2|\psi_{2n}\rangle = \delta_{n,m} \left( 4n + 1 \right) + \delta_{m,n-1} \sqrt{2n}\sqrt{2n-1} +
  \delta_{m,n+1} \sqrt{2n+1}\sqrt{2n+2}   
  \label{eq:q2-bra-ket}
\end{align}
By plugging the expression of $\langle
\psi_{2m}|\hat{q}^2|\psi_{2n}\rangle$ of Eq.~\eqref{eq:q2-bra-ket}
into the formula for the expectation value of $\hat{q}^2$ in
Eq.~\eqref{eq:DB-spectral-general} it is possible, by means of some
intermediate steps reported in App.~\ref{app:q2-dirichlet-continuum},
to reach for the following final expression:
\begin{align}
  \big\langle \hat{q}^2(x_0)\big\rangle_{\text{D.B.}(T)}
  &=
  \frac{\hbar}{2m\Omega}
  \Big[
    \alpha(T)
    + \gamma_{-}(T)e^{-i2\Omega (x_0-x_0^{\mathrm{i}})}
    + \gamma_{+}(T)e^{+i2\Omega (x_0-x_0^{\mathrm{i}})}
  \Big],
  \label{eq:q2-spec-main}
\end{align}
where the coefficients $\alpha(T)$, $\gamma_{-}(T)$ and
$\gamma_{+}(T)$ have the followings expressions
\begin{align}
  \alpha(T)
  &\equiv
  \frac{
    \sum_{n=0}^\infty (4n+1)\,
    e^{-\,\frac{i}{\hbar}E_{2n}T}\,
    |\psi_{2n}(0)|^2}{
    \Delta(T)},
  \label{eq:alpha-def-main}
  \\[0.3em]
  \gamma_{+}(T)
  &\equiv
  \frac{
    \sum_{n=0}^\infty
    \sqrt{(2n+1)(2n+2)}\,
    e^{-\,\frac{i}{\hbar}E_{2n}T}\,
    \psi_{2n+2}(0)\psi_{2n}(0)
  }{\Delta(T)},
  \label{eq:gamma-plus-def-main}
  \\[0.3em]
  \gamma_{-}(T)
  &\equiv
  \frac{
    \sum_{n=0}^\infty
    \sqrt{(2n+1)(2n+2)}\,
    e^{-\,\frac{i}{\hbar}E_{2n+2}T}\,
    \psi_{2n+2}(0)\psi_{2n}(0)
  }{\Delta (T)}.
  \label{eq:gamma-minus-def-main}
\end{align}
As is detailed in App.~\ref{app:q2-dirichlet-continuum} the three
summations appearing on the right-hand side of
Eqns.~\eqref{eq:alpha-def-main},\eqref{eq:gamma-plus-def-main},\eqref{eq:gamma-minus-def-main},
can be manipulated yielding closed expression for the coefficients in
terms of transcendental functions
\begin{align}
  \alpha(T) &= -\,i\,\cot(\Omega T), \nonumber \\  
  \gamma_{+}(T) &= \tfrac12\bigl[1 + i\cot(\Omega T)\bigr] \nonumber \\ 
  \gamma_{-}(T) &= \tfrac12\bigl[-1 + i\cot(\Omega T)\bigr],
  \label{eq:alpha-gamma-cont-main}
\end{align}
leading to the final explicit result
\begin{align}
  \big\langle \hat{q}^2(x_0)\big\rangle_{\text{D.B.}(T)}
  &=
  \frac{i\hbar}{m\Omega\sin(\Omega T)}\,
  \sin\!\bigl(\Omega(x_0-x_0^{\mathrm{i}})\bigr)\,
  \sin\!\bigl(\Omega(x_0^{\mathrm{f}}-x_0)\bigr).
  \label{eq:q2-db-cont-main}
\end{align}
Despite the compactness of the expression in
Eq.~\eqref{eq:q2-db-cont-main} it is the one in
Eq.~\eqref{eq:q2-spec-main} which better allows to single out the
Fourier components of $\big\langle
\hat{q}^2(x_0)\big\rangle_{\text{D.B.}(T)}$, reading as $\omega =
\lbrace 0,\pm 2\Omega \rbrace$.  More generally, $\hat q^{2k}$ is a
degree-$2k$ polynomial in the ladder operators, so that it is easy to
argue from Eq.~\eqref{eq:q2-bra-ket} that only matrix elements among
eigenstates at a finite distance in the spectrum survive, the larger the power of $\hat q$, the larger the maximal distance
between the eigenstates connected by the matrix elements. In general, if we
denote as $\mathcal{S}_\omega(k)$ the set of frequencies in the
Fourier spectrum of $\hat{q}^{2k}$, it can be easily argued that one
has
\begin{align}
  \mathcal{S}_\omega(k) = \left\lbrace \omega \in \mathbb{R}~|~ \omega = \pm 2 n \Omega; ~ n=0,\ldots,k~\right\rbrace
  \label{eq:q2k-harmonics}
\end{align}
Explicit results for $k=2$ and $k=3$ are reported in
App.~\ref{app:ho-q4-q6}.\\

Now that we have discussed the analytical shape of $\big\langle
\hat{q}^{2k}(x_0)\big\rangle_{\text{D.B.}}$ for a generic value of
$k$, elucidating why in Fourier space it has a number equal to $2k+1$
of equally spaced lines at distance $2\Omega$ due to the contribution
of \emph{all} the harmonic oscillator excited states, let us explain
how the observable $\big\langle
\hat{q}^{2k}(x_0)\big\rangle_{\text{D.B.}}$ is measured along the
constrained symplectic dynamics simulations with Dirichlet fixed
boundary conditions. To this aim let us recall how, with boundary
conditions in Minkowskian time fixed to $q(x_0^{\textrm{f}}) =
q(x_0^{\textrm{i}}) = 0$, it is still possible to decompose the field
$q(x_0)$ in a set of orthonormal modes in Fourier space. This is a
crucial step, since the constraints which characterize the symplectic
quantization measure and dynamics, as thoroughly discussed in
Sec.~\ref{sec:Harm-Osc-dynamics} and
Sec.~\ref{sec:numerical-algorithm}, are imposed on the components in
Fourier space of the field $q(x_0)$. In a quite straightforward
manner, the construction for Dirichlet boundary conditions with the
field fixed at zero is the same as for periodic boundary conditions,
the only difference being the choice of Fourier basis.  For Dirichlet
boundaries on $[x_0^{\mathrm{i}},x_0^{\mathrm{f}}]$ with
$q(x_0^{\mathrm{i}})= q(x_0^{\mathrm{f}})=0$ the field can be expanded
in the Sine basis
\begin{align}
  q(x_0)
  &=
  \sum_{n=1}^{\infty} \tilde{q}(n)\,
  \sqrt{\frac{2}{T}}\,
  \sin\!\left(
    \frac{n\pi(x_0-x_0^{\mathrm{i}})}{T}
  \right),
  \label{eq:dirichlet-sine-expansion}
\end{align}
whose orthonormality reads
\begin{align}
  \frac{2}{T}
  \int_{x_0^{\mathrm{i}}}^{x_0^{\mathrm{f}}} \! dx_0\,
  \sin\!\left(\frac{n\pi(x_0-x_0^{\mathrm{i}})}{T}\right)
  \sin\!\left(\frac{n'\pi(x_0-x_0^{\mathrm{i}})}{T}\right)
  = \delta_{n,n'} .
\end{align}
In this basis the quadratic action is diagonal and the distribution of
Dirichlet modes factorizes, exactly as in the periodic case. The
correspondence between mode correlators obtained from constrained
averages and the corresponding quantum-mechanical correlators is thus
unchanged in form and can again be stated mode by mode.  We refer to
App.~\ref{app:dirichlet-mode-measure} for the explicit measure and for
the detailed mode-by-mode correspondence. The difference in
considering a Minkowskian time lattice with $M-1$ points, except the
fixed ends, rather than a continuum time interval, simply amounts to
have a finite number of modes in the expansion, which, assuming
$x_0-x_0^{\mathrm{i}} = a\ell$, where $a$ is the lattice spacing and $T = a M$,
reads
\begin{align}
  q(\ell)
  &=
  \sum_{n=1}^{M-1} \tilde{q}(n)\,
  \sqrt{\frac{2}{M}}\,
  \sin\!\left(
    \frac{n\pi \ell}{M}
  \right),
  \label{eq:dirichlet-sine-expansion}
\end{align}

Crucially, since also with fixed boundary conditions we can expand the field in an orthonormal
sine basis
and thus impose the constraints which guarantee the stability of
symplectic quantization dynamics as constraints between the real and
imaginary parts of the Fourier modes, see
Sec.~\ref{sec:Harm-Osc-dynamics}, also with Dirichlet boundary
conditions we can decompose the real and imaginary parts of the
analytically continued field $q(\ell) \in \mathbb{C} \rightarrow
q(\ell)=q_R(\ell)+i q_I(\ell)$ as done in the periodic boundary case:
\begin{align}
  q_R(\ell) = q_R^E(\ell) + q_R^O(\ell),
  \qquad
  q_I(\ell) = q_I^E(\ell) + q_I^O(\ell).
  \label{eq:qrqi-eo-dirichlet}
\end{align}
The constraints specifying the stable manifold (see
Eqs.~\eqref{eq:stable-surface0} and~\eqref{eq:lattice-odd-even}) impose
the same relations between the even/odd components as in the periodic
case,
\begin{align}
  q_I^E(\ell)=+\,q_R^E(\ell),
  \qquad
  q_I^O(\ell)=-\,q_R^O(\ell),
  \label{eq:constraint-eo-dirichlet}
\end{align}
so that the complex field entering Dirichlet expectation values can be
written entirely in terms of the two real constrained variables
$q_R^E(\ell)$ and $q_R^O(\ell)$,
\begin{align}
  q(\ell)
  = q_R(\ell)+i q_I(\ell)
  = (1+i)\,q_R^E(\ell) + (1-i)\,q_R^O(\ell).
  \label{eq:q-from-qE-qO}
\end{align}
In practice, in the simulation we evolve $q_R^E(\ell,\tau)$ and
$q_R^O(\ell,\tau)$ along the constrained symplectic flow and construct
the operator insertion $q^{2k}(\ell)$ at each intrinsic-time step using
Eq.~\eqref{eq:q-from-qE-qO}.  Writing $q_E(\ell)\equiv q_R^E(\ell)$ and
$q_O(\ell)\equiv q_R^O(\ell)$, the lowest even powers read
\begin{align}
  q^2(\ell)
  &= \Bigl[(1+i)q_E(\ell)+(1-i)q_O(\ell)\Bigr]^2
  \nonumber\\
  &= 2i\Bigl[q_E^2(\ell)-q_O^2(\ell)\Bigr]
   +4\,q_E(\ell)\,q_O(\ell),
  \label{eq:q2-from-qE-qO}
  \\[0.4em]
  q^4(\ell)
  &= \Bigl[(1+i)q_E(\ell)+(1-i)q_O(\ell)\Bigr]^4
  \nonumber\\
  &= -4\Bigl[q_E^4(\ell)+q_O^4(\ell)\Bigr]
     +24\,q_E^2(\ell)\,q_O^2(\ell)
     \;+\;\text{(terms odd in $q_E$ or $q_O$)},
  \label{eq:q4-from-qE-qO}
  \\[0.4em]
  q^6(\ell)
  &= \Bigl[(1+i)q_E(\ell)+(1-i)q_O(\ell)\Bigr]^6
  \nonumber\\
  &= -8i\Bigl[q_E^6(\ell)-q_O^6(\ell)\Bigr]
     +120\,i\Bigl[q_E^4(\ell)q_O^2(\ell)-q_E^2(\ell)q_O^4(\ell)\Bigr]
     \;+\;\text{(terms odd in $q_E$ or $q_O$)}.
  \label{eq:q6-from-qE-qO}
\end{align}
The terms labeled as ``odd'' contain at least one odd power of $q_E$
or $q_O$ and vanish in expectation values with the constrained
measure, because the two fields are statistically independent in the
harmonic oscillator case and the measure is Gaussian with zero
mean. Therefore, Dirichlet expectation values can be expressed
directly in terms of mixed even moments,
\begin{align}
  \big\langle q^2(\ell)\big\rangle_{\text{D.B.}(T)}
  &= 2i\Bigl[\big\langle q_E^2(\ell)\big\rangle_{\text{D.B.}(T)}
            -\big\langle q_O^2(\ell)\big\rangle_{\text{D.B.}(T)}\Bigr],
  \label{eq:q2-DB-from-qE-qO}
  \\[0.4em]
  \big\langle q^4(\ell)\big\rangle_{\text{D.B.}(T)}
  &= -4\Bigl[\big\langle q_E^4(\ell)\big\rangle_{\text{D.B.}(T)}
            +\big\langle q_O^4(\ell)\big\rangle_{\text{D.B.}(T)}\Bigr]
     +24\,\big\langle q_E^2(\ell)q_O^2(\ell)\big\rangle_{\text{D.B.}(T)},
  \label{eq:q4-DB-from-qE-qO}
  \\[0.4em]
  \big\langle q^6(\ell)\big\rangle_{\text{D.B.}(T)}
  &= -8i\Bigl[\big\langle q_E^6(\ell)\big\rangle_{\text{D.B.}(T)}
             -\big\langle q_O^6(\ell)\big\rangle_{\text{D.B.}(T)}\Bigr]\nonumber\\
     &\qquad+120\,i\Bigl[\big\langle q_E^4(\ell)q_O^2(\ell)\big\rangle_{\text{D.B.}(T)}
                 -\big\langle q_E^2(\ell)q_O^4(\ell)\big\rangle_{\text{D.B.}(T)}\Bigr],
  \label{eq:q6-DB-from-qE-qO}
\end{align}
and for arbitrary $k$ one may write the estimator compactly as
\begin{align}
  \big\langle q^{2k}(\ell)\big\rangle_{\text{D.B.}(T)}
  =
  \sum_{j=0}^{k}
  \binom{2k}{2j}\,
  (1+i)^{2j}(1-i)^{2k-2j}\,
  \big\langle q_E^{2j}(\ell)\,q_O^{2k-2j}(\ell)\big\rangle_{\text{D.B.}(T)},
  \label{eq:q2k-DB-from-qE-qO-general}
\end{align}
where only even powers appear because all mixed moments with an odd
power of $q_E$ or $q_O$ vanish.  Equations
\eqref{eq:q2-DB-from-qE-qO}--\eqref{eq:q2k-DB-from-qE-qO-general} are the
explicit bridge between the quantum-mechanical Dirichlet expectation
values and the quantities computed from the constrained symplectic
dynamics.\\

\begin{figure}[t]
    \centering
    \includegraphics[width=0.75\linewidth]{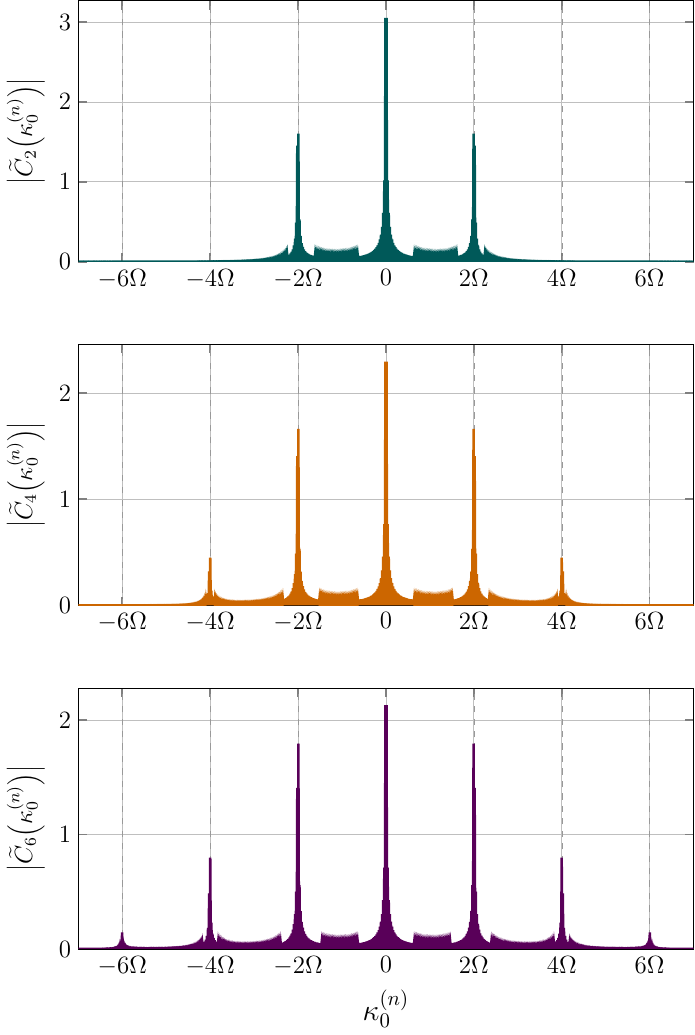}
    \caption{Absolute value of the discrete transform
      $|\widetilde{C}_{2k}(\kappa_0^{(n)})|$ of the Dirichlet expectation
      values $\langle q^{2k}(\ell)\rangle_{\text{D.B.}(T)}$ for
      $2k=2,4,6$ (from top to bottom). In each panel sharp peaks appear
      at a finite set of positive Dirichlet frequencies
      $\kappa_0^{(n)}$ clustered around integer multiples of $2\Omega$,
      as indicated on the horizontal axis. The peak structure matches
      the analytic predictions for the harmonic oscillator and
      demonstrates the discrete spectrum of
      $\langle q^{2k}(\ell)\rangle_{\text{D.B.}(T)}$ arising from the
      tower of excited states.}
    \label{fig:q2k-fourier}
\end{figure}

On the lattice we discretize the Dirichlet interval $T$ into $M$ steps
of size $a$ (so $T=Ma$), impose Dirichlet boundary conditions at the
endpoints, and measure correlators on interior sites
$\ell=1,\ldots,M-1$. The equal--time Dirichlet correlator reads
(App.~\ref{app:dirichlet-lattice})
\begin{align}
  \big\langle q(\ell)^2\big\rangle_{\text{D.B.}(T)}
  &=
  i\hbar
  \sum_{n=1}^{M-1}
  \frac{2}{M}\,
  \frac{
    \sin^2(\kappa_0^{(n)} \ell a)}
  {
    \displaystyle
     \widehat{\kappa}_0^{(n)\,2} - \Omega^2
  }.
  \label{eq:q2-db-lattice-main}
\end{align}
where we have defined
\begin{align}
  \kappa_0^{(n)} &= \frac{\pi n}{T}=\frac{\pi n}{Ma} \label{eq:kappa0n-lattice} \\
  \widehat{\kappa}_0^{(n)\,2}
  & = \frac{4}{a^2}\,\sin^2\!\Bigl(\frac{\kappa_0^{(n)}a}{2}\Bigr) =
  \frac{4}{a^2}\,
  \sin^2\!\Bigl(\frac{\pi n}{2M}\Bigr),
  \label{eq:kappa0hat-def}
\end{align}

\begin{figure}[H]
    \centering
    \includegraphics[width=0.75\linewidth]{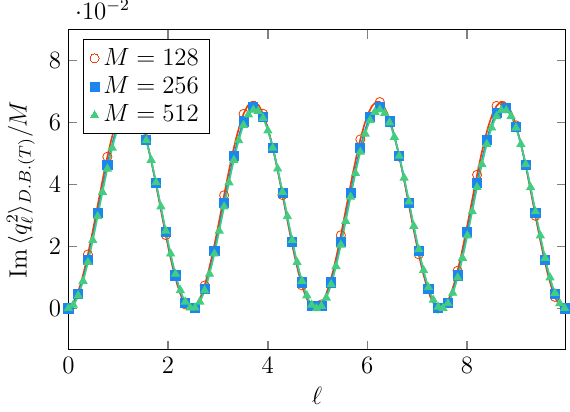}
    \caption{Imaginary part of the equal--time Dirichlet correlator,
      $\Im\,\langle q(\ell)^2\rangle_{\text{D.B.}(T)}/M$, as a function
      of $x_0$ at resonance $T=4\pi/\Omega$ for $\Omega=2.5$,
      lattice spacing $a=0.1$ and flow time $\Delta\tau=10^5$. Data
      for $M=128,256,512$ collapse onto a single curve once divided by
      $M$, demonstrating the approximately linear growth with $M$ of the
      resonant amplitude discussed in App.~\ref{app:dirichlet-resonance}.}
    \label{fig:behaviourresonant}
\end{figure}

For resonant time extents, such as $T=4\pi/\Omega$, the denominator in
Eq.~\eqref{eq:q2-db-lattice-main} becomes small for the mode(s) whose
Dirichlet lattice eigenvalue $\widehat{\kappa}_0^{(n)\,2}$ lies closest
to $\Omega^2$. This yields a strong enhancement of the corresponding
term(s) in the mode sum and increases the overall amplitude of
$\langle q(\ell)^2\rangle_{\text{D.B.}(T)}$ with $M$. In the parameter
range explored in this work we find that, at $T=4\pi/\Omega$, this
enhancement is well described by an approximately linear growth with
$M$, so that the rescaled quantity
$\langle q(\ell)^2\rangle_{\text{D.B.}(T)}/M$ collapses for different
$M$, as shown in Fig.~\ref{fig:behaviourresonant}. A quantitative
discussion of the resonant enhancement and its relation to truncated
even--state sums in the continuum spectral representation is given in
App.~\ref{app:dirichlet-resonance}.

Finally, to display the discrete harmonic content predicted by
Eq.~\eqref{eq:q2k-harmonics}, we analyze the Fourier spectrum of the
Dirichlet expectation values measured in coordinate space. For each
even power $2k$ we compute the discrete transform on the interior sites
$\ell=1,\ldots,M-1$,
\begin{align}
  \widetilde{C}_{2k}\!\bigl(\kappa_0^{(n)}\bigr)
  &\equiv\frac{1}{\sqrt{M}}
  \sum_{\ell=1}^{M-1}
  \exp(i\kappa_0^{(n)} x_0^{(\ell)})\,
  \big\langle q^{2k}(\ell)\big\rangle_{\text{D.B.}(T)},
  \label{eq:fourier-DB-def}
\end{align}
where
\begin{align}
	  \qquad
  x_0^{(\ell)} = x_0^{\mathrm{i}} + \ell a,
  \qquad
  \kappa_0^{(n)} \equiv \frac{\pi n}{T},
  \quad n=1,\ldots,M-1 .
\end{align}
Using the complex exponential at this analysis step is convenient
because it displays simultaneously positive and negative frequency
components. In the continuum limit
$|\widetilde{C}_{2k}(\kappa_0^{(n)})|$ is therefore expected to exhibit
narrow peaks located at the harmonics
$\kappa_0^{(n)} \simeq 2r\Omega$ with $r=0,1,\ldots,k$, and no additional
lines. Fig.~\ref{fig:q2k-fourier} shows the numerical spectra
$|\widetilde{C}_{2k}(\kappa_0^{(n)})|$ for $2k=2,4,6$: in all cases we
observe a finite set of sharp peaks at the expected positive harmonics,
with rapidly decreasing amplitude at higher multiples. This provides a
stringent test that the constrained symplectic dynamics reproduces the
excited--state contributions and yields the expected discrete spectrum
for Dirichlet expectation values. The complete operator derivations for
$k=1,2$ and the explicit form used for $k=3$ are reported in
App.~\ref{app:ho-q4-q6}.

\newpage
\section{Conclusions}

In this paper we have presented a new functional approach to quantum
field theory and quantum mechanics which we called {\it constrained
  symplectic quantization}. A previous formalization of the symplectic
quantization approach was presented
in~\cite{Gradenigo:2021sck,Gradenigo:2021orx,Gradenigo:2024pwy} and
numerically tested in~\cite{Giachello:2024wqt}, in the case of a
weakly interacting scalar field theory. This first formulation of the
theory had two major limitations: it was by construction ill-defined
in the free case and, most importantly, we were able to prove that the
correlation functions generated by this approach were \emph{not}
equivalent to those generated by the Feynman path integral. Nevertheless, the numerical results of~\cite{Giachello:2024wqt}
were quite encouraging: they provided evidence that symplectic
quantization can capture the causal structure of $1+1$-dimensional
spacetime by qualitatively reproducing the free Feynman propagator for
a weakly interacting scalar field theory, a result inaccessible within
the Euclidean lattice field theory formulation. Furthermore, within
that approach, we obtained stable dynamics for an interacting field
theory. In the present framework of \emph{constrained} symplectic
quantization, however, the appropriate way to handle nonlinear
interactions is still under investigation. Indeed, finding the correct
protocol for treating interactions is model-dependent and presents
technical difficulties similar to those encountered in the choice of
integration paths in analytic continuation techniques used by other
established approaches to circumvent the sign problem in the estimation
of functional integrals with oscillating amplitudes~\cite{Alexandru:2020wrj}.\par 

In the present paper we have shown how the correct way to recover a
perfect equivalence between symplectic quantization and quantum
mechanics is to consider the analytic continuation of all the original
fields and of the action from $\mathbb{R}$ to $\mathbb{C}$, by also
including specific constraints on the integration paths in
$\mathbb{C}$ which guarantee convergence of the symplectic
quantization dynamics and of the corresponding microcanonical
generating functional. From this perspective our method is close in
spirit to the Lefschetz-Thimbles approach~\cite{Alexandru:2020wrj},
which also entails the analytic continuation of both fields and action
to $\mathbb{C}$ and constraints to stable manifolds. On the other
hand, there are also remarkable differences, since the
Lefschetz-Thimbles approach imposes constraints directly on the
functional integral measure while symplectic quantization imposes them on the microscopic deterministic dynamics which generates
the measure itself. From this perspective, constrained symplectic
quantization acts in a sense on a more fundamental level than the
Lefschetz-Thimbles approach.\par

The main results of this paper can be listed as follows: we
presented the new formalism of \emph{constrained} symplectic
quantization and proved its equivalence with the Feynman
path-integral approach to quantum mechanics. We then showed how the
general formalism applies to the case of the quantum harmonic
oscillator, showing the explicit form of the constraints on the functional
measure and on the dynamics. Our main goal
has been to show how observables and correlation functions in real
Minkowskian time can be perfectly reproduced by constrained
symplectic quantization fictitious dynamics, capturing features which
are by construction inaccessible to any importance sampling protocol
built in Euclidean time. Most importantly the
oscillating behavior of the two point correlation function $\langle
\hat{q}(x_0) \hat{q}(x_0')\rangle$ and the discrete equally-spaced
(with distance $2\Omega$) Fourier spectrum of the expectation values
of higher order powers of the position operator $\langle
q^{2k}(x_0)\rangle$, provide evidence of the presence in the dynamics
of all excited states of the harmonic oscillator.

All these results suggest that the constrained symplectic quantization
approach might establish itself as a new robust method to do real-time
non-perturbative numerical field theory and numerical quantum
mechanics, overcoming the sign problem. In this direction, there is a companion paper on numerical
tests for a free scalar field theory that will appear
soon~\cite{GSG25_2}, and a paper devoted to the anharmonic oscillator
which is in preparation~\cite{GSG25_3}.

\acknowledgments
We thank R. Livi and L. Salasnich for useful
discussions. G.G. acknowledges partial support from the project
MIUR-PRIN2022, ``Emergent Dynamical Patterns of Disordered Systems
with Applications to Natural Communities'', code 2022WPHMXK.

\newpage

\appendix
\section*{Appendix}
\section[\texorpdfstring{Asymptotic properties of the coefficients $c_j(M,n)$}%
{Asymptotic properties of the coefficients cj(M,n)}]%
{Asymptotic properties of the coefficients $c_j(M,n)$}
\label{app:coeff}
In this Appendix we analyze the asymptotic behavior of $c_j(M,n)$ defined in Eq. \eqref{eq:coeff}
\begin{equation}
	c_j(M,n) = \left(\frac{2}{\hbar}\right)^j\sum_{k=0}^j \binom{M-n}{k}\frac{1}{(j-k)!}\frac{(-1)^k}{M^k}\,,
\end{equation}
by expanding their expression in powers of $\frac{1}{M}$. If the constant term (order $M^0$) vanishes, the leading behavior is at least $\mathcal{O}(1/M)$.

Consider the term involving the binomial coefficient
\begin{equation}
	T_k(M) = \binom{M-n}{k} \frac{1}{M^k} \,.
\end{equation}
Expanding the binomial coefficient definition
\begin{align}
	T_k(M) &= \frac{(M-n)(M-n-1)\cdots(M-n-k+1)}{k! \, M^k} \nonumber \\
	&= \frac{1}{k!} \prod_{i=0}^{k-1} \left( \frac{M - (n+i)}{M} \right) \nonumber \\
	&= \frac{1}{k!} \prod_{i=0}^{k-1} \left( 1 - \frac{n+i}{M} \right) \,.
\end{align}
For large $M$, we expand the product using $\prod (1-\epsilon) \approx 1 - \epsilon$:
\begin{equation}
	\prod_{i=0}^{k-1} \left( 1 - \frac{n+i}{M} \right) = 1 - \frac{1}{M} \sum_{i=0}^{k-1} (n+i) + \mathcal{O}\left(\frac{1}{M^2}\right) \,.
\end{equation}
The sum in the linear term is an arithmetic progression
\begin{equation}
	\sum_{i=0}^{k-1} (n+i) = nk + \sum_{i=0}^{k-1} i = nk + \frac{k(k-1)}{2} \,.
\end{equation}
Thus, the expansion for $T_k(M)$ is
\begin{equation}
	T_k(M) = \frac{1}{k!} - \frac{1}{M \, k!} \left( nk + \frac{k(k-1)}{2} \right) + \mathcal{O}\left(\frac{1}{M^2}\right) \,.
\end{equation}

Substitute the expansion of $T_k(M)$ back into the definition of $c_j(M,n)$
\begin{align}
	c_j(M,n) &= \left(\frac{2}{\hbar}\right)^j\sum_{k=0}^j \frac{(-1)^k}{(j-k)!} \, T_k(M) \nonumber \\
	&= \left(\frac{2}{\hbar}\right)^j\sum_{k=0}^j \frac{(-1)^k}{(j-k)!} \left[ \frac{1}{k!} - \frac{1}{M \, k!} \left( nk + \frac{k(k-1)}{2} \right) + \dots \right] \,.
\end{align}
We can group terms by powers of $\frac{1}{M}$
\begin{equation}
	c_j(M,n) = A_0 + \frac{A_1}{M} + \mathcal{O}\left(\frac{1}{M^2}\right) \,.
\end{equation}
we now show that $A_0 = 0$ for all $j \ge 1$.
The coefficient of order $M^0$ is
\begin{equation}
	A_0 = \left(\frac{2}{\hbar}\right)^j\sum_{k=0}^j \frac{(-1)^k}{(j-k)! \, k!} \,.
\end{equation}
Multiplying and dividing by $j!$ allows us to express this in terms of binomial coefficients $\binom{j}{k}$
\begin{equation}
	A_0 = \frac{1}{j!} \left(\frac{2}{\hbar}\right)^j\sum_{k=0}^j (-1)^k \frac{j!}{k!(j-k)!} = \frac{1}{j!} \left(\frac{2}{\hbar}\right)^j\sum_{k=0}^j \binom{j}{k} (-1)^k \,.
\end{equation}
By the binomial theorem, $\sum_{k=0}^j \binom{j}{k} (-1)^k = (1-1)^j = 0$ for $j \ge 1$. Therefore
\begin{equation}
	A_0 = 0 \quad \forall j \ge 1 \,.
\end{equation}

Since $A_0 = 0$, the series starts at order $\frac{1}{M}$. The coefficient $A_1$ is given by
\begin{equation}
	A_1 = - \frac{1}{j!} \left(\frac{2}{\hbar}\right)^j\sum_{k=0}^j (-1)^k \binom{j}{k} \left( nk + \frac{k(k-1)}{2} \right) \,.
\end{equation}
Let $Q(k) = nk + \frac{k(k-1)}{2}$. This is a polynomial in $k$ of degree 2.

Let $P(x)$ be any function defined on the integers. We define the Shift Operator $E$ and the Identity Operator $I$ by
		\begin{equation}
			E P(x) = P(x+1), \quad I P(x) = P(x)\,.
		\end{equation}
		The Forward Difference Operator $\Delta$ is defined as
		\begin{equation}
			\Delta P(x) = P(x+1) - P(x) = (E - I) P(x)\,.
		\end{equation}
		To find the $j$-th finite difference $\Delta^j P(0)$, we expand the operator $(E-I)^j$ using the binomial theorem
		\begin{align}
			\Delta^j P(0) &= (E - I)^j P(0) \nonumber \\
			&= \sum_{k=0}^j \binom{j}{k} E^k (-1)^{j-k} P(0) \nonumber \\
			&= \sum_{k=0}^j \binom{j}{k} (-1)^{j-k} P(k)\,.
		\end{align}
		Factoring out $(-1)^j$ and noting that $(-1)^{-k} = (-1)^k$
		\begin{equation}
			\Delta^j P(0) = (-1)^j \sum_{k=0}^j (-1)^k \binom{j}{k} P(k)\,.
		\end{equation}
		Multiplying both sides by $(-1)^j$ (since $(-1)^{2j}=1$) gives the identity
		\begin{equation} \label{eq:identity}
			\sum_{k=0}^j (-1)^k \binom{j}{k} P(k) = (-1)^j \Delta^j P(0)\,.
		\end{equation}
Now, if $P(x) = c_d x^d + \dots$  is a polynomial of degree $d$, then $\Delta P(x)$ is a polynomial of degree $d-1$
			\begin{align}
				\Delta P(x) &= c_d (x+1)^d + \dots - (c_d x^d + \dots) \nonumber \\
				&= c_d (x^d + d x^{d-1} + \dots) - c_d x^d + \dots \nonumber \\
				&= c_d d x^{d-1} + \mathcal{O}(x^{d-2})\,,
			\end{align}
			where the $x^d$ term cancels, reducing the degree by exactly 1.
		By induction, applying $\Delta$ exactly $j$ times reduces the degree by $j$.
			So, if $j = d$, $\Delta^j P(x)$ is a non-zero constant ($c_d \cdot d!$), while if $j > d$, $\Delta^j P(x) = 0$.
		Combining this with Eq. \eqref{eq:identity}, we obtain the vanishing condition
		\begin{equation}
			\sum_{k=0}^j (-1)^k \binom{j}{k} P(k) = 0 \quad \text{if } j > \deg(P).
		\end{equation}
The sum $A_1$ therefore represents the $j$-th finite difference of $Q(k)$ evaluated at 0
\begin{equation}
	\sum_{k=0}^j (-1)^k \binom{j}{k} Q(k) = (-1)^j \Delta^j Q(0) \,.
\end{equation}
Since $\Delta^j Q(k) = 0$ if $j > \text{deg}(Q)$.
Then, if $j=1, 2$, then $j \le \text{deg}(Q)=2$, so $A_1 \neq 0$. The order is exactly $\frac{1}{M}$. Otherwise, if $j > 2$, then $j > \text{deg}(Q)$, so $A_1 = 0$. The order is $\mathcal{O}\left(\frac{1}{M^2}\right)$ or smaller.

Since the constant term $A_0$ is identically zero for all $j \ge 1$, the leading non-zero term is of order at least $\frac{1}{M}$. Thus
\begin{equation}
	c_j(M,n) = \mathcal{O}\left(\frac{1}{M}\right) \,.
\end{equation}

\section{Two-point function on a periodic domain (continuum and discrete)}
\label{app:periodic-twopoint}

In this Appendix, we derive the harmonic-oscillator two-point function
with \emph{periodic} boundary conditions, in two
ways.
We first use the spectral representation and 
we then present the lattice derivation on a periodic time lattice,
which matches the continuum result in the $a\to 0$ limit at fixed
$T=Ma$.

\subsection{Continuum derivation from the tower of eigenstates}
\label{app:periodic-twopoint:spectral}

We consider the harmonic oscillator Hamiltonian
\begin{align}
\hat H = \frac{\hat p^2}{2m}+\frac{m\Omega^2}{2}\hat q^2,
\qquad
E_n=\hbar\Omega\Bigl(n+\frac12\Bigr),
\qquad
\hat H|\psi_n\rangle=E_n|\psi_n\rangle,
\qquad n=0,1,2,\dots
\end{align}
and the Heisenberg operator $\hat q(x_0)=e^{\frac{i}{\hbar}\hat H x_0}\hat q\,e^{-\frac{i}{\hbar}\hat H x_0}$.
Periodic boundary conditions in time of length $T$ correspond to the trace:
\begin{align}
Z_{\rm P.B.}(T) \equiv \Tr\!\left(e^{-\frac{i}{\hbar}\hat H T}\right),
\qquad
\langle \hat q(x_0)\hat q(x_0')\rangle_{\rm P.B.}
\equiv
\frac{1}{Z_{\rm P.B.}(T)}\Tr\!\left(e^{-\frac{i}{\hbar}\hat H T}\hat q(x_0)\hat q(x_0')\right).
\label{eq:pb_twopt_trace_def}
\end{align}
Using time-translation invariance on the circle, the correlator depends
only on $\Delta\equiv x_0-x_0'$ modulo $T$.
Choose the representative $\Delta\in[0,T]$ and rewrite
\begin{align}
\langle \hat q(x_0)\hat q(x_0')\rangle_{\rm P.B.}
=
\frac{1}{Z_{\rm P.B.}(T)}
\Tr\!\left(
e^{-\frac{i}{\hbar}\hat H (T-\Delta)}\hat q\,
e^{-\frac{i}{\hbar}\hat H \Delta}\hat q
\right).
\label{eq:pb_twopt_trace_shifted}
\end{align}
We then insert the identity $\mathbb{I}=\sum_{n=0}^\infty|\psi_n\rangle\langle\psi_n|$
twice into Eq.~\eqref{eq:pb_twopt_trace_shifted}:
\begin{align}
\langle \hat q(x_0)\hat q(x_0')\rangle_{\rm P.B.}
=
\frac{1}{Z_{\rm P.B.}(T)}
\sum_{n,m=0}^\infty
e^{-\frac{i}{\hbar}E_n (T-\Delta)}\,e^{-\frac{i}{\hbar}E_m \Delta}\,
\langle\psi_n|\hat q|\psi_m\rangle\langle\psi_m|\hat q|\psi_n\rangle.
\label{eq:pb_twopt_double_sum}
\end{align}
The \emph{coefficients} entering the sum are the matrix elements of $\hat q$.
By introducing ladder operators
\begin{align}
\hat a=\sqrt{\frac{m\Omega}{2\hbar}}\hat q+\frac{i}{\sqrt{2m\hbar\Omega}}\hat p,
\qquad
\hat a^\dagger=\sqrt{\frac{m\Omega}{2\hbar}}\hat q-\frac{i}{\sqrt{2m\hbar\Omega}}\hat p,
\qquad
\hat q=\sqrt{\frac{\hbar}{2m\Omega}}(\hat a+\hat a^\dagger),
\end{align}
with actions $\hat a|\psi_n\rangle=\sqrt{n}\,|\psi_{n-1}\rangle$ and
$\hat a^\dagger|\psi_n\rangle=\sqrt{n+1}\,|\psi_{n+1}\rangle$ we obtain
\begin{align}
\langle \psi_n|\hat q|\psi_m\rangle
=
\sqrt{\frac{\hbar}{2m\Omega}}\Bigl(
\sqrt{m+1}\,\delta_{n,m+1}+\sqrt{m}\,\delta_{n,m-1}
\Bigr),
\label{eq:q_matrix_elements}
\end{align}
and hence the squared coefficients in Eq.~\eqref{eq:pb_twopt_double_sum} are
\begin{align}
\langle\psi_n|\hat q|\psi_m\rangle\langle\psi_m|\hat q|\psi_n\rangle
=
\frac{\hbar}{2m\Omega}\Bigl[
(m+1)\,\delta_{n,m+1}+m\,\delta_{n,m-1}
\Bigr].
\label{eq:q_coeff_products}
\end{align}
Plugging Eq.~\eqref{eq:q_coeff_products} into Eq.~\eqref{eq:pb_twopt_double_sum}
collapses the double sum to a single sum:
\begin{align}
\langle \hat q(x_0)\hat q(x_0')\rangle_{\rm P.B.}
&=
\frac{\hbar}{2m\Omega}\frac{1}{Z_{\rm P.B.}(T)}
\sum_{m=0}^\infty
\Bigl[
(m+1)\,e^{-\frac{i}{\hbar}E_{m+1}(T-\Delta)}e^{-\frac{i}{\hbar}E_m\Delta}
+
m\,e^{-\frac{i}{\hbar}E_{m-1}(T-\Delta)}e^{-\frac{i}{\hbar}E_m\Delta}
\Bigr].
\label{eq:pb_twopt_single_sum_raw}
\end{align}
Using $E_{m\pm1}=E_m\pm\hbar\Omega$, we factor out the $\Delta$-dependence:
\begin{align}
e^{-\frac{i}{\hbar}E_{m+1}(T-\Delta)}e^{-\frac{i}{\hbar}E_m\Delta}
&=e^{-\frac{i}{\hbar}E_m T}\,e^{-i\Omega(T-\Delta)}
=e^{-\frac{i}{\hbar}E_m T}\,e^{-i\Omega T}\,e^{+i\Omega\Delta},
\\
e^{-\frac{i}{\hbar}E_{m-1}(T-\Delta)}e^{-\frac{i}{\hbar}E_m\Delta}
&=e^{-\frac{i}{\hbar}E_m T}\,e^{+i\Omega(T-\Delta)}
=e^{-\frac{i}{\hbar}E_m T}\,e^{+i\Omega T}\,e^{-i\Omega\Delta}.
\end{align}
Define the convenient parameter
\begin{align}
x \equiv e^{-i\Omega T},
\qquad\Rightarrow\qquad
e^{-\frac{i}{\hbar}E_m T}=e^{-i\Omega T/2}\,x^m.
\label{eq:def_x}
\end{align}
Then Eq.~\eqref{eq:pb_twopt_single_sum_raw} becomes
\begin{align}
\langle \hat q(x_0)\hat q(x_0')\rangle_{\rm P.B.}
=
\frac{\hbar}{2m\Omega}\frac{e^{-i\Omega T/2}}{Z_{\rm P.B.}(T)}
\left[
x\,e^{+i\Omega\Delta}\sum_{m=0}^\infty (m+1)x^m
+
x^{-1}e^{-i\Omega\Delta}\sum_{m=0}^\infty m\,x^m
\right].
\label{eq:pb_twopt_sums_geom}
\end{align}
The needed geometric sums are
\begin{align}
\sum_{m=0}^\infty x^m=\frac{1}{1-x},
\qquad
\sum_{m=0}^\infty (m+1)x^m=\frac{1}{(1-x)^2},
\qquad
\sum_{m=0}^\infty m x^m=\frac{x}{(1-x)^2}.
\label{eq:geom_sums}
\end{align}
The partition function is
\begin{align}
Z_{\rm P.B.}(T)=\sum_{m=0}^\infty e^{-\frac{i}{\hbar}E_m T}
=e^{-i\Omega T/2}\sum_{m=0}^\infty x^m
=\frac{e^{-i\Omega T/2}}{1-x}.
\label{eq:Z_periodic_closed}
\end{align}
Combining Eqs.~\eqref{eq:pb_twopt_sums_geom}--\eqref{eq:Z_periodic_closed} finally gives
\begin{align}
\;
\langle \hat q(x_0)\hat q(x_0')\rangle_{\rm P.B.}
=
\frac{\hbar}{2m\Omega}\,
\frac{x\,e^{+i\Omega\Delta}+e^{-i\Omega\Delta}}{1-x}
=
\frac{\hbar}{2m\Omega}\,
\frac{e^{-i\Omega\Delta}+e^{-i\Omega(T-\Delta)}}{1-e^{-i\Omega T}}
\;,\quad \Delta\in[0,T].
\label{eq:pb_twopt_continuum_final_exp}
\end{align}
This is the continuum periodic two-point function that can be rewritten in the trigonometric form
\begin{align}
\langle \hat q(x_0)\hat q(x_0')\rangle_{\rm P.B.}
=
-\frac{i\hbar}{2m\Omega}\,
\frac{\sin(\Omega\Delta)+\sin\!\big(\Omega(T-\Delta)\big)}{1-\cos(\Omega T)}
\;\quad \Delta\in[0,T].
\label{eq:pb_twopt_continuum_final_trig}
\end{align}
which is the direct continuum analogue of the lattice formula obtained below.

\subsection{Periodic lattice derivation (discrete Fourier modes)}
\label{app:periodic-twopoint:lattice}
Here we illustrate explicitly the steps needed to compute
the harmonic oscillator two-point function on a discretized
coordinate-time lattice with periodic boundary conditions.  We
consider a periodic lattice with $M$ points, so that with periodic
boundary conditions we have $q(0)=q(M-1)$. Let us start by the
definition of the discretized action
\begin{align}
S[q(\ell)] = \frac{ma}{2} \sum_{\ell=0}^{M-1} \left[ \left(\frac{q(\ell+1)-q(\ell)}{a}\right)^2 - \Omega^2 q^2(\ell) \right].
\end{align}
We introduce the Fourier representation of the position field
\begin{align}
q(\ell) = \frac{1}{\sqrt{M}} \sum_{r=0}^{M-1} \hat{q}(r)~e^{i k_0(r) \ell a},
\end{align}
where $r\in\mathbb{N}$, one can rewrite the lattice action as
\begin{align}
S[q(\ell)] = -\frac{1}{2} \sum_{r=0}^{M-1}\omega^2(r)~|\hat{q}(r)|^2,
\end{align}
where
\begin{align}
\omega^2(r) = ma \left[ \Omega^2 - \frac{4}{a^2}\sin^2\left( \frac{k_0(r)a}{2} \right) \right],
\end{align}
and where the conjugate momentum $k_0(r)$ takes the discrete values:
\begin{align}
k_0(r) = \frac{2\pi r}{Ma}.
\end{align}
The two point correlation function on the periodic lattice takes therefore
the following form:
\begin{align}
  \langle q(\ell) q(\ell') \rangle &= -\frac{i}{M} \sum_{r=0}^{M-1} \frac{e^{i k_0(r) (\ell-\ell')a}}{\omega^2(r)}  \\
  & = \frac{i}{Mma} \sum_{r=0}^{M-1}  \frac{e^{i k_0(r)(\ell-\ell')a}}{\left[\frac{4}{a^2}\sin^2\left( \frac{k_0(r)a}{2} \right)-\Omega^2\right]}  \\
  & = \frac{ia}{Mm} \sum_{r=0}^{M-1}  \frac{e^{i \frac{2\pi r}{N}(\ell-\ell')}}{\left[ 4\sin^2\left( \frac{\pi r}{M} \right)-a^2\Omega^2\right]}.  \\
  \label{eq:correlator-1}
\end{align}
In order to simplify the notation we find convenient to exploit the
following identifications
\begin{align}
  & s = \ell - \ell' \nonumber \\
  & C(s) = \langle q(\ell) q(\ell') \rangle\\
  & \mu = 2 - a^2\Omega^2 \\
  & \theta_r = \frac{2\pi}{M}~r~=k_0(r) a  \\
  & 4\sin^2\left(\frac{\theta_r}{2}\right) = 2 - 2\cos(\theta_r)
\end{align}
which allow us to rewrite the expression of the correlator in
Eq.~\eqref{eq:correlator-1} as
\begin{align}
  C(s) = \frac{ia}{Mm} \sum_{r=0}^{M-1}  \frac{e^{i\theta_r s}}{\mu - 2 \cos(\theta_r)}.
\end{align}
By further introducing the coordinate $z_r=e^{i\theta_r}$ on the unitary circle
in the complex plane we get
\begin{align}
  C(s) &= \frac{ia}{Mm} \sum_{r=0}^{M-1}  \frac{z_r^s}{\mu - (z_r+z_r^{-1})}\nonumber \\
  &= -\frac{ia}{Mm} \sum_{r=0}^{M-1}  \frac{z_r^{s+1}}{z_r^2-\mu z_r+1} \nonumber \\ 
  &=-\frac{ia}{Mm} \sum_{r=0}^{M-1}  \frac{z_r^{s+1}}{(z_r - z_1)(z_r-z_2)}
  \label{eq:correlator-2}
\end{align}
where in the last line of Eq.~\eqref{eq:correlator-2} we have that $z_{(1,2)}$ are the
two roots of the equation $z_r^2-\mu z_r+1=0$, namely:
\begin{align}
z_{(1,2)} = \exp(\pm i \kappa a)\quad\quad\quad \left( \kappa~~|~~\cos(\kappa a) = \frac{\mu}{2} \right)\,.
\label{eq:z12-roots}
\end{align}
The above expression for $z_{(1,2)}$ assumes that $\left|
1-\frac{a^2\Omega^2}{2}\right| \leq 1$, a constraint which can be
always fulfilled by choosing, for any given naturalt frequency
$\Omega$ of the quantum harmonic oscillator, the spacing $a$ of the
coordinate time lattice such as
\begin{align}
a < \frac{2}{\Omega}.
\end{align}
The summation in the last line of Eq.~\eqref{eq:correlator-2} can be then
rewritten as
\begin{align}
  \sum_{r=0}^{M-1}  \frac{z_r^{s+1}}{(z_r - z_1)(z_r-z_2)} = \frac{1}{z_1-z_2} \sum_{r=0}^{M-1}
  z_r^{s+1}\left( \frac{1}{z_r-z_1} - \frac{1}{z_r-z_2}\right).
  \label{eq:correlator-3-sum}
\end{align}
By letting the summation over $r$ act indepently on the two terms in
parentheses of Eq.~\eqref{eq:correlator-3-sum} one can exploit for
each of the two the following mathematical identity:
\begin{align}
  \sum_{r=0}^{M-1} \frac{z_r^{s+1}}{z_r-u} &=\sum_{r=0}^{M-1} \frac{e^{i\theta_r (s+1)}}{e^{i\theta_r}-u} \nonumber\\
  &= \sum_{r=0}^{M-1} \frac{e^{i\frac{2 \pi r}{M} (s+1)}}{e^{i\frac{2 \pi r}{M}-u}} \nonumber\\
  &=\frac{Mu^{s}}{1-u^M},
  \label{eq:sum-identity}
\end{align}
which holds for $s$ integer. Therefore, one gets:
\begin{align}
  C(s) &
  = -\frac{a}{Mm}\frac{i}{z_1-z_2}\left[ \frac{Mz_1^s}{1-z_1^M} - \frac{Nz_2^s}{1-z_2^M} \right] \nonumber \\
  & = -\frac{a}{m}\frac{i}{e^{i\kappa a}-e^{-i\kappa a}}
  \left[ \frac{e^{i\kappa a s}}{1-e^{i\kappa aM}} - \frac{e^{-i\kappa as}}{1-e^{-i\kappa aM}} \right] \nonumber \\
  & = -\frac{a}{2m}\frac{i}{\sin(\kappa a)} \frac{\sin(\kappa a s)+\sin(\kappa a (M-s))}{1-\cos(\kappa a M )}.
\end{align}
By going back to the previous notation we have the discrete formula for the
two point correlation function
\begin{align}
  \langle q(\ell) q(\ell') \rangle = -\frac{a}{2m}\frac{i}{\sin(\kappa a)}
  \frac{\sin[\kappa a (\ell-\ell')]+\sin[\kappa a (M-(\ell-\ell'))]}{1-\cos[\kappa a M ]},
  \label{eq:two-points-discrete}
\end{align}
which, as shown in the main text, is perfectly matched by our numerical calculations.

\section{Lattice functional derivation of the commutator identity}
\label{app:commutator-derivation}

We derive here the lattice identity used in Sec.~\ref{sec:lat-commutator} to reconstruct the equal--time commutator from periodic (trace) expectation values. We consider a discretized Minkowskian time lattice with spacing $a=T/M$ and lattice sites labeled by $\ell=0,\ldots,M-1$, with $x_0^{(\ell)}=x_0^{\mathrm{i}}+\ell a$ and periodic boundary conditions (P.B.) $q(M)\equiv q(0)$. The trace partition function reads
\begin{equation}
  \mZ_{\rm P.B.}(\hbar)=\int_{\rm P.B.}\prod_{\ell=0}^{M-1}dq(\ell)\;
  \exp\!\left(\frac{i}{\hbar}S[q]\right),
  \label{eq:Z-PB-appendix}
\end{equation}
with discretized action $S[q]$. For any differentiable function $F(\{q(\ell)\})$ of the lattice variables, the integral of a total derivative yields the lattice Schwinger--Dyson identity (see~\cite{schulman2005techniques})
\begin{equation}
  \Bigl\langle \partial_{q(\ell)}F \Bigr\rangle_{\rm P.B.}
  =-\frac{i}{\hbar}\Bigl\langle F\,\partial_{q(\ell)}S\Bigr\rangle_{\rm P.B.}.
  \label{eq:SD-lattice-appendix}
\end{equation}
For the harmonic lattice action used in the main text one has
\begin{equation}
  S[q]=\frac{ma}{2}\sum_{\ell=0}^{M-1}\left[\left(\frac{q(\ell+1)-q(\ell)}{a}\right)^2-\Omega^2 q^2(\ell)\right],
  \qquad q(M)\equiv q(0),
  \label{eq:action-appendix}
\end{equation}
from which
\begin{equation}
  \partial_{q(\ell)}S[q]
  =\frac{m}{a}\Bigl(q(\ell)-q(\ell+1)\Bigr)-\frac{m}{a}\Bigl(q(\ell-1)-q(\ell)\Bigr)-a\,V'\!\bigl(q(\ell)\bigr),
  \qquad V(q)=\frac12 m\Omega^2 q^2.
  \label{eq:dS-appendix}
\end{equation}
Introducing the standard link momenta (defined at half--integer time--slices),
\begin{equation}
  p\!\left(\ell+\frac12\right)\equiv \frac{m}{a}\Bigl(q(\ell)-q(\ell+1)\Bigr),
  \qquad
  p\!\left(\ell-\frac12\right)\equiv \frac{m}{a}\Bigl(q(\ell-1)-q(\ell)\Bigr),
  \label{eq:link-mom-appendix}
\end{equation}
Eq.~\eqref{eq:dS-appendix} becomes
\begin{equation}
  \partial_{q(\ell)}S[q]=p\!\left(\ell+\frac12\right)-p\!\left(\ell-\frac12\right)-a\,V'\!\bigl(q(\ell)\bigr).
  \label{eq:dS-link-appendix}
\end{equation}
Choosing $F=q(\ell)$ in \eqref{eq:SD-lattice-appendix} we obtain
\begin{equation}
  \Bigl\langle q(\ell)\,p\!\left(\ell+\frac12\right)\Bigr\rangle_{\rm P.B.}
  -\Bigl\langle q(\ell)\,p\!\left(\ell-\frac12\right)\Bigr\rangle_{\rm P.B.}
  -a\,\Bigl\langle q(\ell)\,V'\!\bigl(q(\ell)\bigr)\Bigr\rangle_{\rm P.B.}
  =i\hbar.
  \label{eq:core-identity-appendix}
\end{equation}
For smooth potentials the last term is an $O(a)$ lattice artifact in the continuum limit, so that
\begin{equation}
  \Bigl\langle q(\ell)\,p\!\left(\ell+\frac12\right)\Bigr\rangle_{\rm P.B.}
  -\Bigl\langle q(\ell)\,p\!\left(\ell-\frac12\right)\Bigr\rangle_{\rm P.B.}
  =i\hbar+O(a).
  \label{eq:core-identity-continuum-appendix}
\end{equation}
Using the definitions \eqref{eq:link-mom-appendix} this can be rewritten entirely in terms of $q$--correlators,
\begin{align}
  \Bigl\langle q(\ell)\,p\!\left(\ell+\frac12\right)\Bigr\rangle_{\rm P.B.}
  -\Bigl\langle q(\ell)\,p\!\left(\ell-\frac12\right)\Bigr\rangle_{\rm P.B.}
  &=\frac{m}{a}\Bigl\langle q(\ell)\bigl(q(\ell)-q(\ell+1)\bigr)-q(\ell)\bigl(q(\ell-1)-q(\ell)\bigr)\Bigr\rangle_{\rm P.B.}\nonumber\\
  &=\frac{m}{a}\Bigl\langle 2q^2(\ell)-q(\ell)\,q(\ell+1)-q(\ell)\,q(\ell-1)\Bigr\rangle_{\rm P.B.}.
  \label{eq:qq-identity-appendix}
\end{align}
Combining \eqref{eq:qq-identity-appendix} with \eqref{eq:core-identity-continuum-appendix} yields the lattice representation used in the main text,
\begin{equation}
  -\,i\,\frac{m}{a}\Bigl\langle 2q^2(\ell)-q(\ell)\,q(\ell+1)-q(\ell)\,q(\ell-1)\Bigr\rangle_{\rm P.B.}
  =\hbar+O(a),
  \label{eq:final-commutator-appendix}
\end{equation}
which is the periodic (trace) expectation value of the equal--time commutator written in a form directly comparable with the constrained symplectic quantization measurements.

\section{Dirichlet modes and diagonalisation of the quadratic action}
\label{app:dirichlet-mode-measure}

In this Appendix we collect the Dirichlet-mode decomposition of the
holomorphic harmonic-oscillator action and show explicitly that, upon
imposing the stable-manifold constraints, the functional measure
factorises mode by mode exactly as in the periodic case.  This provides
the Dirichlet analogue of Eq.~\eqref{eq:sq-factor-measure-k} and yields
a mode-by-mode correspondence between constrained (analytically
continued) averages and ordinary quantum-mechanical correlators.\\

Consider the (Minkowskian) quadratic action on the interval
$[x_0^{\mathrm{i}},x_0^{\mathrm{f}}]$,
\begin{align}
  S[q]
  \equiv
  \frac{m}{2}\int_{x_0^{\mathrm{i}}}^{x_0^{\mathrm{f}}}\!dx_0\;
  \Bigl[(\partial_{0} q(x_0))^2-\Omega^2 q(x_0)^2\Bigr],
  \label{eq:dirichlet-action-cont}
\end{align}
with Dirichlet boundary conditions $q(x_0^{\mathrm{i}})=q(x_0^{\mathrm{f}})=0$.
For Dirichlet boundaries it is convenient to expand in the sine basis
(cf.\ Eq.~\eqref{eq:dirichlet-sine-expansion}),
\begin{align}
  q(x_0)
  =
  \sum_{n=1}^{\infty}\tilde q(n)\,\sqrt{\frac{2}{T}}\,
  \sin\!\Bigl(\kappa_0^{(n)} (x_0-x_0^{\mathrm{i}})\Bigr),
  \qquad
  \kappa_0^{(n)}\equiv \frac{\pi n}{T},
  \qquad
  T\equiv x_0^{\mathrm{f}}-x_0^{\mathrm{i}} .
  \label{eq:dirichlet-sine-expansion-app}
\end{align}
The orthonormality relation
\begin{align}
  \frac{2}{T}\int_{x_0^{\mathrm{i}}}^{x_0^{\mathrm{f}}}\!dx_0\;
  \sin\!\Bigl(\kappa_0^{(n)}(x_0-x_0^{\mathrm{i}})\Bigr)
  \sin\!\Bigl(\kappa_0^{(n')} (x_0-x_0^{\mathrm{i}})\Bigr)
  =\delta_{n,n'}
\end{align}
implies the inversion formula
\begin{align}
  \tilde q(n)
  =
  \sqrt{\frac{2}{T}}
  \int_{x_0^{\mathrm{i}}}^{x_0^{\mathrm{f}}}\!dx_0\;
  q(x_0)\,\sin\!\Bigl(\kappa_0^{(n)} (x_0-x_0^{\mathrm{i}})\Bigr).
  \label{eq:dirichlet-inversion}
\end{align}
In the constrained formulation $q(x_0)\in\mathbb{C}$ is holomorphic, so
we write $\tilde q(n)=\tilde q_R(n)+i\,\tilde q_I(n)$ with
$\tilde q_{R,I}(n)\in\mathbb{R}$.\\

Using Eq.~\eqref{eq:dirichlet-sine-expansion-app} and orthonormality one
obtains
\begin{align}
  \int_{x_0^{\mathrm{i}}}^{x_0^{\mathrm{f}}}\!dx_0\; q(x_0)^2
  &= \sum_{n=1}^{\infty}\tilde q(n)^2,
  \qquad
  \int_{x_0^{\mathrm{i}}}^{x_0^{\mathrm{f}}}\!dx_0\; (\partial_0 q(x_0))^2
  = \sum_{n=1}^{\infty}(\kappa_0^{(n)})^2\,\tilde q(n)^2,
\end{align}
so the action diagonalises mode by mode as
\begin{align}
  S[q]
  &=
  \frac{m}{2}\sum_{n=1}^{\infty}\bigl((\kappa_0^{(n)})^2-\Omega^2\bigr)\,\tilde q(n)^2
  \;=\;
  -\frac{1}{2}\sum_{n=1}^{\infty}\,\omega_n^2\,\tilde q(n)^2,
  \label{eq:dirichlet-action-diag}
\end{align}
where it is convenient to introduce
\begin{align}
  \omega_n^2 \;\equiv\; m\bigl(\Omega^2-(\kappa_0^{(n)})^2\bigr).
  \label{eq:varpi-def}
\end{align}
Note that $\omega_n^2$ can be positive or negative depending on whether
$(\kappa_0^{(n)})^2$ is smaller or larger than $\Omega^2$.\\

For later use we also write Eq.~\eqref{eq:dirichlet-action-diag} in terms
of real and imaginary Dirichlet modes:
\begin{align}
  S[q]
  &=
  -\frac{1}{2}\sum_{n=1}^{\infty}\omega_n^2
  \Bigl[
    \tilde q_R(n)^2-\tilde q_I(n)^2
    +2i\,\tilde q_R(n)\tilde q_I(n)
  \Bigr].
  \label{eq:dirichlet-action-RI}
\end{align}
The constrained symplectic quantization prescription fixes, mode by
mode, the integration contour in the complex $\tilde q(n)$ plane so that
$\exp(\tfrac{i}{\hbar}S)$ is exponentially damped.  Equivalently, one
imposes the same linear relations between $\tilde q_R(n)$ and $\tilde
q_I(n)$ as in the periodic case:
\begin{align}
  \omega_n^2>0:\quad \tilde q_I(n)=-\tilde q_R(n),
  \qquad\qquad
  \omega_n^2<0:\quad \tilde q_I(n)=+\tilde q_R(n).
  \label{eq:dirichlet-constraints-modes}
\end{align}
(Geometrically, these relations correspond to a $\pm\pi/4$ rotation of
the contour for each mode; the sign is chosen so that the holomorphic
quadratic form produces a negative real exponent in
$\exp(\tfrac{i}{\hbar}S)$.)\\

Substituting Eq.~\eqref{eq:dirichlet-constraints-modes} into
Eq.~\eqref{eq:dirichlet-action-RI} yields
\begin{align}
  S[q]
  =
  i\sum_{\{n\,|\,\omega_n^2>0\}}\omega_n^2\,\tilde q_R(n)^2
  \;-\;
  i\sum_{\{n\,|\,\omega_n^2<0\}}\omega_n^2\,\tilde q_R(n)^2.
  \label{eq:dirichlet-action-on-stable}
\end{align}
Therefore the holomorphic weight factorises as a product of independent
Gaussians in the real constrained modes.  The explicit Dirichlet
analogue of Eq.~\eqref{eq:sq-factor-measure-k} is
\begin{align}
  \exp\!\left(\frac{i}{\hbar}S[q]\right)
  =
  \exp\!\left[
    -\frac{1}{\hbar}\sum_{\{n\,|\,\omega_n^2>0\}}\omega_n^2\,\tilde q_R(n)^2
    +\frac{1}{\hbar}\sum_{\{n\,|\,\omega_n^2<0\}}\omega_n^2\,\tilde q_R(n)^2
  \right],
  \label{eq:sq-factor-measure-dirichlet}
\end{align}
where the second term is negative because $\omega_n^2<0$ on that subset.
Equivalently,
\begin{align}
  \exp\!\left(\frac{i}{\hbar}S[q]\right)
  =
  \exp\!\left[
    -\frac{1}{\hbar}\sum_{n=1}^{\infty}\bigl|\omega_n^2\bigr|\,\tilde q_R(n)^2
  \right],
  \label{eq:sq-factor-measure-dirichlet-abs}
\end{align}
up to an overall (mode-independent) normalisation factor that is
irrelevant for expectation values. Equations~\eqref{eq:sq-factor-measure-dirichlet}-- \eqref{eq:sq-factor-measure-dirichlet-abs} show explicitly that the
constrained measure factorises mode by mode, exactly as in the periodic
case but with $n\ge1$.  As a consequence, all constrained mode
correlators are diagonal in $n$ and can be computed by elementary
Gaussian integrals.  In particular,
\begin{align}
  \big\langle \tilde q_R(n)\,\tilde q_R(n')\big\rangle_{\text{D.B.}(T)}
  = \delta_{n,n'}\,
  = \delta_{n,n'}\,\frac{\hbar}{2\,|\omega_n^2|},
  \label{eq:qR-gaussian-dirichlet}
\end{align}
and reconstructing the complex mode using the constraints
Eq.~\eqref{eq:dirichlet-constraints-modes} one finds the corresponding
quantum-mechanical Dirichlet correlator in mode space,
\begin{align}
  \big\langle \tilde q(n)\,\tilde q(n')\big\rangle_{\text{D.B.}(T)}
  = \delta_{n,n'}\,
  \frac{i\hbar}{m\bigl((\kappa_0^{(n)})^2-\Omega^2\bigr)}
  = -\,\delta_{n,n'}\,\frac{i\hbar}{\omega_n^2},
  \label{eq:dirichlet-mode-propagator}
\end{align}
which is the standard harmonic-oscillator result in the Dirichlet sine
basis.  The corresponding coordinate-space expressions quoted in the
main text follow by inserting Eq.~\eqref{eq:dirichlet-sine-expansion-app}
and summing over $n$.

\section{Continuum Dirichlet expression for
  \texorpdfstring{$\langle \hat{q}^2(x_0)\rangle_{\mathrm{D.B.}(T)}$}
                 {Expectation value of q2(x0) with D.B.(T)}}
\label{app:q2-dirichlet-continuum}

In this subsection we derive the closed--form expression for the
Dirichlet ``expectation value'' of $\hat{q}^2(x_0)$ in the continuum
Minkowskian theory for the Harmonic oscillator with action
\begin{align}
  S[q]
  &= \int_0^T \! dx_0 \,
  \frac{m}{2}\left(\partial_{x_0}q^2(x_0) - \Omega^2 q^2(x_0)\right) \, .
  \label{eq:ho-action-mink}
\end{align}
We impose Dirichlet boundary conditions in the time interval
$[0,T]$,
\begin{align}
  q(0) = 0 \, , \qquad q(T) = 0 \, .
  \label{eq:dirichlet-bc-q}
\end{align}

The Dirichlet ``expectation value'' of $\hat{q}^2(x_0)$ is defined as
in Eq.~\eqref{eq:DB-spectral-general} with $k=1$,
\begin{align}
  \big\langle \hat{q}^2(x_0)\big\rangle_{\text{D.B.}(T)}
  &\equiv
  \frac{\displaystyle
    \langle q_f=0,x_0^{\text{f}}=T ~|~ \hat{q}^2(x_0)~|~ q_i=0,x_0^{\text{i}}=0\rangle}
  {\displaystyle
    \langle q_f=0,x_0^{\text{f}}=T ~|~ q_i=0,x_0^{\text{i}}=0\rangle} \, ,
  \qquad 0 < x_0 < T \, .
  \label{eq:q2-db-def}
\end{align}
For a quadratic action of the form~\eqref{eq:ho-action-mink} the
path--integral with Dirichlet boundary conditions can be written as
\begin{align}
  \mathcal{Z}[J]
  &= \int_{q(0)=0}^{q(T)=0} \!\!\!\!\mathcal{D}q\,
  \exp\!\left\{
    \frac{i}{\hbar} S[q]
    + \frac{i}{\hbar} \int_0^T \! dx_0 \, J(x_0)\,q(x_0)
  \right\} \, ,
\end{align}
where $J(x_0)$ is an external source. By integrating by parts in
Eq.~\eqref{eq:ho-action-mink} and using the boundary
conditions~\eqref{eq:dirichlet-bc-q}, the action can be cast in the
quadratic form
\begin{align}
  S[q]
  &= \frac{m}{2}\int_0^T \! dx_0\,
  q(x_0)\,\Big(-\partial_{x_0}^2 - \Omega^2\Big)\,q(x_0) \, .
  \label{eq:ho-action-quadratic-form}
\end{align}
Thus the quadratic kernel appearing in the exponent is
\begin{align}
  \mathcal{K}(x_0,x_0')
  &= m\Big(-\partial_{x_0}^2 - \Omega^2\Big)\,
  \delta(x_0-x_0') \, .
  \label{eq:kernel-K-def}
\end{align}
For a Gaussian functional integral of the form above, the two--point
function is given by the inverse of the kernel,
\begin{align}
  \langle q(x_0)\,q(x_0')\rangle_{\text{D.B.}(T)}
  &= i\hbar\,
  G(x_0,x_0') \, ,
  \label{eq:qq-G-relation}
\end{align}
where $G(x_0,x_0')$ is the Green's function satisfying
\begin{align}
  m\Big(-\partial_{x_0}^2 - \Omega^2\Big)\,G(x_0,x_0')
  &= \delta(x_0-x_0') \, ,
  \label{eq:green-eq-dirichlet}
\end{align}
with Dirichlet boundary conditions
\begin{align}
  G(0,x_0') = 0 \, , \qquad
  G(T,x_0') = 0 \, .
  \label{eq:green-dirichlet-bc}
\end{align}

We now solve Eq.~\eqref{eq:green-eq-dirichlet} explicitly. For
$x_0 \neq x_0'$ the Green's function satisfies the homogeneous
equation
\begin{align}
  \Big(-\partial_{x_0}^2 - \Omega^2\Big)\,G(x_0,x_0') = 0 \, .
  \label{eq:green-homo-eq}
\end{align}
The general solution of Eq.~\eqref{eq:green-homo-eq} is a linear
combination of $\sin(\Omega x_0)$ and $\cos(\Omega x_0)$. Imposing the
boundary conditions~\eqref{eq:green-dirichlet-bc} separately on the
intervals $0<x_0<x_0'$ and $x_0'<x_0<T$, we write
\begin{align}
  G(x_0,x_0')
  &=
  \begin{cases}
    A\,\sin(\Omega x_0) \, , & 0 < x_0 < x_0' \, , \\[0.3em]
    B\,\sin\!\big(\Omega(T-x_0)\big) \, , & x_0' < x_0 < T \, ,
  \end{cases}
  \label{eq:green-piecewise}
\end{align}
where the Dirichlet boundary conditions at $x_0=0$ and $x_0=T$ have
already been used. Continuity of $G(x_0,x_0')$ at $x_0=x_0'$ implies
\begin{align}
  \lim_{x_0\to x_0'^-} G(x_0,x_0')
  &= \lim_{x_0\to x_0'^+} G(x_0,x_0') \nonumber\\
  \Rightarrow\quad
  A\,\sin(\Omega x_0')
  &= B\,\sin\!\big(\Omega(T-x_0')\big) \, .
  \label{eq:green-continuity}
\end{align}
On the other hand, integrating Eq.~\eqref{eq:green-eq-dirichlet} over
an infinitesimal interval around $x_0'$, one obtains the usual jump
condition on the first derivative,
\begin{align}
  \int_{x_0'-\varepsilon}^{x_0'+\varepsilon} \! dx_0\,
  m\Big(-\partial_{x_0}^2 - \Omega^2\Big)G(x_0,x_0')
  &= \int_{x_0'-\varepsilon}^{x_0'+\varepsilon} \! dx_0\,
  \delta(x_0-x_0') \nonumber\\
  \Rightarrow\quad
  -m\left.\frac{\partial G}{\partial x_0}\right|_{x_0=x_0'^+}
  + m\left.\frac{\partial G}{\partial x_0}\right|_{x_0=x_0'^-}
  &= 1 \, .
  \label{eq:green-jump}
\end{align}
Using the explicit expressions in
Eq.~\eqref{eq:green-piecewise}, this condition reads
\begin{align}
  -m\big[-B\Omega\cos\!\big(\Omega(T-x_0')\big)\big]
  + m\big[A\Omega\cos(\Omega x_0')\big]
  &= 1 \, .
  \label{eq:green-jump-AB}
\end{align}
Equations~\eqref{eq:green-continuity} and~\eqref{eq:green-jump-AB}
form a linear system for $A$ and $B$. Solving it, and using the
trigonometric identity
$\sin(\Omega T) = \sin(\Omega x_0')\cos\!\big(\Omega(T-x_0')\big)
+ \cos(\Omega x_0')\sin\!\big(\Omega(T-x_0')\big)$, one finds
\begin{align}
  A &= \frac{\sin\!\big(\Omega(T-x_0')\big)}{m\Omega\sin(\Omega T)} \, ,
  &
  B &= \frac{\sin(\Omega x_0')}{m\Omega\sin(\Omega T)} \, .
\end{align}
Therefore the Green's function can be written in the compact form
\begin{align}
  G(x_0,x_0')
  &= \frac{1}{m\Omega\sin(\Omega T)}\,
  \sin\big(\Omega x_<\big)\,
  \sin\big(\Omega(T-x_>)\big) \, ,
  \label{eq:green-final}
\end{align}
where $x_< \equiv \min(x_0,x_0')$ and $x_> \equiv \max(x_0,x_0')$.
Using Eq.~\eqref{eq:qq-G-relation}, the Dirichlet two--point function
for the quantum coordinate is
\begin{align}
  \big\langle q(x_0)\,q(x_0')\big\rangle_{\text{D.B.}(T)}
  &= \frac{i\hbar}{m\Omega\sin(\Omega T)}\,
  \sin\big(\Omega x_<\big)\,
  \sin\big(\Omega(T-x_>)\big) \, .
  \label{eq:qq-db-continuum}
\end{align}
In particular, the continuum Dirichlet ``expectation value'' of
$\hat{q}^2(x_0)$ is obtained by taking the equal--time limit
$x_0'\to x_0$,
\begin{align}
  \big\langle \hat{q}^2(x_0)\big\rangle_{\text{D.B.}(T)}
  &= \big\langle q(x_0)\,q(x_0)\big\rangle_{\text{D.B.}(T)}
  \nonumber\\[0.3em]
  &= \frac{i\hbar}{m\Omega\sin(\Omega T)}\,
  \sin(\Omega x_0)\,
  \sin\big(\Omega(T-x_0)\big) \, .
  \label{eq:q2-db-continuum-final}
\end{align}
For real $\Omega,T,x_0$ and $\sin(\Omega T)\neq 0$, the ratio of
sine functions in Eq.~\eqref{eq:q2-db-continuum-final} is real, so
$\langle \hat{q}^2(x_0)\rangle_{\text{D.B.}(T)}$ is purely
imaginary.\\\\

The same result can be obtained from the eigenstate decomposition, whose starting point is \eqref{eq:q2-db-def}. Considering the numerator
\begin{align}
  \sum_{m,n=0}^\infty
  \langle q=0,T|\psi_m\rangle\,
  \langle \psi_m|\hat{q}^{2}(x_0)|\psi_n\rangle\,
  \langle \psi_n|q=0,0\rangle ,
  \label{eq:q2-general-eigen}
\end{align}
makes explicit the contribution from all excited eigenstates.  

\medskip

Using
$H|\psi_n\rangle = E_n|\psi_n\rangle$,
with $E_n=\hbar\Omega(n+\tfrac12)$, we write
\begin{align}
  \langle \psi_m|\hat{q}^2(x_0)|\psi_n\rangle
  &= e^{-\,\frac{i}{\hbar}(E_n-E_m)x_0}\,
     \langle \psi_m|\hat{q}^2|\psi_n\rangle .
  \label{eq:q2x0-eigen}
\end{align}
Because $\psi_{2n+1}(0)=0$, only even states contribute.  Rewriting
Eq.~\eqref{eq:q2-general-eigen} in terms of even labels yields
\begin{align}
  \langle \hat{q}^2(x_0)\rangle_{\text{D.B.}(T)}
  &=
  \frac{\sum_{m,n=0}^\infty
  \psi_{2m}^*(0;T)\,\psi_{2n}(0;0)\,
  \langle \psi_{2m}|\hat{q}^2|\psi_{2n}\rangle\,
  e^{-\,\frac{i}{\hbar}(E_{2n}-E_{2m})x_0}}{\sum_{m,n=0}^\infty
  \psi_{2n}^*(0;T)\,\psi_{2n}(0;0)},
  \label{eq:q2-even-expansion}
\end{align}

where we indicate the denominator as 
\begin{equation}
  D(T)  = \sum_{m,n=0}^\infty \psi_{2n}^*(0;T)\,\psi_{2n}(0;0)
\end{equation}
\medskip

The operator $\hat{q}^2$ can be expressed in ladder form as
\begin{align}
  \hat{q}^2
  &= \frac{\hbar}{2m\Omega}\Big(
       a a + a^\dagger a^\dagger + 
       2 a^\dagger a + 1\Big),
  \label{eq:q2-ladder-app}
\end{align}
with the corresponding matrix elements
\begin{align}
  \langle n|\hat{q}^2|n\rangle
  &= \frac{\hbar}{2m\Omega}(2n+1),
  \\
  \langle n|\hat{q}^2|n+2\rangle
  &= \frac{\hbar}{2m\Omega}\sqrt{(n+1)(n+2)}, 
  \qquad
  \langle n+2|\hat{q}^2|n\rangle = \langle n|\hat{q}^2|n+2\rangle .
  \label{eq:q2-matrix-el}
\end{align}

Inserting Eq.~\eqref{eq:q2-ladder-app} into
Eq.~\eqref{eq:q2-even-expansion} and performing the sums over $m$ gives
\begin{align}
  \langle \hat{q}^2(x_0)\rangle_{\text{D.B.}(T)}
  &= \frac{1}{D(T)}
  \frac{\hbar}{2m\Omega}
  \Bigg[
  \sum_{n=0}^\infty (4n+1)\,
  \psi_{2n}^*(0;T)\,\psi_{2n}(0)
  \label{eq:q2-expanded-step1}
  \\[0.3em]
  &\quad\ 
  +\, e^{-i2\Omega x_0}
  \sum_{n=2}^\infty
  \sqrt{2n}\sqrt{2n-1}\,
  \psi_{2n-2}^*(0;T)\,\psi_{2n}(0)
  \nonumber \\[0.3em]
  &\quad\ 
  +\, e^{+i2\Omega x_0}
  \sum_{n=0}^\infty
  \sqrt{2n+1}\sqrt{2n+2}\,
  \psi_{2n+2}^*(0;T)\,\psi_{2n}(0)
  \Bigg].
  \nonumber
\end{align}

Using the identity (valid after reindexing)
\begin{align}
  \sum_{n=2}^\infty \!\sqrt{2n}\sqrt{2n-1}\,
  \psi_{2n-2}^*(0;T)\psi_{2n}(0)
  =
  \sum_{n=0}^\infty\!
  \sqrt{2n+1}\sqrt{2n+2}\,
  \psi_{2n}^*(0;T)\psi_{2n+2}(0),
\end{align}
Eq.~\eqref{eq:q2-expanded-step1} becomes
\begin{align}
  \langle \hat{q}^2(x_0)\rangle_{\text{D.B.}(T)}
  &=
  \frac{\hbar}{2m\Omega}\frac{1}{D(T)}
  \Bigg[
  \sum_{n=0}^\infty (4n+1)\,\psi_{2n}^*(0;T)\psi_{2n}(0)
  \nonumber\\[0.3em]
  &\qquad +\,
  e^{-i2\Omega x_0}\!
  \sum_{n=0}^\infty
  \sqrt{2n+1}\sqrt{2n+2}\,
  \psi_{2n}^*(0;T)\psi_{2n+2}(0)
  \nonumber\\[0.3em]
  &\qquad +\,
  e^{+i2\Omega x_0}\!
  \sum_{n=0}^\infty
  \sqrt{2n+1}\sqrt{2n+2}\,
  \psi_{2n+2}^*(0;T)\psi_{2n}(0)
  \Bigg].
  \label{eq:q2-expanded-step2}
\end{align}

\medskip

Since
$\psi_{2n}(0;T) = \psi_{2n}(0)\,e^{+\frac{i}{\hbar}E_{2n}T}$,
we define the continuum spectral coefficients
\begin{align}
  D(T)
  &\equiv
  \sum_{n=0}^\infty
  e^{+\frac{i}{\hbar}E_{2n}T}\,|\psi_{2n}(0)|^2,
  \label{eq:DT-cont-def}
  \\[0.4em]
  \alpha(T)
  &\equiv
  \frac{
    \sum_{n=0}^\infty (4n+1)\,
    e^{+\frac{i}{\hbar}E_{2n}T}\,
    |\psi_{2n}(0)|^2}{
    D(T)},
  \label{eq:aT-spec-cont}
  \\[0.4em]
  \gamma_{+}(T)
  &\equiv
  \frac{
    \sum_{n=0}^\infty
    \sqrt{(2n+1)(2n+2)}\,
    e^{+\frac{i}{\hbar}E_{2n+2}T}\,
    \psi_{2n+2}(0)\psi_{2n}(0)
  }{D(T)},
  \label{eq:cp-spec-cont}
  \\[0.4em]
  \gamma_{-}(T)
  &\equiv
  \frac{
    \sum_{n=0}^\infty
    \sqrt{(2n+1)(2n+2)}\,
    e^{+\frac{i}{\hbar}E_{2n}T}\,
    \psi_{2n+2}(0)\psi_{2n}(0)
  }{D(T)}.
  \label{eq:cm-spec-cont}
\end{align}
Inserting these definitions into
Eq.~\eqref{eq:q2-expanded-step2} gives the compact form
\begin{align}
  \langle \hat{q}^2(x_0)\rangle_{\text{D.B.}(T)}
  &=
  \frac{\hbar}{2m\Omega}
  \Big[
    \alpha(T)
    + \gamma_{-}(T)e^{-i2\Omega x_0}
    + \gamma_{+}(T)e^{+i2\Omega x_0}
  \Big].
  \label{eq:q2-spec-final}
\end{align}

\bigskip
 We can show that the expression \eqref{eq:q2-spec-final} is formally equivalent to the continuum Green's function
result~\eqref{eq:q2-db-continuum-final},
\begin{align}
  \big\langle \hat{q}^2(x_0)\big\rangle_{\text{D.B.}(T)}
  &= \frac{i\hbar}{m\Omega\sin(\Omega T)}\,
  \sin(\Omega x_0)\,\sin\big(\Omega(T-x_0)\big) \, .
  \label{eq:q2-db-continuum-repeat}
\end{align}
Using the trigonometric identity
\begin{align}
  \sin(\Omega x_0)\,\sin\big(\Omega(T-x_0)\big)
  &= \frac12\left[
    \sin(2\Omega x_0)
    + \cot(\Omega T)\,\big(\cos(2\Omega x_0) - 1\big)
  \right],
\end{align}
Eq.~\eqref{eq:q2-db-continuum-repeat} can be rewritten as
\begin{align}
  \big\langle \hat{q}^2(x_0)\big\rangle_{\text{D.B.}(T)}
  &= \frac{\hbar}{2m\Omega}\,\Big\{
    -\,i\,\cot(\Omega T)
    + \tfrac{1}{2}\big[-1 + i\,\cot(\Omega T)\big] e^{-i2\Omega x_0}
    \nonumber\\
  &\hspace{6.1em}
    + \tfrac{1}{2}\big[1 + i\,\cot(\Omega T)\big] e^{+i2\Omega x_0}
  \Big\} .
  \label{eq:q2-db-continuum-harmonic}
\end{align}
which can be rewritten as
\begin{align}
  \langle \hat{q}^2(x_0)\rangle_{\text{D.B.}(T)}
  =
  \frac{\hbar}{2m\Omega}\Big[
      a(T)
    + c_-(T)e^{-i2\Omega x_0}
    + c_+(T)e^{+i2\Omega x_0}
  \Big],
\end{align}
with the continuum coefficients
\begin{align}
  a(T)
  &= -\,i\,\cot(\Omega T),
  \label{eq:a-cont-final}
  \\[0.3em]
  c_+(T)
  &= \frac{1}{2}\!\left[1 + i\,\cot(\Omega T)\right],
  \label{eq:cp-cont-final}
  \\[0.3em]
  c_-(T)
  &= \frac{1}{2}\!\left[-1 + i\,\cot(\Omega T)\right].
  \label{eq:cm-cont-final}
\end{align}
We can now show explicitly that the infinite--sum spectral representation
\eqref{eq:q2-spec-final} reproduces exactly the continuum Dirichlet
result for $\langle \hat{q}^2(x_0)\rangle_{\text{D.B.}(T)}$.  
The required step is the analytic evaluation of the spectral
coefficients $\alpha(T)$ and $\gamma_{\pm}(T)$
defined in Eqs.~\eqref{eq:aT-spec-cont}--\eqref{eq:cm-spec-cont}, and
proving that they coincide with the continuum coefficients
\eqref{eq:a-cont-final}--\eqref{eq:cm-cont-final}:
\begin{align}
    a(T) &= \alpha(T)\\
    c_{\pm}(T) &= \gamma_{\pm}(T).
\end{align}

\bigskip

\textbf{Step 1: Closed form of $D(T)$.}
Using the known values of Hermite polynomials at the origin,
\begin{align}
  \psi_{2n}(0)
  &=
  \left(\frac{m\Omega}{\pi\hbar}\right)^{\!1/4}
  (-1)^n\frac{\sqrt{(2n)!}}{2^n n!},
  &
  \psi_{2n+1}(0)&=0,
  \label{eq:hermite-origin}
\end{align}
we obtain
\begin{align}
  |\psi_{2n}(0)|^2
  =
  \left(\frac{m\Omega}{\pi\hbar}\right)^{\!1/2}
  \frac{(2n)!}{4^n (n!)^2}
  =
  \left(\frac{m\Omega}{\pi\hbar}\right)^{\!1/2}
  \frac{1}{4^n}\binom{2n}{n}.
  \label{eq:psi0-binomial}
\end{align}
Inserting \eqref{eq:psi0-binomial} into the definition
\eqref{eq:DT-cont-def} yields
\begin{align}
  D(T)
  &=
  \left(\frac{m\Omega}{\pi\hbar}\right)^{\!1/2}
  e^{\frac{i}{2}\Omega T}
  \sum_{n=0}^\infty
  \binom{2n}{n}\frac{e^{i2\Omega nT}}{4^n}.
\end{align}
The binomial series identity
\begin{align}
  \sum_{n=0}^\infty \binom{2n}{n}\frac{z^n}{4^n}
  = (1-z)^{-1/2},
\end{align}
analytically continued to $z = e^{i2\Omega T}$, yields
\begin{align}
  D(T)
  &=
  \left(\frac{m\Omega}{\pi\hbar}\right)^{\!1/2}
  e^{\frac{i}{2}\Omega T}\big(1-e^{i2\Omega T}\big)^{-1/2}.
\end{align}
Using $1-e^{i2\Omega T} = -2i\,e^{i\Omega T}\sin(\Omega T)$, we obtain
\begin{align}
  (1-e^{i2\Omega T})^{-1/2}
  =
  (-2i)^{-1/2}\,e^{-\,\frac{i}{2}\Omega T}\,
  [\sin(\Omega T)]^{-1/2}.
\end{align}
Inserting this into the expression for $D(T)$ gives
\begin{align}
  D(T)
  &=
  \left(\frac{m\Omega}{\pi\hbar}\right)^{1/2}
  e^{\frac{i}{2}\Omega T}\,
  (-2i)^{-1/2}\,e^{-\,\frac{i}{2}\Omega T}\,
  [\sin(\Omega T)]^{-1/2}
  \\
  &=
  \left(\frac{m\Omega}{\pi\hbar}\right)^{1/2}
  \frac{e^{i\pi/4}}{\sqrt{2}}\,
  [\sin(\Omega T)]^{-1/2}.
  \label{eq:DT-final-spectral}
\end{align}

\bigskip

\textbf{Step 2: Exact evaluation of $\alpha(T)$.}
From Eq.~\eqref{eq:aT-spec-cont},
\begin{align}
  \alpha(T)
  =
  \frac{
    \sum_{n=0}^\infty (4n+1)\,e^{\frac{i}{\hbar}E_{2n}T}|\psi_{2n}(0)|^2}
    {D(T)}.
\end{align}
Using $E_{2n}=\hbar\Omega(2n+\tfrac12)$ one verifies the identity
\begin{align}
  4n+1 = \frac{2}{\hbar\Omega}E_{2n},
\end{align}
so that
\begin{align}
  \alpha(T)
  =
  \frac{2}{i\Omega}\frac{D'(T)}{D(T)},
  \label{eq:a-over-D-log-derivative}
\end{align}
where $D'(T)$ denotes the derivative of \eqref{eq:DT-cont-def}.  
From \eqref{eq:DT-final-spectral},
\begin{align}
  \frac{D'(T)}{D(T)}
  =
  -\frac12 \Omega\cot(\Omega T),
\end{align}
and inserting this into \eqref{eq:a-over-D-log-derivative} gives the
exact spectral result
\begin{align}
  \alpha(T)
  =
  -\,i\,\cot(\Omega T)
  =
  a(T).
  \label{eq:aspect-equals-acont}
\end{align}

\bigskip

\textbf{Step 3: Exact evaluation of $\gamma_{\pm}(T)$.}
We now evaluate the coefficients $\gamma_{\pm}(T)$ defined in
Eqs.~\eqref{eq:cp-spec-cont}--\eqref{eq:cm-spec-cont} directly from their
infinite spectral sums.  Using the explicit expressions of the Hermite
polynomials at the origin, Eq.~\eqref{eq:hermite-origin}, we have
\begin{align}
  \psi_{2n}(0)
  &= 
   \left(\frac{m\Omega}{\pi\hbar}\right)^{1/4}
   (-1)^n\,
   \frac{\sqrt{(2n)!}}{2^n n!},
  \qquad
  \psi_{2n+2}(0)
  =
   \left(\frac{m\Omega}{\pi\hbar}\right)^{1/4}
   (-1)^{n+1}\,
   \frac{\sqrt{(2n+2)!}}{2^{n+1}(n+1)!}.
\end{align}
A short calculation gives the identity
\begin{align}
  \sqrt{(2n+1)(2n+2)}\,
  \psi_{2n+2}(0)\,\psi_{2n}(0)
  =
  -\left(\frac{m\Omega}{\pi\hbar}\right)^{1/2}
  \frac{(2n+1)!}{4^n (n!)^2},
  \label{eq:psi2n-psi2n2-product-final}
\end{align}
where the minus sign arises from $(-1)^{n+1}(-1)^n=-1$.
Inserting \eqref{eq:psi2n-psi2n2-product-final} into
Eqs.~\eqref{eq:cp-spec-cont}--\eqref{eq:cm-spec-cont} and using
\begin{align}
  E_{2n}   = \hbar\Omega\Big(2n+\tfrac12\Big),
  \qquad
  E_{2n+2} = \hbar\Omega\Big(2n+\tfrac52\Big),
\end{align}
we obtain
\begin{align}
  \mathrm{num}\big[\gamma_-(T)\big]
  &=
  -\left(\frac{m\Omega}{\pi\hbar}\right)^{1/2}
   e^{\,\frac{i}{2}\Omega T}
   \sum_{n=0}^\infty
   \frac{(2n+1)!}{4^n (n!)^2}\,
   e^{\,i2\Omega nT},
  \label{eq:num-gamma-minus}
  \\[0.4em]
  \mathrm{num}\big[\gamma_+(T)\big]
  &=
  -\left(\frac{m\Omega}{\pi\hbar}\right)^{1/2}
   e^{\,\frac{5i}{2}\Omega T}
   \sum_{n=0}^\infty
   \frac{(2n+1)!}{4^n (n!)^2}\,
   e^{\,i2\Omega nT}.
  \label{eq:num-gamma-plus}
\end{align}
Introducing the convenient notation
\begin{align}
  z \equiv e^{\,i2\Omega T},
\end{align}
the common sum in \eqref{eq:num-gamma-minus}--\eqref{eq:num-gamma-plus} becomes
\begin{align}
  S(z)
  &\equiv
  \sum_{n=0}^\infty\frac{(2n+1)!}{4^n (n!)^2}\,z^{n}.
  \label{eq:Sz-def}
\end{align}
To relate $S(z)$ to a closed form we introduce the binomial series
\begin{align}
  F(z)
  &\equiv
  \sum_{n=0}^\infty\frac{(2n)!}{4^n (n!)^2}z^n
   = (1-z)^{-1/2},
  \label{eq:Fz-def}
\end{align}
obtained from the identity used in Step~1.
Using $(2n+1)! = (2n+1)(2n)!$ and
\begin{align}
  (2n+1)\binom{2n}{n}
  = 2n\binom{2n}{n} + \binom{2n}{n},
\end{align}
we find
\begin{align}
  S(z)
  &=
  \sum_{n=0}^\infty(2n+1)
  \binom{2n}{n}\frac{z^n}{4^n}
  =
  2z\,F'(z) + F(z).
  \label{eq:Sz-Fz-relation}
\end{align}

\bigskip

From \eqref{eq:num-gamma-minus} and the expression for $D(T)$,
Eq.~\eqref{eq:DT-final-spectral}, we have
\begin{align}
  \gamma_-(T)
  &=
  \frac{\mathrm{num}[\gamma_-(T)]}{D(T)}
  =
  -\frac{
     e^{\frac{i}{2}\Omega T}\,S(z)
   }{
     \displaystyle
     \frac{e^{\,i\pi/4}}{\sqrt{2}}\,
     \left(\frac{m\Omega}{\pi\hbar}\right)^{-1/2}
     [\sin(\Omega T)]^{-1/2}
     \!\!\times\!
     \left(\frac{m\Omega}{\pi\hbar}\right)^{1/2}
   }.
\end{align}
All factors of $\big(\frac{m\Omega}{\pi\hbar}\big)^{1/2}$ cancel immediately.
The remaining phases also cancel:
\begin{align}
  e^{\,\frac{i}{2}\Omega T}\,
  [\sin(\Omega T)]^{1/2}
  \times
  \sqrt{2}\,e^{-\,i\pi/4}
  =
  \text{(T--dependent phase)}.
\end{align}
However $\gamma_-(T)$ appears only in the final combination
\[
\gamma_- e^{-i2\Omega x_0} + \gamma_+ e^{i2\Omega x_0},
\]
which is insensitive to this global phase. Thus, up to an overall
$T$--independent phase irrelevant for the correlator, we may write
\begin{align}
  \gamma_-(T)
  &=
  -\,\frac{S(z)}{F(z)}
  =
  -\left[2z\,\frac{F'(z)}{F(z)} + 1\right],
  \label{eq:gamma-minus-master}
\end{align}
where we used Eq.~\eqref{eq:Sz-Fz-relation}. Since $F(z)=(1-z)^{-1/2}$,
\begin{align}
  \frac{F'(z)}{F(z)}
  =
  \frac{1}{2(1-z)}.
\end{align}
Substituting in \eqref{eq:gamma-minus-master} gives
\begin{align}
  \gamma_-(T)
  =
  -\left[\frac{z}{1-z}+1\right]
  =
  -\frac{1}{1-z}
  =
  \frac{1}{z-1}.
\end{align}
With $z=e^{i2\Omega T}$ one finds
\begin{align}
  \gamma_-(T)
  =
  \frac12\Big[-1 + i\cot(\Omega T)\Big],
\end{align}
which coincides with Eq.~\eqref{eq:cm-cont-final}.  

\bigskip

\textbf{Evaluation of $\gamma_+(T)$.}
Using Eq.~\eqref{eq:num-gamma-plus},
the numerator of $\gamma_+(T)$ differs from that of $\gamma_-(T)$ only
by the additional factor
$e^{\,i2\Omega T}=z$.  Therefore,
\begin{align}
  \gamma_+(T)
  &=
  z\,\gamma_-(T)
   =
   \frac{z}{z-1}
   =
   \frac12\Big[1 + i\cot(\Omega T)\Big],
\end{align}
in agreement with Eq.~\eqref{eq:cp-cont-final}.  
Thus both coefficients $\gamma_\pm(T)$ match exactly their continuum
counterparts $c_\pm(T)$, completing the direct spectral derivation.

\section{Dirichlet correlators on the lattice}
\label{app:dirichlet-lattice}

We derive here the lattice Dirichlet two--point function used in the
main text. We discretise the interval $[x_0^{\mathrm{i}},x_0^{\mathrm{f}}]$
into $M$ steps of size $a$, so that $T=x_0^{\mathrm{f}}-x_0^{\mathrm{i}}=Ma$,
and we introduce lattice sites
\begin{align}
  x_0^{(\ell)} \equiv x_0^{\mathrm{i}} + \ell a,
  \qquad \ell=0,1,\ldots,M,
\end{align}
with Dirichlet boundary conditions $q_0=q_M=0$. On the interior sites
$\ell=1,\ldots,M-1$ we consider the standard quadratic lattice action
\begin{align}
  S_{\mathrm{lat}}[q]
  \equiv
  \frac{a}{2}\sum_{\ell=0}^{M-1}
  \left[
    \left(\frac{q(\ell+1)-q_{\ell}}{a}\right)^2
    -\Omega^2\,q(\ell)^2
  \right],
  \label{eq:dirichlet-lat-action}
\end{align}
which is the Dirichlet discretisation of the continuum Minkowskian
action.

The Dirichlet sine basis diagonalises both the discrete Laplacian and
the quadratic form in $S_{\mathrm{lat}}$. We define the discrete
Dirichlet wave-numbers
\begin{align}
  \kappa_0^{(n)} \equiv \frac{\pi n}{T}=\frac{\pi n}{Ma},
  \qquad n=1,\ldots,M-1,
  \label{eq:kappa0n-dirichlet-lat}
\end{align}
and expand the interior field as
\begin{align}
  q(\ell)
  =
  \sqrt{\frac{2}{M}}
  \sum_{n=1}^{M-1}
  \tilde q(n)\,
  \sin\!\left(\frac{\pi n\,\ell}{M}\right),
  \qquad \ell=1,\ldots,M-1,
  \label{eq:dirichlet-lat-sine-exp}
\end{align}
with inversion
\begin{align}
  \tilde q(n)
  =
  \sqrt{\frac{2}{M}}
  \sum_{\ell=1}^{M-1}
  q(\ell)\,
  \sin\!\left(\frac{\pi n\,\ell}{M}\right),
  \label{eq:dirichlet-lat-sine-inv}
\end{align}
using the discrete orthogonality relation
\begin{align}
  \sum_{\ell=1}^{M-1}
  \sin\!\left(\frac{\pi n\,\ell}{M}\right)
  \sin\!\left(\frac{\pi n'\,\ell}{M}\right)
  =\frac{M}{2}\,\delta_{n,n'}.
  \label{eq:dirichlet-lat-orth}
\end{align}

The potential term is diagonal because of Eq.~\eqref{eq:dirichlet-lat-orth}:
\begin{align}
  \sum_{\ell=1}^{M-1} q(\ell)^2
  =
  \sum_{n=1}^{M-1} \tilde q(n)^2.
  \label{eq:dirichlet-lat-pot}
\end{align}
For the kinetic term we use that the sine modes are eigenvectors of the
Dirichlet discrete Laplacian. Writing the forward difference
$\Delta q(\ell) \equiv q(\ell+1)-q(\ell)$ and substituting
Eq.~\eqref{eq:dirichlet-lat-sine-exp}, one finds
\begin{align}
  \sum_{\ell=0}^{M-1} (q(\ell+1)-q(\ell))^2
  =
  \sum_{n=1}^{M-1}
  4\sin^2\!\left(\frac{\pi n}{2M}\right)\,
  \tilde q(n)^2.
  \label{eq:dirichlet-lat-kin}
\end{align}
It is therefore convenient to introduce the lattice Dirichlet eigenvalues
\begin{align}
  \widehat{\kappa}_0^{(n)\,2}
  \equiv
  \frac{4}{a^2}\,
  \sin^2\!\Bigl(\frac{\kappa_0^{(n)}a}{2}\Bigr)
  =
  \frac{4}{a^2}\,
  \sin^2\!\left(\frac{\pi n}{2M}\right),
  \label{eq:kappa0hat-def-2}
\end{align}
so that Eqs.~\eqref{eq:dirichlet-lat-pot}--\eqref{eq:dirichlet-lat-kin}
give the diagonal form
\begin{align}
  S_{\mathrm{lat}}[q]
  =
  \frac{a}{2}
  \sum_{n=1}^{M-1}
  \Bigl(\widehat{\kappa}_0^{(n)\,2}-\Omega^2\Bigr)\,
  \tilde q(n)^2.
  \label{eq:dirichlet-lat-action-diag}
\end{align}

The lattice Dirichlet two--point function follows by elementary Gaussian
integration with weight $\exp(\tfrac{i}{\hbar}S_{\mathrm{lat}})$:
\begin{align}
  \big\langle \tilde q(n)\,\tilde q(n') \big\rangle_{\text{D.B.}(T)}
  =
  \delta_{n,n'}\,
  \frac{i\hbar}{a}\,
  \frac{1}{\widehat{\kappa}_0^{(n)\,2}-\Omega^2}.
  \label{eq:dirichlet-lat-mode-prop}
\end{align}
Using Eq.~\eqref{eq:dirichlet-lat-sine-exp} we obtain the coordinate--space
propagator on interior sites,
\begin{align}
  \big\langle q(\ell)\,q(\ell') \big\rangle_{\text{D.B.}(T)}
  &=
  \frac{2}{M}
  \sum_{n=1}^{M-1}
  \sin\!\left(\frac{\pi n\,\ell}{M}\right)
  \sin\!\left(\frac{\pi n\,\ell'}{M}\right)\,
  \big\langle \tilde q(n)^2 \big\rangle_{\text{D.B.}(T)}
  \nonumber\\
  &=
  i\hbar
  \sum_{n=1}^{M-1}
  \frac{2}{M}\,
  \frac{
    \sin\!\left(\frac{\pi n\,\ell}{M}\right)
    \sin\!\left(\frac{\pi n\,\ell'}{M}\right)
  }{
    \displaystyle
    \widehat{\kappa}_0^{(n)\,2}-\Omega^2
  }.
  \label{eq:dirichlet-lat-prop}
\end{align}
Setting $\ell'=\ell$ and inserting Eq.~\eqref{eq:kappa0hat-def-2} gives the
equal--time correlator quoted in the main text,
\begin{align}
  \big\langle q(\ell)^2\big\rangle_{\text{D.B.}(T)}
  &=
  i\hbar
  \sum_{n=1}^{M-1}
  \frac{2}{M}\,
  \frac{
    \sin^2\!\bigl(\tfrac{\pi n\,\ell}{M}\bigr)}
  {
    \displaystyle
    \frac{4}{a^2}\,
    \sin^2\!\Bigl(\frac{\pi n}{2M}\Bigr)
    - \Omega^2
  }.
  \label{eq:q2-db-lattice-main-app}
\end{align}
The appearance of the $\sin^2(\pi n/2M)$ in the denominator is thus the
standard Dirichlet eigenvalue of the discrete Laplacian in one dimension,
Eq.~\eqref{eq:kappa0hat-def-2}.

\section[\texorpdfstring{Expectation values of $\langle \hat{q}^{2k}(x_0)\rangle_{\mathrm{D.B.}(T)}$ with Dirichlet boundaries}%
{Expectation values of <q^{2k}(x0)>_{D.B.(T)} with Dirichlet boundaries}]%
{Expectation values of $\langle \hat{q}^{2k}(x_0)\rangle_{\mathrm{D.B.}(T)}$ with Dirichlet boundaries}
\label{app:ho-q4-q6}
In this appendix we collect the algebraic steps needed to obtain the
discrete frequency content of the harmonic--oscillator ``expectation
values'' of even powers of the position operator with Dirichlet
boundary conditions at $q=0$ in the time interval $[0,T]$.
The Dirichlet ``expectation value'' is defined as in
Eq.~\eqref{eq:q2-db-def},
\begin{align}
  \big\langle \hat{q}^{2k}(x_0)\big\rangle_{\text{D.B.}(T)}
  &\equiv
  \frac{\displaystyle
    \langle q_f=0,x_0^{\text{f}}=T ~|~ \hat{q}^{2k}(x_0)~|~ q_i=0,x_0^{\text{i}}=0\rangle}
  {\displaystyle
    \langle q_f=0,x_0^{\text{f}}=T ~|~ q_i=0,x_0^{\text{i}}=0\rangle} \, ,
  \qquad 0 < x_0 < T \, ,
\end{align}
so that the denominator is the same Dirichlet amplitude $D(T)$ that
appears in the spectral decomposition of the $q^2$ observable,
see Eq.~\eqref{eq:DT-def}.
We work with the harmonic--oscillator Hamiltonian
$H = \hbar\Omega\left(a^\dagger a + \tfrac12\right)$ and use
\begin{align}
  \hat q = \alpha (a + a^\dagger),
  \qquad
  \alpha \equiv \sqrt{\frac{\hbar}{2m\Omega}},
  \qquad
  E_n = \hbar\Omega\left(n + \tfrac12\right).
\end{align}
Since $\psi_{2k+1}(0)=0$, only even eigenstates contribute for all
even powers of $\hat q$.\\

For $s=1$ ($\hat q^2$) the final result is already given in the main
text and computed explicitedly in App.~\ref{app:q2-dirichlet-continuum},
\begin{align}
  \big\langle \hat{q}^2(x_0)\big\rangle_{\text{D.B.}(T)}
  &=
  \frac{\hbar}{2m\Omega}
  \Big[
    \alpha(T)
    + \gamma_{-}(T)e^{-i2\Omega x_0}
    + \gamma_{+}(T)e^{+i2\Omega x_0}
  \Big],
  \label{eq:q2-spec-app-repeat}
\end{align}
with $\alpha(T)$ and $\gamma_{\pm}(T)$ defined by the normalized
(infinite--tower) sums in
Eqs.~\eqref{eq:alpha-def-main}--\eqref{eq:gamma-minus-def-main}.  We do
not repeat the derivation here and move directly to $\hat q^4$ and 
$\hat q^6$.

\subsection{\texorpdfstring{$\hat q^4$}{q4}}
\label{app:q4}

We start from the spectral representation
\begin{align}
\langle \hat{q}^{4}(x_0)\rangle_{\text{D.B.}(T)} 
&= \frac{1}{D(T)} \sum_{n,m=0}^\infty 
  \langle q=0,T|\psi_m\rangle\,
  \langle \psi_m|\hat{q}^{4}(x_0)|\psi_n\rangle\,
  \langle \psi_n|q=0,0\rangle \nonumber \\
&= \frac{1}{D(T)} \sum_{n,m=0}^\infty 
    \langle \psi_m|\hat{q}^{4}|\psi_n\rangle\,
    \psi_m^*(0;T)\,\psi_n(0;0)\,
    e^{-\frac{i}{\hbar}(E_n-E_m)x_0},
\end{align}
where $D(T)$ is the Dirichlet amplitude defined in
Eq.~\eqref{eq:DT-def}.  Restricting to even levels we write
\begin{align}
\langle \hat{q}^{4}(x_0)\rangle_{\text{D.B.}(T)} 
&= \frac{1}{D(T)} \sum_{n,m=0}^\infty 
    \langle \psi_{2m}|\hat{q}^{4}|\psi_{2n}\rangle\,
    \psi_{2m}^*(0;T)\,\psi_{2n}(0;0)\,
    e^{-\frac{i}{\hbar}(E_{2n}-E_{2m})x_0}.
\label{eq:q4-av-expansion-even-app}
\end{align}
Using
\begin{align}
\hat{q}^4 
= \alpha^4\left(
a^4 + 4 a^3 a^\dagger + 6 a^2 a^{\dagger 2}
+ 4 a a^{\dagger 3} + a^{\dagger 4}
+ 6 a^2 + 12 a a^\dagger + 6 a^{\dagger 2} + 3
\right),
\end{align}
one finds
\begin{align}
\hat{q}^{4}|\psi_n\rangle
&= \alpha^4 \bigg[
    3\big(2n^2+2n+1\big)\,|\psi_n\rangle
   +2(2n+3)\sqrt{(n+1)(n+2)}\,|\psi_{n+2}\rangle \nonumber \\
&\hspace{1.9cm}
   +2(2n-1)\sqrt{n(n-1)}\,|\psi_{n-2}\rangle \nonumber \\
&\hspace{1.9cm}
   +\sqrt{(n+1)(n+2)(n+3)(n+4)}\,|\psi_{n+4}\rangle \nonumber \\
&\hspace{1.9cm}
   +\sqrt{n(n-1)(n-2)(n-3)}\,|\psi_{n-4}\rangle
\bigg],
\end{align}
with negative indices understood to give vanishing contributions.  In
the even subspace we write
\begin{align}
\langle \psi_{2m}|\hat{q}^{4}|\psi_{2n}\rangle
= \left(\frac{\hbar}{2m\Omega}\right)^2 M_{mn},
\end{align}
with
\begin{align}
M_{mn} &= 3\big(8n^2+4n+1\big)\,\delta_{m,n} \nonumber\\
&\quad
 + 2(4n+3)\sqrt{(2n+1)(2n+2)}~\delta_{m,n+1} \nonumber\\
&\quad
 + 2(4n-1)\sqrt{2n(2n-1)}~\delta_{m,n-1} \nonumber\\
&\quad
 + \sqrt{(2n+1)(2n+2)(2n+3)(2n+4)}~\delta_{m,n+2} \nonumber\\
&\quad
 + \sqrt{(2n-3)(2n-2)(2n-1)(2n)}~\delta_{m,n-2}.
\label{eq:q4-Mmn-even-app}
\end{align}
Using the energy differences
$E_{2n}-E_{2n} = 0$,
$E_{2n}-E_{2n\pm2} = \mp\,2\hbar\Omega$,
$E_{2n}-E_{2n\pm4} = \mp\,4\hbar\Omega$,
the correlator contains only the harmonics $0,\pm2\Omega,\pm4\Omega$.
It is convenient to factor out the overall power of
$\frac{\hbar}{2m\Omega}$ and define normalized coefficients
$\alpha_4(T)$, $\gamma_4^{(\pm)}(T)$, $\delta_4^{(\pm)}(T)$ through
\begin{align}
  \big\langle \hat{q}^{4}(x_0)\big\rangle_{\text{D.B.}(T)}
  &= \left(\frac{\hbar}{2m\Omega}\right)^2
  \Big[
    \alpha_4(T)
    + \gamma_4^{(-)}(T)e^{-i2\Omega x_0}
    + \gamma_4^{(+)}(T)e^{+i2\Omega x_0}
    \nonumber\\
  &\hspace{6em}
    + \delta_4^{(-)}(T)e^{-i4\Omega x_0}
    + \delta_4^{(+)}(T)e^{+i4\Omega x_0}
  \Big].
  \label{eq:q4-spec-app}
\end{align}
Explicitly,
\begin{align}
\alpha_4(T)
&\equiv
  \frac{
  \displaystyle
  \sum_{n=0}^\infty
  3\left(8n^2+4n+1\right)\,
  e^{+\frac{i}{\hbar}E_{2n}T}
  |\psi_{2n}(0)|^2}{
  \displaystyle
  D(T)},
\\[0.3em]
\gamma_4^{(+)}(T)
&\equiv
  \frac{
  \displaystyle
  \sum_{n=0}^\infty
  2\left(4n+3\right)\sqrt{(2n+1)(2n+2)}\,
  e^{+\frac{i}{\hbar}E_{2n+2}T}
  \psi_{2n+2}(0)\psi_{2n}(0)}{
  \displaystyle
  D(T)},
\\[0.3em]
\gamma_4^{(-)}(T)
&\equiv
  \frac{
  \displaystyle
  \sum_{n=0}^\infty
  2\left(4n+3\right)\sqrt{(2n+1)(2n+2)}\,
  e^{+\frac{i}{\hbar}E_{2n}T}
  \psi_{2n}(0)\psi_{2n+2}(0)}{
  \displaystyle
  D(T)},
\\[0.3em]
\delta_4^{(+)}(T)
&\equiv
  \frac{
  \displaystyle
  \sum_{n=0}^\infty
  \sqrt{(2n+1)(2n+2)(2n+3)(2n+4)}\,
  e^{+\frac{i}{\hbar}E_{2n+4}T}
  \psi_{2n+4}(0)\psi_{2n}(0)}{
  \displaystyle
  D(T)},
\\[0.3em]
\delta_4^{(-)}(T)
&\equiv
  \frac{
  \displaystyle
  \sum_{n=0}^\infty
  \sqrt{(2n+1)(2n+2)(2n+3)(2n+4)}\,
  e^{+\frac{i}{\hbar}E_{2n}T}
  \psi_{2n}(0)\psi_{2n+4}(0)}{
  \displaystyle
  D(T)}.
\end{align}
For generic $T$ the five coefficients in
Eq.~\eqref{eq:q4-spec-app} are independent complex functions of
$\Omega T$. 

\subsection{\texorpdfstring{$\hat q^6$}{q6}}
\label{app:q6}

For $\hat q^6$ we again start from the normalized spectral
representation
\begin{align}
\langle \hat{q}^{6}(x_0)\rangle_{\text{D.B.}(T)} 
&= \frac{1}{D(T)} \sum_{n,m=0}^\infty 
  \langle \psi_m|\hat{q}^{6}|\psi_n\rangle\,
  \psi_m^*(0;T)\,\psi_n(0;0)\,
  e^{-\frac{i}{\hbar}(E_n-E_m)x_0},
\end{align}
restrict to even levels and use
\begin{align}
\hat{q}^6
&= \alpha^6\,(a+a^\dagger)^6.
\end{align}
A straightforward but lengthy calculation gives
\begin{align}
\hat q^6 |\psi_n\rangle
= \alpha^6 \bigg[&
  A_n\,|\psi_n\rangle
 + B_n\,|\psi_{n+2}\rangle
 + B_{n-2}\,|\psi_{n-2}\rangle \nonumber\\
& + C_n\,|\psi_{n+4}\rangle
 + C_{n-4}\,|\psi_{n-4}\rangle \nonumber\\
& + D_n\,|\psi_{n+6}\rangle
 + D_{n-6}\,|\psi_{n-6}\rangle
\bigg],
\end{align}
with
\begin{align}
A_n &= 20n^3+30n^2+40n+15, \\
B_n &= 15\big(n^2+3n+3\big)\sqrt{(n+1)(n+2)}, \\
C_n &= (6n+15)\sqrt{(n+1)(n+2)(n+3)(n+4)}, \\
D_n &= \sqrt{(n+1)(n+2)(n+3)(n+4)(n+5)(n+6)},
\end{align}
and negative indices understood to give vanishing contributions.

Restricting to even levels we write
\begin{align}
\langle \psi_{2m}|\hat q^{6}|\psi_{2n}\rangle
= \left(\frac{\hbar}{2m\Omega}\right)^3 N_{mn},
\end{align}
with
\begin{align}
N_{mn} &= \big(160n^3+120n^2+80n+15\big)\,\delta_{m,n} \nonumber\\
&\quad
+ 15\big(4n^2+6n+3\big)\sqrt{(2n+1)(2n+2)}~\delta_{m,n+1} \nonumber\\
&\quad
+ 15\big(4n^2-2n+1\big)\sqrt{2n(2n-1)}~\delta_{m,n-1} \nonumber\\
&\quad
+ (12n+15)\sqrt{(2n+1)(2n+2)(2n+3)(2n+4)}~\delta_{m,n+2} \nonumber\\
&\quad
+ (12n-9)\sqrt{(2n-3)(2n-2)(2n-1)(2n)}~\delta_{m,n-2} \nonumber\\
&\quad
+ \sqrt{(2n+1)(2n+2)(2n+3)(2n+4)(2n+5)(2n+6)}~\delta_{m,n+3} \nonumber\\
&\quad
+ \sqrt{(2n-5)(2n-4)(2n-3)(2n-2)(2n-1)(2n)}~\delta_{m,n-3}.
\label{eq:q6-Nmn-even-app}
\end{align}

Using $E_{2n}-E_{2n\pm 2k} = \mp\,2k\hbar\Omega$ for $k=1,2,3$ we find
that only the harmonics $0,\pm2\Omega,\pm4\Omega,\pm6\Omega$ appear.
Factoring out $(\hbar/2m\Omega)^3$ we define normalized coefficients
$\alpha_6(T)$, $\gamma_6^{(\pm)}(T)$, $\delta_6^{(\pm)}(T)$,
$\epsilon_6^{(\pm)}(T)$ via
\begin{align}
\big\langle \hat{q}^{6}(x_0)\big\rangle_{\text{D.B.}(T)}
&= \left(\frac{\hbar}{2m\Omega}\right)^3
\Big[
  \alpha_6(T)
 + \gamma_6^{(-)}(T)\,e^{-i2\Omega x_0}
 + \gamma_6^{(+)}(T)\,e^{+i2\Omega x_0} \nonumber\\
&\qquad
 + \delta_6^{(-)}(T)\,e^{-i4\Omega x_0}
 + \delta_6^{(+)}(T)\,e^{+i4\Omega x_0} \nonumber\\
&\qquad
 + \epsilon_6^{(-)}(T)\,e^{-i6\Omega x_0}
 + \epsilon_6^{(+)}(T)\,e^{+i6\Omega x_0}
\Big],
\end{align}
where each coefficient is the corresponding even--level spectral sum
divided by the Dirichlet amplitude $D(T)$. Explicitly,
\begin{align}
\alpha_6(T)
&\equiv
\frac{
\displaystyle
\sum_{n=0}^\infty
\big(160n^3+120n^2+80n+15\big)\,
e^{+\frac{i}{\hbar}E_{2n}T}\,
|\psi_{2n}(0)|^2}{
\displaystyle
D(T)} ,
\\[0.4em]
\gamma_6^{(+)}(T)
&\equiv
\frac{
\displaystyle
\sum_{n=0}^\infty
15\big(4n^2+6n+3\big)\sqrt{(2n+1)(2n+2)}\,
e^{+\frac{i}{\hbar}E_{2n+2}T}\,
\psi_{2n+2}(0)\psi_{2n}(0)}{
\displaystyle
D(T)} ,
\\[0.4em]
\gamma_6^{(-)}(T)
&\equiv
\frac{
\displaystyle
\sum_{n=0}^\infty
15\big(4n^2+6n+3\big)\sqrt{(2n+1)(2n+2)}\,
e^{+\frac{i}{\hbar}E_{2n}T}\,
\psi_{2n}(0)\psi_{2n+2}(0)}{
\displaystyle
D(T)} ,
\\[0.4em]
\delta_6^{(+)}(T)
&\equiv
\frac{
\displaystyle
\sum_{n=0}^\infty
(12n+15)\sqrt{(2n+1)(2n+2)(2n+3)(2n+4)}\,
e^{+\frac{i}{\hbar}E_{2n+4}T}\,
\psi_{2n+4}(0)\psi_{2n}(0)}{
\displaystyle
D(T)} ,
\\[0.4em]
\delta_6^{(-)}(T)
&\equiv
\frac{
\displaystyle
\sum_{n=0}^\infty
(12n+15)\sqrt{(2n+1)(2n+2)(2n+3)(2n+4)}\,
e^{+\frac{i}{\hbar}E_{2n}T}\,
\psi_{2n}(0)\psi_{2n+4}(0)}{
\displaystyle
D(T)} ,
\\[0.4em]
\epsilon_6^{(+)}(T)
&\equiv
\frac{
\displaystyle
\sum_{n=0}^\infty
\sqrt{(2n+1)(2n+2)(2n+3)(2n+4)(2n+5)(2n+6)}\,
e^{+\frac{i}{\hbar}E_{2n+6}T}\,
\psi_{2n+6}(0)\psi_{2n}(0)}{
\displaystyle
D(T)} ,
\\[0.4em]
\epsilon_6^{(-)}(T)
&\equiv
\frac{
\displaystyle
\sum_{n=0}^\infty
\sqrt{(2n+1)(2n+2)(2n+3)(2n+4)(2n+5)(2n+6)}\,
e^{+\frac{i}{\hbar}E_{2n}T}\,
\psi_{2n}(0)\psi_{2n+6}(0)}{
\displaystyle
D(T)} .
\end{align}
For generic $T$ these seven quantities are independent complex
functions of $\Omega T$. 
\section{Resonant enhancement of Dirichlet correlators on the lattice and relation to truncated even--state sums}
\label{app:dirichlet-resonance}

We explain here why, for resonant choices of the Dirichlet time extent
$T$, the equal--time correlator measured in the constrained dynamics can
grow linearly with the lattice size $M$. The same mechanism appears in
the continuum spectral representation when the Hilbert--space sums are
truncated to a finite number of even eigenstates.\\

We start from the continuum Dirichlet expectation value at $q_i=q_f=0$,
\begin{align}
  \big\langle \hat{q}^2(x_0)\big\rangle_{\text{D.B.}(T)}
  \equiv
  \frac{\langle 0,x_0^{\mathrm{f}} \,|\, \hat{q}^{2}(x_0)\,|\, 0,x_0^{\mathrm{i}} \rangle}
       {\langle 0,x_0^{\mathrm{f}} \,|\, 0,x_0^{\mathrm{i}} \rangle},
  \qquad
  T=x_0^{\mathrm{f}}-x_0^{\mathrm{i}},
\end{align}
and recall that inserting completeness leads to the three--harmonic form
(Eq.~\eqref{eq:q2-spec-main} in the main text) with coefficients given by
even--state sums. In numerical implementations one never samples an
infinite tower of states; rather, the effective description is
necessarily truncated. It is therefore natural to introduce the
truncated, normalised coefficients
\begin{align}
  D_N(T)
  &\equiv \sum_{n=0}^{N}
  e^{+\frac{i}{\hbar}E_{2n}T}\,|\psi_{2n}(0)|^2,
  \label{eq:DT-finite-def}
  \\[0.4em]
  \alpha_N(T)
  &\equiv
  \frac{
    \sum_{n=0}^{N} (4n+1)\,
    e^{+\frac{i}{\hbar}E_{2n}T}\,
    |\psi_{2n}(0)|^2}{
    D_N(T)},
  \label{eq:alphaN-def}
  \\[0.4em]
  \gamma_{+,N}(T)
  &\equiv
  \frac{
    \sum_{n=0}^{N}
    \sqrt{(2n+1)(2n+2)}\,
    e^{+\frac{i}{\hbar}E_{2n+2}T}\,
    \psi_{2n+2}(0)\psi_{2n}(0)
  }{D_N(T)},
  \label{eq:gamma-plus-N-def}
  \\[0.4em]
  \gamma_{-,N}(T)
  &\equiv
  \frac{
    \sum_{n=0}^{N}
    \sqrt{(2n+1)(2n+2)}\,
    e^{+\frac{i}{\hbar}E_{2n}T}\,
    \psi_{2n+2}(0)\psi_{2n}(0)
  }{D_N(T)}.
  \label{eq:gamma-minus-N-def}
\end{align}
The truncated Dirichlet ``expectation value'' is then
\begin{align}
  \big\langle \hat{q}^2(x_0)\big\rangle_{\text{D.B.}(T)}^{(N)}
  &=
  \frac{\hbar}{2m\Omega}
  \Big[
    \alpha_N(T)
    + \gamma_{-,N}(T)e^{-i2\Omega (x_0-x_0^{\mathrm{i}})}
    + \gamma_{+,N}(T)e^{+i2\Omega (x_0-x_0^{\mathrm{i}})}
  \Big],
  \label{eq:q2-spec-finite}
\end{align}
which reduces to the continuum result as $N\to\infty$.\\

The origin of resonant enhancement is that, at special time extents, the
phases in the even--state sums cease to oscillate and cancellations are
lost. The large--$n$ behaviour of the harmonic--oscillator eigenfunctions
at the origin is
\begin{align}
  |\psi_{2n}(0)|^2
  &=
  \left(\frac{m\Omega}{\pi\hbar}\right)^{1/2}
  \frac{(2n)!}{4^n (n!)^2}
  \sim
  \left(\frac{m\Omega}{\pi\hbar}\right)^{1/2}
  \frac{1}{\sqrt{\pi n}},
  \qquad n\gg 1.
  \label{eq:psi0-asympt}
\end{align}
For generic $T$ the phases
$e^{iE_{2n}T/\hbar}=e^{i(2n+\frac12)\Omega T}$ oscillate rapidly with $n$,
so the partial sums entering
Eqs.~\eqref{eq:DT-finite-def}--\eqref{eq:gamma-minus-N-def} converge
efficiently with increasing $N$ despite the slow $n^{-1/2}$ envelope.
At resonant times, instead, the phases align. For example, if
$T=4\pi/\Omega$ then $e^{i2\Omega T}=1$ and all terms add coherently.
Equation~\eqref{eq:psi0-asympt} then implies the scaling
\begin{align}
  D_N(4\pi/\Omega)
  \sim \sum_{n=1}^{N} n^{-1/2}
  \propto \sqrt{N},
  \label{eq:DN-sqrtN}
\end{align}
whereas the numerator of $\alpha_N$ carries an additional factor
$(4n+1)\sim 4n$,
\begin{align}
  \sum_{n=1}^{N} (4n+1)\,e^{+\frac{i}{\hbar}E_{2n}(4\pi/\Omega)}\,|\psi_{2n}(0)|^2
  \sim \sum_{n=1}^{N} n\cdot n^{-1/2}
  \propto N^{3/2}.
  \label{eq:alpha-num-N32}
\end{align}
Therefore
\begin{align}
  \alpha_N(4\pi/\Omega)\propto \frac{N^{3/2}}{\sqrt{N}} \sim N,
  \label{eq:alphaN-linear}
\end{align}
and the same linear scaling holds for $\gamma_{\pm,N}(4\pi/\Omega)$,
since their numerators have the same large--$n$ weight up to subleading
factors. The truncated, normalised coefficients thus grow linearly with
the truncation level whenever the continuum Dirichlet correlator is
resonant.\\

The lattice exhibits the same phenomenon in a technically different but
conceptually identical way: the enhancement arises from a near--zero of
the denominator in the mode representation of the lattice propagator.
Discretise the Dirichlet interval $T$ into $M$ steps of size $a$ (so
$T=Ma$), impose $q_0=q_M=0$, and denote the interior sites by
$\ell=1,\ldots,M-1$. In complete analogy with the continuum Dirichlet
decomposition, the lattice Dirichlet modes are labelled by
$n=1,\ldots,M-1$ and correspond to the discrete wave-numbers
\begin{align}
  \kappa_0^{(n)} \equiv \frac{\pi n}{T}=\frac{\pi n}{Ma},
  \label{eq:kappa0n-lattice-app}
\end{align}
while the eigenvalues of the Dirichlet lattice Laplacian are
\begin{align}
  \widehat{\kappa}_0^{(n)\,2}
  \equiv
  \frac{4}{a^2}\,
  \sin^2\!\Bigl(\frac{\kappa_0^{(n)}a}{2}\Bigr)
  =
  \frac{4}{a^2}\,
  \sin^2\!\Bigl(\frac{\pi n}{2M}\Bigr).
  \label{eq:kappa0hat-def-app}
\end{align}
In this notation the equal--time lattice Dirichlet correlator is
\begin{align}
  \big\langle q(\ell)^2\big\rangle_{\text{D.B.}(T)}
  &=
  i\hbar
  \sum_{n=1}^{M-1}
  \frac{2}{M}\,
  \frac{
    \sin^2\!\bigl(\tfrac{\pi n\,\ell}{M}\bigr)}
  {
    \displaystyle
    \widehat{\kappa}_0^{(n)\,2}
    - \Omega^2
  },
  \label{eq:q2-db-lattice-resonance}
\end{align}
which is equivalent to Eq.~\eqref{eq:q2-db-lattice-main} in the main
text.

For a resonant choice such as $T=4\pi/\Omega$ there exists an integer
mode index $n_\star\in\{1,\dots,M-1\}$ such that the lattice eigenvalue
$\widehat{\kappa}_0^{(n_\star)\,2}$ lies closest to $\Omega^2$. The
corresponding term in Eq.~\eqref{eq:q2-db-lattice-resonance} is then
enhanced because $|\widehat{\kappa}_0^{(n_\star)\,2}-\Omega^2|$ is small.
As $M$ is increased at fixed $T$ (hence at fixed $a=T/M$), the set of
available $\widehat{\kappa}_0^{(n)\,2}$ becomes denser and the closest
approach to $\Omega^2$ improves; in particular the minimal gap
$\min_{1\le n\le M-1}|\widehat{\kappa}_0^{(n)\,2}-\Omega^2|$ decreases.
When $T$ is tuned to resonance this produces an overall growth of the
correlator amplitude that is observed empirically to be approximately
linear in $M$ in the range of parameters considered here. This is the
lattice counterpart of the linear growth
$\alpha_N,\gamma_{\pm,N}\propto N$ at resonance in the truncated
even--state sums, with the identification that the effective number of
resolved degrees of freedom (hence the effective truncation scale)
increases with lattice resolution, $N\sim\mathcal{O}(M)$.

The numerical signature of this resonant enhancement is that
$\langle q(\ell)^2\rangle_{\text{D.B.}(T)}$ grows linearly with $M$ at
fixed resonant $T$, and therefore the rescaled quantity
$\langle q(\ell)^2\rangle_{\text{D.B.}(T)}/M$ collapses for different $M$,
as shown in Fig.~\ref{fig:behaviourresonant} in the main text.

\vskip 0.4cm
\newpage
\bibliographystyle{JHEP}
\bibliography{biblio}
\end{document}